\newcommand*\figpath{./}

\documentclass[12pt,a4]{article}

\usepackage{lmodern,sansmath}
\usepackage{textcomp}
\usepackage{microtype}
\usepackage{amsmath, amssymb, amsthm, dsfont, bm}         
\usepackage{amsfonts}         
\usepackage{array}

\usepackage[sort]{natbib}
\usepackage[colorlinks=true,linkcolor=red,citecolor=blue]{hyperref}

\usepackage{mathtools}
\newcolumntype{M}[1]{>{\raggedright}m{#1}}
\newcolumntype{C}[1]{>{\centering\arraybackslash}m{#1}}

\usepackage{color}
\usepackage[dvipsnames,svgnames]{xcolor}

\usepackage{booktabs}
\usepackage{graphicx}

\usepackage{caption}
\usepackage{subcaption}
\usepackage{multirow}
\usepackage{todonotes}

\renewcommand{\vec}[1]{\boldsymbol{#1}}

\newcommand {\R} {{\mathds R}}
\newcommand{\N}{{\mathds N}}

\newcommand{\ie}{{\em i.\thinspace{}e. }}

\newcommand{\eg}{{\em e.\thinspace{}g. }}

\newcommand{\ds}{\displaystyle}
\newcommand{\pl}{\partial}

\newcommand{\hb}{\hbar}

\newcommand{\TF} {{\hbox{\tiny TF}}}

\newcommand{\trap}{{\rm trap}}

\newcommand{\ff}{FreeFem{\small +$\!$+}  }

\newcommand{\C}{\ensuremath{\mathbb C}}

\usepackage{braket}

\usepackage{xcolor}
\usepackage{textcomp}
\usepackage{listings}
\makeatletter
\renewcommand*{\eqref}[1]{%
	\hyperref[{#1}]{\textup{\tagform@{\ref*{#1}}}}%
}
\makeatother

\begin{document}




\title{A finite element toolbox for the Bogoliubov-de Gennes stability analysis of Bose-Einstein condensates}


%
%
%
%
%
%
%
%

\author{Georges Sadaka$^1$, Victor Kalt$^1$, Ionut Danaila$^{1, *}$ and Fr{\'e}d{\'e}ric Hecht$^2$
	\\ \\
	$^1$Univ Rouen Normandie, CNRS,  \\
	Laboratoire de Math{\'e}matiques Rapha{\"e}l Salem,\\
	UMR 6085, F-76000 Rouen, France\\
	\\ 
		$^2$Sorbonne Universit{\'e}, CNRS,\\
		Laboratoire Jacques-Louis Lions,\\
		 UMR 7598, F-75005, Paris, France.
	\\ 
	$^*$ Corresponding author: ionut.danaila@univ-rouen.fr
}

\date{\today}
\maketitle

\begin{abstract}
We present a finite element toolbox for the computation of Bogoliubov-de Gennes  modes used to assess the linear stability of stationary solutions of  the Gross-Pitaevskii (GP) equation. Applications concern one (single GP equation) or two-component (a system of coupled GP equations) Bose-Einstein condensates in one, two and three dimensions of space.  An implementation using the free software \ff is distributed with this paper. 
For the computation of the GP stationary (complex or real) solutions we use a Newton algorithm coupled with a continuation method exploring the parameter space  (the chemical potential or the interaction constant).
Bogoliubov-de Gennes equations are then solved using dedicated libraries for the associated eigenvalue problem.  Mesh adaptivity is proved to considerably reduce the computational time for cases implying complex vortex states. Programs are validated through comparisons with known theoretical results for simple cases and numerical results reported in the literature.
\end{abstract}





\noindent
{\bf Programm summary}\\
{\em Program Title:}  {FFEM\_BdG\_toolbox.zip}\\
{\em Catalogue identifier:}\\ 
{\em Program summary URL:}\\
{\em Program obtainable from:}\\
{\em Licensing provisions:} \\
{\em No. of lines in distributed program, including test data, etc.: }  35 296 \\
{\em No. of bytes in distributed program, including test data, etc.: } 125 303\\
{\em Distribution format:} .zip\\
{\em Programming language:} \ff (v 4.12) free software (www.freefem.org)\\
{\em Computer:} PC, Mac, Super-computer.\\
{\em Operating system:} Mac OS, Linux, Windows.\\
{\em Nature of problem:} The software computes  Bogoliubov-de Gennes (BdG) complex modes of Bose-Einstein condensates described by the Gross-Pitaevskii (GP) equation. BdG equations are obtained by linearizing the GP equation (or the system of coupled GP equations) around a stationary solution. Obtained BdG modes are used to assess on the stability of stationary states.\\
{\em Solution method:} 
Stationary states of the GP equation are obtained by a Newton algorithm.  Parameter space is explored using a continuation on the chemical potential. Once the stationary (complex or real) state is captured accurately, BdG modes are computed by solving the associated eigenvalue problem with the ARPACK library. Complex eigenvalues and eigenvectors are computed and stored. The wave function is discretized by P2 (piece-wise quadratic) Galerkin triangular (in 2D) or tetrahedral (in 3D) finite elements.
Mesh adaptation is implemented to reduce the computational time. 
Examples are given for stationary states in one- and two-component Bose-Einstein condensates.\\
{\em Running time:} From seconds to hours depending on the mesh resolution and space dimension.\\



\section{Introduction}

Since their first observation \citep{anderson1995observation,davis1995bose}, Bose-Einstein condensates have become a powerful experimental framework for the study of waves and excitations in superfluids and nonlinear systems. The study of wave related structures (solitons, vortices) and their stability is an active area of research, and many efforts have been devoted to the developments of new experimental techniques for the creation and the study of new excited states. We can mention the use of rotation  \citep{BEC-physV-2000-Madison-a,BEC-physV-2001-haljan}, imprinting techniques manipulating the phase of the wave function  \citep{BEC-physV-2002-imprint,becker2008oscillations}, counterflows techniques \citep{yan2011multiple}, the use of anisotropic potentials  \citep{theocharis2010multiple} or multicomponent BEC settings  \citep{wang2017two}. A large variety of wave or vortex-related states could be thus obtained. Basic examples refer to single vortex lines (with I-, U- or S- shape)   in rotating BEC \citep{dan-2003-aft}, vortex rings and one-dimensional solitons. More complex states with multiple vortex rings, vortex stars, hopfions and solitons can be created \citep{dan-2004-cras,bisset2015robust,wang2017single}. In multicomponent BECs, dark-bright \citep{charalampidis2020bifurcation} and dark-antidark \citep{dan-2016-PRA} states can be also  obtained.

The main application of the programs presented in this paper is the study of the stability of such solitary waves or vortex states that are 
theoretically or numerically found as stationary solutions to  the Gross-Pitaevskii (GP) equation \citep{gross1961structure,pitaevskii1961vortex}.
The linearization of the GP equation around a given stationary solution results in the Bogoliubov-de Gennes (BdG) system of equations \citep{pitaevskii2003bose}. Solving the BdG eigenvalue problem provides linear modes, for which their  stability could be studied.  The present toolbox thus contains two distinct parts: the computation of stationary solution of the GP equation (or a system of coupled GP equations) and the computation of complex eigenvalues and modes for the associated BdG system.

Concerning the computation of stationary states of the GP equation, a large variety of discretization methods were suggested in the open literature: spectral methods \citep{BEC-CPC-2007-dion-cances,BEC-CPC-2013-Caliari,BEC-CPC-2014-antoine-duboscq}, finite-elements \citep{BEC-CPC-2016-FEM,dan-2016-CPC} or finite-differences \citep{BEC-CPC-2009-Muruganandam,BEC-CPC-2012-Vudragovic,BEC-CPC-2013-Caplan,BEC-CPC-2014-simplectic,BEC-CPC-2014-Hohenester,BEC-CPC-2019-rotating}. Programs written in Fortran \citep{BEC-CPC-2007-dion-cances,BEC-CPC-2009-Muruganandam}, C \citep{BEC-CPC-2012-Vudragovic,BEC-CPC-2013-Caplan}, Matlab \citep{BEC-CPC-2013-Caliari,BEC-CPC-2013-Caplan,BEC-CPC-2014-antoine-duboscq,BEC-CPC-2014-Hohenester}, FreeFem++ \citep{dan-2016-CPC} or C and Fortran with OpenMP \citep{BEC-CPC-2019-rotating} have been shared.
All these works use a common approach to the problem, which is to find the stationary states of the GP equation under the constraint of the total mass conservation (\ie the L2-norm of the wave function is fixed). When studying the stability of excitations in the BdG framework, another approach is commonly adopted: the chemical potential is used as a convenient parameter to explore all possible states (and bifurcations) and thus the total mass varies from one state to another. This second approach has been already applied with finite elements \citep{dan-2016-PRA,boulle2020deflation,charalampidis2020bifurcation,carreterogonzalez2016vortex}, finite differences \citep{charalampidis2020bifurcation,wang2017single,bisset2015robust} or spectral methods \citep{wang2017single} and will be used in this work. 
To solve the BdG eigenvalue problem, specialized libraries are generally used: \texttt{ARPACK} \citep{lehoucq1998arpack}, \texttt{SLEPc} \citep{hernandez2005slepc} or \texttt{FEAST} \citep{polizzi2009density}. Such libraries offer the flexibility to be easily interfaced with different types of discretization, since only final matrices of the eigenvalue problem are needed. 
A mathematical study of the BdG equation with numerical comparisons between spectral and finite difference discretizations has been recently reported in \cite{gao2020numerical}. 

To the best of our knowledge, FACt \citep[Fluctuations in Atomic Condensates,][]{arko2020fact} is the only publicly accessible code for the BdG problem. It considers thermal excitations of the BECs at non-zero temperatures by solving the BdG equation in two component BECs with a pseudo-spectral method. The present \ff toolbox uses a different (PDE-oriented) formulation of the problem and has the advantage to hide all technicalities related to the implementation of the finite-element method and the interface with eigenvalue libraries (such as ARPACK) \citep{hecht-2012-JNM}. The user can thus focus on the physical and mathematical model, and eventually on the numerical algorithm solving the problem. The high level programming language offered in \ff  and the syntax close to mathematical formulations make the implementation of numerical algorithms very easy. Finite elements algorithms were successfully used to solve the GP equation \citep{dan-2010-JCP,dan-2016-CPC} or the BdG problem \citep{dan-2016-PRA} and recently to identify vortices in a quantum field \citep{dan-2023-CPC-postproc}. Another advantage of the present toolbox is to use mesh adaptivity to reduce the mesh size and the computational time. Solving the BdG problem for complex cases becomes thus possible using personal computers.

The structure of the paper is as follows. In Sect. \ref{sec-theory}, we introduce the GP and BdG models. Sections \ref{sec-num-meth-gp} and \ref{sec-num-meth-bdg} describe the numerical methods used for the computation the stationary states and BdG modes. We present various benchmarks used for the validation of our codes in Sects. \ref{sec-valid1c} and \ref{sec-valid2c}. The architecture of the programs and a description of parameter and output files are given in Sect. \ref{sec-desc-prog}. Finally, we summarize the main features of the toolbox and present some possible extensions in Sect. \ref{sec-conclusions}.


\section{The Gross-Pitaevskii model and Bogoliubov-de Gennes equations}\label{sec-theory}

\subsection{The Gross-Pitaevskii equation} \label{sec-GP}

In the zero temperature limit, the Gross-Pitaevskii equation describes the time-evolution of the complex-valued macroscopic wave function $\psi:{\cal D} \rightarrow \C$, with ${\cal D} \in \R^d$ the domain of the $d$-dimensional condensate ($d=1, 2, 3$):
\begin{equation}
\label{eq-GP}
i\hb\frac{\partial \psi}{\partial t} = -\frac{\hb^2}{2m} \nabla^2 \psi + V_\trap \psi + g |\psi|^2 \psi,
\end{equation}
where $V_\text{trap}({\vec x})$ is the external trapping potential, $\hb$ the reduced Planck constant and $m$ the atomic mass. The nonlinear term models the interaction between  atoms, and for $d=3$ (3D condensate) $g = \frac{4\pi\hbar^2 a_s}{m}$, with $a_s$ the scattering length. For $d=1$ or $2$ (1D or 2D condensates), the nonlinear interaction constant $g$ is specified according to the dimension reduction \citep{frantzeskakis2010dark,bao2013mathematical}. We consider the case of harmonic trapping potentials:
\begin{equation}
	\label{eq-GP-V}
	V_\trap(\vec{x})=\frac{m}{2}\left(
	\omega_{x}^2 {x}^2 + \omega_y^2 {y}^2 + \omega_z^2 {z}^2 \right),
\end{equation}
where $\omega_{x}, \omega_{x}, \omega_{x}$ are  trapping frequencies.
The atomic density $n({\vec x})=|\psi({\vec x})|^2$ vanishes outside the condensate due to the trapping, which implies that homogeneous Dirichlet boundary conditions should be imposed for the wave function ($\psi=0$ on $\partial \cal D$). The corresponding GP energy is:
\begin{equation}\label{eq-NRJ}
{\mathcal E}(\psi) = \int_{\cal D} \left(\frac{\hb^2}{2m}|\nabla\psi({\vec x},t)|^2 + V_\trap({\vec x})|\psi({\vec x},t)|^2 + \frac{g}{2}|\psi({\vec x},t)|^4 \right)d{\vec x},
\end{equation}
and  the total number of atoms:
\begin{equation}
\label{eq-GP-N}
N(\psi)=\int_{\cal D} \psi\overline{ \psi}\, d{\vec x}=\int_{\cal D} |\psi|^2 d{\vec x},
\end{equation}
where $\overline{\psi}$ denotes the complex conjugate.

Stationary solutions to the GP equation \eqref{eq-GP} are obtained by imposing the form 
\begin{equation}
	\label{eq-GP-phi}
	\psi({\vec x},t) = \phi({\vec x})e^{-\frac{i}{\hb}\mu t},
\end{equation}
with $\mu$ the chemical potential. The stationary wave function $\phi$ is then solution of the stationary GP equation:
\begin{equation}
\label{eq-GP-stat}
 -\frac{\hb^2}{2m} \nabla^2  \phi + V_\trap  \phi + g |\phi|^2 \phi = \mu \phi,
\end{equation}
which is a nonlinear eigenvalue problem. Note that from \eqref{eq-GP-phi} we infer that $|\psi|^2=|\phi|^2$  and thus $N(\phi)=N(\psi)$ and ${\mathcal E}(\psi)={\mathcal E}(\phi)$. 
The chemical potential is related to the number of atoms by the relation:
\begin{equation}
	\label{eq-GP-mu}
\mu = \frac{1}{N(\phi)}\left( \mathcal E(\phi) + \frac{g}{2}\int_{\cal D} |\phi|^4 d\vec{x}\right).
\end{equation}

In this work, we compute stationary solutions for fixed values of the chemical potential $\mu$. Branches of solutions are followed by continuation on $\mu$. To catch such branches, two limits associated to the value of $\mu$ can be considered as initial condition. 
In the case of low density (corresponding to a small number of particles), the nonlinear term in \eqref{eq-GP} can be neglected to obtain the linear GP equation. In the case of an harmonic potential \eqref{eq-GP-V}, $\psi$ can then be described as an eigenstate of the quantum harmonic oscillator. Using a separation of variables, these eigenstates can be written as a product of Hermite and Laguerre polynomials or spherical harmonics, depending on the dimension and the coordinate system. As an example, for the 2D BEC with trapping potential \eqref{eq-GP-V} with $\omega_x = \omega_y = \omega_\perp$, the eigenstates formulated in Cartesian coordinates are $\ket{k,l} \propto H_k(\sqrt{\omega_\perp}x)H_l(\sqrt{\omega_\perp}y)e^{-\frac{1}{2}\omega_\perp(x^2+y^2)}$, where $H_k, H_l$ are Hermite polynomials; $k$ and $l$ index the eigenstates and correspond to the number of cuts in the condensate along the $x$ and $y$ axes respectively. Solutions in the linear limit  corresponding to various exited states of the condensate have been analysed in many studies  \citep[\eg][]{dan-2004-cras,boulle2020deflation}.

The other limit is the Thomas-Fermi limit, associated to large values of $\mu$. In this case, the kinetic energy  becomes negligible when compared to the nonlinear term. The stationary GP equation \eqref{eq-GP-stat} reduces to:
\begin{equation}
	\mu\phi = V_\trap \phi + g|\phi|^2\phi,
\end{equation}
which gives an analytical expression for the atomic density:
\begin{equation}\label{eq-TF}
	n_\TF = |\psi_\TF|^2 = \frac{1}{g}(\mu - V_\trap)_+.
\end{equation}

\subsection{The Bogoliubov-de Gennes equation}\label{sec-BdG}

The Bogoliubov-de Gennes model is based on the linearisation of  \eqref{eq-GP} assuming that:
\begin{equation}
	\psi(\vec{x},t) =(\phi(\vec{x}) + \delta \phi(\vec{x},t))e^{-\frac{i}{\hb}\mu t},
	\label{eq-BdG-psi}
\end{equation}
where  $\phi(\vec{x})$ is a stationary state satisfying Eq. \eqref{eq-GP-stat} and $\delta \phi$ a small perturbation. Inserting \eqref{eq-BdG-psi} in  \eqref{eq-GP}, we obtain, after  neglecting second order terms in $\delta \phi$, an evolution equation for the perturbation $\delta \phi$:
\begin{equation}
		\label{eq-BdG-dpsi}
i\hb\frac{\pl \delta\phi}{\pl t} = {\mathcal H}\delta\phi    -\mu  \delta\phi + 2 g|\phi|^2\delta\phi + g\phi^2\overline{\delta\phi},
\end{equation}
where we denoted by ${\mathcal H} \equiv  -\frac{\hb^2}{2m} \nabla^2  + V_\trap$ the linear part of the Hamiltonian.
Considering perturbations of the form 
\begin{equation}
		\label{eq-BdG-dphi}
	\delta \phi (\vec{x},t) = A(\vec{x}) e^{-i\omega t} + \overline{B}(\vec{x}) e^{i\overline{\omega}t},
\end{equation}
we obtain, after separating terms in $e^{-i\omega t}$ and $e^{i\bar{\omega}t}$, the Bogoliubov-de Gennes (BdG) system of equations \citep{pitaevskii2003bose,castin-BdG}:
\begin{equation}
	\label{eq-BdG}
\begin{pmatrix}
{\mathcal H} -\mu + 2 g|\phi|^2 & g\phi^2\\
-g{\overline{\phi}}^2 & -({\mathcal H} -\mu + 2 g|\phi|^2)\\
\end{pmatrix}
\begin{pmatrix} A \\ B \end{pmatrix} = 
\hb\omega \begin{pmatrix} A \\ B \end{pmatrix}.
\end{equation}
Note that the BdG equation \eqref{eq-BdG} is a linear eigenvalue problem, since $\phi$ is fixed and ${\mathcal H}$  is a linear real operator. 
The present toolbox computes, for a given complex stationary state $\phi \in \C$, solutions $(\omega, A, B)$ to the BdG equation \eqref{eq-BdG}, with $\omega$ denoting complex eigenvalues and $(A, B)$ complex eigenvectors. 

The following properties of the BdG eigenvalue problem can be obtained by elementary manipulations and will be useful to check the accuracy of calculations:
\begin{enumerate}
	\item If $(\omega, A, B)$ is solution to  \eqref{eq-BdG}, then $(-\overline{\omega}, \overline{B}, \overline{A})$ is also a solution. This property is obtained by taking the conjugate of \eqref{eq-BdG}.
	\item $\omega=0$ is always an eigenvalue (the zero-energy mode). It can be checked that the full solution is $(0, \alpha \phi, -\alpha  \overline{\phi}), \alpha \in \C$ \citep{pitaevskii2003bose} and  represents following \eqref{eq-BdG-dphi} a time invariant (small) excitation $\delta \phi=(\alpha - \overline{\alpha}) \phi$. This corresponds in \eqref{eq-BdG-psi} to a gauge transformation and, consequently, it does not add any physical excitation to the system.
	\item $\overline{\omega}$ is also an eigenvalue.  If $\phi \in \R$, it is easy to see from \eqref{eq-BdG} that $(\overline{\omega},\overline{A},\overline{B})$ is also a solution. For the general case of $\phi \in \C$ this property also holds and it can be proved using the Hamiltonian nature of the problem \citep{castin-BdG}. 
	\item If we multiply the first equation of the system \eqref{eq-BdG} by $\overline{A}$ and the second by $\overline{B}$, integrate over the domain $\cal D$ and then sum the two equations, we obtain that:
	\begin{equation}
\delta\mathcal{E} = \hb\omega \int_{\cal D} \left(|A|^2 - |B|^2\right) d\vec{x} \in \R, 
	\label{eq-BdG-norm}
	\end{equation}
	which is generally presented in the literature in the equivalent form  \citep{BEC-book-2008-panos,pitaevskii2003bose,gao2020numerical}:
		\begin{equation}
	(\omega- \overline{\omega}) \int_{\cal D} \left(|A|^2 - |B|^2\right) d\vec{x} =0. 
		\label{eq-BdG-norm2}
	\end{equation} 
\end{enumerate}

From \eqref{eq-BdG-norm} or \eqref{eq-BdG-norm2} we can draw two main observations that are important to interpret the results of the  BdG analysis:
\begin{enumerate}
	\item  If the BdG modes are normalized such that $ \int_{\cal D} \left(|A|^2 - |B|^2\right) d\vec{x}  \neq 0$, then  we infer from  \eqref{eq-BdG-norm2} that only real eigenvalues are possible. These correspond to elementary excitations. A mathematical study of the properties of the BdG modes when $\omega$ and $\phi$ are real is offered in \cite{gao2020numerical}. Moreover, the quantity $\delta\mathcal{E}$ in \eqref{eq-BdG-norm}  represents  the energy difference between the stationary $\psi$ and the perturbed state $\psi+\delta \psi$ \citep{pitaevskii2003bose}. The Krein signature $K$ was introduced  as  the sign of the energy difference, $K = sign(\delta \mathcal{E})$ \citep{chernyavsky2018krein}. If $K>0$ for all modes, then $\phi$ is the global minimum of the energy, \ie the ground state. On the contrary, if there exists a mode with $K < 0$, then the excitation reduces the energy of the system and the stationary state is thus energetically unstable, \ie excited state (or local minimum of the energy).
	
	\item If $ \int_{\cal D} \left(|A|^2 - |B|^2\right) d\vec{x}  =0$, complex eigenvalues $\omega=\omega_r+i \omega_i$ are possible. If $\omega_i \ne 0$, then the BdG mode is dynamically unstable.
	
\end{enumerate}


\subsection{Two-component BECs}

Mixtures of BECs have been experimentally created, either by considering different atomic species or by using hyperfine states of a single isotope. We consider two-component BECs described by the following system of two coupled GP equations for  wave functions $\psi_1$ and $\psi_2$:
\begin{equation}
	\label{eq-GP2c}
\begin{dcases}
i\hb\frac{\partial \psi_1}{\partial t} = \left(-\frac{\hb^2}{2m} \nabla^2 + V_\trap  + g_{11} |\psi_1|^2  + g_{12} |\psi_2|^2\right) \psi_1,\\
i\hb\frac{\partial \psi_2}{\partial t} = \left(-\frac{\hb^2}{2m} \nabla^2 + V_\trap  + g_{21} |\psi_1|^2 + g_{22} |\psi_2|^2\right) \psi_2.\\
\end{dcases}
\end{equation}
Coefficients $g_{11}$ and $g_{22}$ represent interactions between atoms of the same species while $g_{12}$ and $g_{21}$ describe interactions between different species. The total energy is the sum of the GP energy of each component:
\begin{equation}
	\label{eq-GP2c-energ} 
	\ds {\cal E}(\psi_1,\psi_2) = \ds \int_{\cal D} \sum_{i=1}^2 \left(\frac{\hbar^2}{2m} |\nabla \psi_i|^2 + V_\trap\, |\psi_i|^2 +
	\frac{1}{2}\sum_{j=1}^2  g_{ij}  |\psi_i|^2 |\psi_j|^2 \right)  d{\vec x}. 
\end{equation}
Similarly to the one component case, stationary states are sought as  $\psi_1 = \phi_1 e^{-\frac{i}{\hb}\mu_1 t}$ and $\psi_2 = \phi_2 e^{-\frac{i}{\hb}\mu_2 t}$, with chemical potentials $\mu_1$ and $\mu_2$.
We obtain the following system of equations:
\begin{equation}
	\label{eq-GP2c-stat}
\begin{dcases}
\mu_1\phi_1= \left(-\frac{\hb^2}{2m} \nabla^2 + V_\trap + g_{11} |\phi_1|^2  + g_{12} |\phi_2|^2\right) \phi_1,\\
\mu_2\phi_2 = \left(-\frac{\hb^2}{2m} \nabla^2 + V_\trap + g_{21} |\phi_1|^2+ g_{22} |\phi_2|^2\right)  \phi_2.
\end{dcases}
\end{equation}
For the linear stability analysis we consider perturbations of the form:
\begin{equation}
		\label{eq-GP2c-phi}
\begin{cases}
\delta \phi_1 (\vec{x}) &=A(\vec{x}) e^{-i\omega t} + \overline{B}(\vec{x}) e^{i\overline{\omega}t},\\
\delta \phi_2 (\vec{x}) &= C(\vec{x}) e^{-i\omega t} + \overline{D}(\vec{x}) e^{i\overline{\omega}t},
\end{cases}
\end{equation}
and obtain the BdG equations for the two-component case:
\begin{equation}
	\label{eq-BdG2c-M}
M\begin{pmatrix} A \\ B \\C \\D \end{pmatrix} = 
\hb\omega \begin{pmatrix} A \\ B \\C \\D \end{pmatrix},
\end{equation}
where the matrix $M$ can be presented in the form:
\begin{equation}
M = 
\begin{pmatrix}
M_{11} & g_{11}\phi_1^2 & g_{12}\phi_1\overline{\phi_2} & g_{12}\phi_1\phi_2\\
-g_{11}\overline{\phi_1}^2 & M_{22} & -g_{12}\overline{\phi_1}\overline{\phi_2} & -g_{12}\overline{\phi_1}\phi_2\\
g_{21}\overline{\phi_1}\phi_2 & g_{21}\phi_1\phi_2 & M_{33} & g_{22}\phi_2^2\\
-g_{21}\overline{\phi_1}\overline{\phi_2} & -g_{21}\phi_1\overline{\phi_2} & -g_{22}\overline{\phi_2}^2 & M_{44}
\end{pmatrix},
\end{equation}
with:
\begin{equation}
		\label{eq-BdG2c-M1}
		\begin{cases}
	M_{11} &= {\mathcal H} - \mu_1 + 2g_{11}|\phi_1|^2 + g_{12}|\phi_2|^2,\\ 
	M_{22} &=  -M_{11},\\
	M_{33} &= {\mathcal H} - \mu_2 + g_{21}|\phi_1|^2 + 2g_{22}|\phi_2|^2,\\
	M_{44} &=  -M_{33}.
\end{cases}
\end{equation}

\subsection{Scaling}

Various forms of scaling are used in the literature (see \cite{dan-2023-Singapore} for a unified form of the GP scaling). 
To allow one to  switch between different forms, we first introduce a reference (trapping) frequency $\omega_s$ which will define  a time scale $t_s$ and a length scale $x_s$ (the corresponding harmonic oscillator length):
\begin{equation}
	\label{eq-scal-tsxs}
	t_s=\frac{1}{\omega_s}, \quad x_s=\sqrt{\frac{\hb}{m \omega_s}}.
\end{equation}
We then introduce a reference value $\psi_s$ for the wave function and scale variables as: 
\begin{equation}
	\label{eq-scal-gen}
	{\vec{x}} \rightarrow  \frac{\vec x}{x_s}, \quad {t} \rightarrow \frac{t}{t_s}, \quad {\psi} \rightarrow \frac{\psi}{\psi_s}.
\end{equation}
The same scaling is used for the stationary state $\phi$. The dimensionless form of the time-dependent stationary GP equation \eqref{eq-GP} becomes:
\begin{equation}
	\label{eq-scal-GP}
	i \frac{\partial \psi}{\partial t}=-\frac{1}{2} \nabla^2  \psi + C_\trap  \psi + \beta |\psi|^2  \psi,
\end{equation}
with
\begin{equation}
	\label{eq-scal-Coeff}
	C_\trap(\vec{x})= \frac{1}{\hb \omega_s}\, V_\trap(\vec{x}), \quad \beta = \frac{g \psi_s^2}{\hb \omega_s}.
\end{equation}
Note that the coefficient $1/2$ in front of the Laplacian in Eq. \eqref{eq-scal-GP} comes from the choice  \eqref{eq-scal-tsxs} for the reference length, since all terms were divided by the quantity (of the dimension of an energy):
\begin{equation}
	\label{eq-scal-homega}
\hb \omega_s = m x_s^2 \omega_s^2 = \frac{\hb^2}{m x_s^2}.
\end{equation}
From \eqref{eq-GP-V} we infer that the non-dimensional trapping potential takes the form:
\begin{equation}
	\label{eq-scal-GP-V}
	C_\trap(x,y,z)=\frac{1}{2}\left(
	\omega_{x}^2 {x}^2 + \omega_y^2 {y}^2 + \omega_z^2 {z}^2 \right), \,\, \text{where}\,\, \omega_{x,y,z} \rightarrow \frac{\omega_{x,y,z}}{\omega_s}.
\end{equation}
Similarly, the stationary GP equation \eqref{eq-GP-stat} becomes:
\begin{equation}
	\label{eq-scal-GP-stat}
	 -\frac{1}{2} \nabla^2  \phi + C_\trap   \phi + \beta |\phi|^2  = \mu  \phi,\,\, \text{where}\,\,   {\mu} \rightarrow  \frac{\mu}{\hb \omega_s}.
\end{equation}
Finally, the BdG system of equations takes the non-dimensional form:
\begin{equation}
	\label{eq-scal-BdG}
\begin{pmatrix}
	{\mathcal H} -\mu + 2 \beta|\phi|^2 & g\phi^2\\
	-g{\overline{\phi}}^2 & -({\mathcal H} -\mu + 2 \beta|\phi|^2)\\
\end{pmatrix}
	\begin{pmatrix} A \\ B \end{pmatrix} = 
	\omega \begin{pmatrix} A \\ B \end{pmatrix},
\end{equation}
where ${\mathcal H} \equiv  -\frac{1}{2} \nabla^2  + C_\trap$ is dimensionless,  $\omega \rightarrow \omega/\omega_s$ and $A \rightarrow A/\psi_s$, $B \rightarrow B/\psi_s$.

 In the two component case, the wave functions are scaled as ${\psi_1} \rightarrow  \frac{\psi_1}{\psi_{s,1}}$ and ${\psi_2} \rightarrow  \frac{\psi_2}{\psi_{s,2}}$, and the system \eqref{eq-GP2c-stat} for the stationary state becomes:
\begin{equation}
	\label{eq-scal-GP2c-stat}
\begin{dcases}
	\mu_1\phi_1= \left(-\frac{1}{2} \nabla^2 + C_\trap + \beta_{11} |\phi_1|^2  + \beta_{12} |\phi_2|^2\right) \phi_1,\\
	\mu_2\phi_2 = \left(-\frac{1}{2} \nabla^2 + C_\trap + \beta_{21} |\phi_1|^2+ \beta_{22} |\phi_2|^2\right)  \phi_2,
\end{dcases}
\end{equation}
where $\beta_{ij} \rightarrow \frac{g_{ij}}{\hb\omega_s}\psi_{s,j}^2$, $\mu_i \rightarrow \frac{\mu_{i}}{\hb\omega_s}$. It follows that the non-dimensional form of the  BdG system for the two-component BEC is obtained from \eqref{eq-BdG2c-M}-\eqref{eq-BdG2c-M1}, by replacing coefficients $g_{ij}$ with $\beta_{ij}$ and using the non-dimensional form of ${\mathcal H}$.

\section{Computing stationary solutions of the GP equation}\label{sec-num-meth-gp}

\subsection{Newton method for a single component BEC}

Stationary solutions of Eq. \eqref{eq-scal-GP-stat} are computed using a Newton method. Considering that $\phi = \phi_r+i\phi_i$, we obtain the following system of equations after separating real and imaginary parts:
\begin{equation}
	\label{eq-num-GP-stat}
\begin{dcases}
-\frac{1}{2} \nabla^2 \phi_r + C_\trap \phi_r + \beta f(\phi_r,\phi_i)\phi_r -\mu\phi_r &=0,\\
-\frac{1}{2} \nabla^2 \phi_i + C_\trap \phi_i + \beta f(\phi_r,\phi_i)\phi_i - \mu\phi_i &=0,
\end{dcases}
\end{equation}
where $f(\phi_r,\phi_i)$ implements the non-linear (interaction) term. Here $f(\phi_r,\phi_i)=|\phi|^2=\phi_r^2+\phi_i^2$, but the method is described (and programmed) for a general expression of $f$ that can be easily changed in the toolbox.

We consider homogeneous Dirichlet boundary conditions for $\phi_r$ and $\phi_i$, \ie $\phi_r=\phi_i=0$ on $\pl \mathcal{D}$, and set the classical Sobolev spaces $V=H^1_0(\mathcal{D})$ for $\phi_r$ and $\phi_i$. The weak formulation of \eqref{eq-num-GP-stat} can be thus written as follows: find $(\phi_r,\phi_i) \in V\times V=V^2$, such that for all test functions $(v_r,v_i) \in V^2$:
\begin{equation}
	\label{eq-num-GP-stat-weak}
\begin{dcases}
\begin{aligned}
\mathcal F_r(\phi_r,\phi_i,v_r) &= \int_{\mathcal D} (C_\trap - \mu)\phi_r v_r +\int_{\mathcal D}\frac{1}{2}\nabla \phi_r\cdot\nabla v_r + \int_{\mathcal D} \beta f(\phi_r,\phi_i)\phi_r v_r&=0,\\
\mathcal F_i(\phi_r,\phi_i, v_i) &= \int_{\mathcal D}(C_\trap - \mu)\phi_i v_i +\int_{\mathcal D}\frac{1}{2}\nabla \phi_i\cdot\nabla v_i + \int_{\mathcal D}\beta f(\phi_r,\phi_i)\phi_i v_i&=0.
\end{aligned}
\end{dcases}
\end{equation}

Starting from an initial guess $(\phi_r^{0},\phi_i^{0})$, solution increments
\begin{equation}
	q = \phi_r^k-\phi_r^{k+1},\quad s = \phi_i^k-\phi_i^{k+1}, \quad k \geq 0,
\end{equation}
are computed using the Newton algorithm:
\begin{eqnarray}
\label{eq-num-GP-stat-diff-sys}
\begin{pmatrix}
\left(\dfrac{\pl \mathcal F_r}{\pl \phi_r}\right)_{\phi_r=\phi_r^k,\phi_i = \phi_i^k} & \left(\dfrac{\pl \mathcal F_r}{\pl \phi_i}\right)_{\phi_r=\phi_r^k,\phi_i = \phi_i^k} \\ \left(\dfrac{\pl \mathcal F_i}{\pl \phi_r}\right)_{\phi_r=\phi_r^k,\phi_i = \phi_i^k} & \left(\dfrac{\pl \mathcal F_i}{\pl \phi_i}\right)_{\phi_r=\phi_r^k,\phi_i = \phi_i^k}
\end{pmatrix}
\begin{pmatrix}
q \\ s
\end{pmatrix}
= \begin{pmatrix}
\mathcal F_r(\phi_r^k,\phi_i^k, v_r)\\
\mathcal F_i(\phi_r^k,\phi_i^k, v_i)
\end{pmatrix},
\end{eqnarray}
with the corresponding weak formulation:
{\small
\begin{equation}\label{eq-num-GP-stat-Newton}
\begin{dcases}
\begin{aligned}
&\int_{\mathcal D}(C_\trap - \mu) q v_r+\int_{\mathcal D}\frac{1}{2}\nabla  q\cdot \nabla v_r + \int_{\mathcal D} \beta\left(\dfrac{\pl f}{\pl\phi_r}(\phi_r^k,\phi_i^k)\,\phi_r^k q + \dfrac{\pl f}{\pl\phi_i}(\phi_r^k,\phi_i^k)\, \phi_r^k s + f(\phi_r^k,\phi_i^k) q\right) v_r\\
&\hspace{1cm}=\int_{\mathcal D}(C_\trap - \mu)\phi_r^k v_r + \int_{\mathcal D}\frac{1}{2}\nabla \phi_{r}^k\cdot \nabla v_r + \int_{\mathcal D} \beta f(\phi_r^k,\phi_i^k)\phi_r^k v_r,\\
&\int_{\mathcal D}(C_\trap - \mu) s v_i+\int_{\mathcal D}\frac{1}{2}\nabla  s\cdot\nabla v_i + \int_{\mathcal D} \beta\left(\dfrac{\pl f}{\pl\phi_r}(\phi_r^k,\phi_i^k)\,\phi_i^k q + \dfrac{\pl f}{\pl\phi_i}(\phi_r^k,\phi_i^k)\,  \phi_i^k s + f(\phi_r^k,\phi_i^k) s\right) v_i\\
&\hspace{1cm} =\int_{\mathcal D}(C_\trap-\mu) \phi_i^k v_i
+\int_{\mathcal D}\frac{1}{2}\nabla \phi_i^k\cdot\nabla v_i + \int_{\mathcal D} \beta f(\phi_r^k,\phi_i^k)\phi_i^k v_i.
\end{aligned}
\end{dcases}
\end{equation}
}Note that the metalanguage used in \ff enables the implementation of Eqs. \eqref{eq-num-GP-stat-Newton} in a form very similar to mathematical formulae, which is appreciable to rapidly build bug-free numerical codes.

\subsection{Newton method for a two-component BEC}

For a two-component BEC, we solve the following system, obtained from \eqref{eq-scal-GP2c-stat} after separating real and imaginary parts for $\phi_1 = \phi_{1r}+i\phi_{1i}$ and $\phi_{2} = \phi_{2r} + i\phi_{2i}$:
\begin{equation}
	\label{eq-num-GP2c-stat}
\begin{dcases}
- \frac{1}{2} \nabla^2 \phi_{1r} + (C_\trap-\mu_1) \phi_{1r}
+ \beta_{11} f(\phi_{1r},\phi_{1i})\phi_{1r} + \beta_{12} f(\phi_{2r},\phi_{2i})\phi_{1r}&= 0,\\
- \frac{1}{2} \nabla^2 \phi_{1i} + (C_\trap-\mu_1) \phi_{1i} + \beta_{11}f(\phi_{1r},\phi_{1i})\phi_{1i} + \beta_{12} f(\phi_{2r},\phi_{2i})\phi_{1i}&=0,\\
- \frac{1}{2} \nabla^2 \phi_{2r} + (C_\trap-\mu_2) \phi_{2r}
+ \beta_{21} f(\phi_{1r},\phi_{1i})\phi_{2r} + \beta_{22} f(\phi_{2r},\phi_{2i})\phi_{2r} &=0,\\
- \frac{1}{2} \nabla^2 \phi_{2i} + (C_\trap-\mu_2) \phi_{2i} + \beta_{21} f(\phi_{1r},\phi_{1i})\phi_{2i} + \beta_{22} f(\phi_{2r},\phi_{2i})\phi_{2i} &=0.
\end{dcases}
\end{equation}
We consider again homogeneous Dirichlet boundary conditions, \ie $\phi_{1r}=\phi_{1i} = \phi_{2r} = \phi_{2i}=0$ on $\pl \mathcal{D}$. The weak formulation of \eqref{eq-num-GP2c-stat} can be written as follows: find $(\phi_{1r},\phi_{1i},\phi_{2r},\phi_{2i}) \in V^4$, such that for all test functions $ (v_{1r}, v_{1i}, v_{2r}, v_{2i}) \in V^4$:
{\small
\begin{equation}
		\label{eq-num-GP2c-stat-weak}
\begin{dcases}
\begin{aligned}
\mathcal F_{1r} = &\int_{\mathcal D}(C_\trap - \mu_1)\phi_{1r} v_{1r} +\int_{\mathcal D}\frac{1}{2} \nabla \phi_{1r}\cdot\nabla v_{1r} + \int_{\mathcal D}(\beta_{11} f(\phi_{1r},\phi_{1i})+\beta_{12} f(\phi_{2r},\phi_{2i}))\phi_{1r} v_{1r} &= 0,\\
\mathcal F_{1i} = &\int_{\mathcal D}(C_\trap - \mu_1)\phi_{1i} v_{1i} + \int_{\mathcal D}\frac{1}{2} \nabla \phi_{1i}\cdot\nabla v_{1i} + \int_{\mathcal D}(\beta_{11}f(\phi_{1r},\phi_{1i})+\beta_{12} f(\phi_{2r},\phi_{2i}))\phi_{1i} v_{1i} &= 0,\\
\mathcal F_{2r} = &\int_{\mathcal D}(C_\trap -\mu_2)\phi_{2r} v_{2r} + \int_{\mathcal D}\frac{1}{2} \nabla \phi_{2r}\cdot\nabla v_{2r} + \int_{\mathcal D}(\beta_{21} f(\phi_{1r},\phi_{1i}) + \beta_{22} f(\phi_{2r},\phi_{2i}))\phi_{2r} v_{2r} &= 0,\\
\mathcal F_{2i} = &\int_{\mathcal D}(C_\trap-\mu_2)\phi_{2i} v_{2i} + \int_{\mathcal D}\frac{1}{2} \nabla \phi_{2i}\cdot\nabla v_{2i} + \int_{\mathcal D}(\beta_{21} f(\phi_{1r},\phi_{1i}) + \beta_{22} f(\phi_{2r},\phi_{2i}))\phi_{2i} v_{2i} &= 0.\\
\end{aligned}
\end{dcases}
\end{equation}
}
The Newton step for increments
\begin{equation}
	q_1 = \phi_{1r}^k-\phi_{1r}^{k+1}, \quad s_1 = \phi_{1i}^k-\phi_{1i}^{k+1}, \quad q_2 = \phi_{2r}^k-\phi_{2r}^{k+1}, \quad s_2 = \phi_{2i}^k-\phi_{2i}^{k+1},
\end{equation}
consists then in solving the following four equations:
{\small
\begin{equation}
	\label{eq-num-GP2c-stat-Newton-1}
\begin{aligned}
& \int_{\mathcal D}(C_\trap-\mu_1)q_1  v_{1r} + \int_{\mathcal D} \frac{1}{2} \nabla q_1\cdot \nabla  v_{1r} + \int_{\mathcal D} (\beta_{11} f(\phi_{1r}^k,\phi_{1i}^k) + \beta_{12} f(\phi_{2r}^k,\phi_{2i}^k))q_1v_{1r}\\
&\hspace{1cm}+ \int_{\mathcal D} \beta_{11}\left(\dfrac{\pl f}{\pl\phi_r}(\phi_{1r}^k,\phi_{1i}^k)\phi_{1r}^k q_1 + \dfrac{\pl f}{\pl\phi_i}(\phi_{1r}^k,\phi_{1i}^k)\phi_{1r}^k s_1\right) v_{1r}\\
&\hspace{1cm}+ \int_{\mathcal D} \beta_{12}\left(\dfrac{\pl f}{\pl\phi_r}(\phi_{2r}^k,\phi_{2i}^k)\phi_{1r}^k q_2 + \dfrac{\pl f}{\pl\phi_i}(\phi_{2r}^k,\phi_{2i}^k)\phi_{1r}^k s_2\right) v_{1r}\\
&\hspace{.5cm}= \int_{\mathcal D}(C_\trap - \mu_1)\phi_{1r}^k v_{1r} +\int_{\mathcal D}\frac{1}{2} \nabla \phi_{1r}^k\cdot\nabla v_{1r} + \int_{\mathcal D}(\beta_{11} f(\phi_{1r}^k,\phi_{1i}^k)+\beta_{12} f(\phi_{2r}^k,\phi_{2i}^k))\phi_{1r}^k v_{1r},
\end{aligned}
\end{equation}
\begin{equation}
	\label{eq-num-GP2c-stat-Newton-2}
\begin{aligned}
& \int_{\mathcal D}(C_\trap-\mu_1)s_1 v_{1i} + \int_{\mathcal D}\frac{1}{2} \nabla s_1\cdot\nabla v_{1i} + \int_{\mathcal D} (\beta_{11} f(\phi_{1r}^k,\phi_{1i}^k) + \beta_{12} f(\phi_{2r}^k,\phi_{2i}^k))s_1v_{1i}\\
&\hspace{1cm}+ \int_{\mathcal D} \beta_{11}\left(\dfrac{\pl f}{\pl\phi_r}(\phi_{1r}^k,\phi_{1i}^k)\phi_{1i}^k q_1 + \dfrac{\pl f}{\pl\phi_i}(\phi_{1r}^k,\phi_{1i}^k)\phi_{1i}^k s_1\right) v_{1i}\\
&\hspace{1cm}+ \int_{\mathcal D} \beta_{12}\left(\dfrac{\pl f}{\pl\phi_r}(\phi_{2r}^k,\phi_{2i}^k)\phi_{1i}^k q_2 + \dfrac{\pl f}{\pl\phi_i}(\phi_{2r}^k,\phi_{2i}^k)\phi_{1i}^k s_2\right) v_{1i}\\
&\hspace{.5cm} = \int_{\mathcal D}(C_\trap - \mu_1)\phi_{1i}^k v_{1i} + \int_{\mathcal D}\frac{1}{2} \nabla \phi_{1i}^k\cdot\nabla v_{1i} + \int_{\mathcal D}(\beta_{11}f(\phi_{1r}^k,\phi_{1i}^k)+\beta_{12} f(\phi_{2r}^k,\phi_{2i}^k))\phi_{1i}^k v_{1i},
\end{aligned}
\end{equation}
\begin{equation}
	\label{eq-num-GP2c-stat-Newton-3}
\begin{aligned}
& \int_{\mathcal D}(C_\trap-\mu_2)q_2 v_{2r} +\int_{\mathcal D} \frac{1}{2} \nabla q_2\cdot\nabla v_{2r} + \int_{\mathcal D} (\beta_{22} f(\phi_{2r}^k,\phi_{2i}^k) + \beta_{21} f(\phi_{1r}^k,\phi_{1i}^k))q_2v_{2r}\\
&\hspace{1cm}+ \int_{\mathcal D} \beta_{21}\left(\dfrac{\pl f}{\pl\phi_r}(\phi_{1r}^k,\phi_{1i}^k)\phi_{2r}^k q_1 + \dfrac{\pl f}{\pl\phi_i}(\phi_{1r}^k,\phi_{1i}^k)\phi_{2r}^k s_1\right) v_{2r}\\
&\hspace{1cm}+ \int_{\mathcal D} \beta_{22}\left(\dfrac{\pl f}{\pl\phi_r}(\phi_{2r}^k,\phi_{2i}^k)\phi_{2r}^k q_2 + \dfrac{\pl f}{\pl\phi_i}(\phi_{2r}^k,\phi_{2i}^k)\phi_{2r}^k s_2\right) v_{2r}\\
&\hspace{.5cm} = \int_{\mathcal D}(C_\trap -\mu_2)\phi_{2r}^k v_{2r} + \int_{\mathcal D}\frac{1}{2} \nabla \phi_{2r}^k\cdot\nabla v_{2r} + \int_{\mathcal D}(\beta_{21} f(\phi_{1r}^k,\phi_{1i}^k) + \beta_{22} f(\phi_{2r}^k,\phi_{2i}^k))\phi_{2r}^k v_{2r},
\end{aligned}
\end{equation}
\begin{equation}
	\label{eq-num-GP2c-stat-Newton-4}
\begin{aligned}
& \int_{\mathcal D}(C_\trap-\mu_2)s_2 v_{2i} +\int_{\mathcal D} \frac{1}{2} \nabla s_2\cdot\nabla v_{2i} + \int_{\mathcal D} (\beta_{22} f(\phi_{2r}^k,\phi_{2i}^k) + \beta_{21} f(\phi_{1r}^k,\phi_{1i}^k))s_2v_{2i}\\
&\hspace{1cm}+ \int_{\mathcal D} \beta_{21}\left(\dfrac{\pl f}{\pl\phi_r}(\phi_{1r}^k,\phi_{1i}^k)\phi_{2i}^k q_1 + \dfrac{\pl f}{\pl\phi_i}(\phi_{1r}^k,\phi_{1i}^k)\phi_{2i}^k s_1\right) v_{2i}\\
&\hspace{1cm}+ \int_{\mathcal D} \beta_{22}\left(\dfrac{\pl f}{\pl\phi_r}(\phi_{2r}^k,\phi_{2i}^k)\phi_{2i}^k q_2 + \dfrac{\pl f}{\pl\phi_i}(\phi_{2r}^k,\phi_{2i}^k)\phi_{2i}^k s_2\right) v_{2i}\\
&\hspace{.5cm} = \int_{\mathcal D}(C_\trap -\mu_2)\phi_{2i}^k v_{2i} + \int_{\mathcal D}\frac{1}{2} \nabla \phi_{2i}^k\cdot\nabla v_{2i} + \int_{\mathcal D}(\beta_{21} f(\phi_{1r}^k,\phi_{1i}^k) +\beta_{22} f(\phi_{2r}^k,\phi_{2i}^k))\phi_{2i}^k v_{2i}.
\end{aligned}
\end{equation}
}Again, the implementation of Eqs.  \eqref{eq-num-GP2c-stat-Newton-1}-\eqref{eq-num-GP2c-stat-Newton-4} with \ff is very similar to the mathematical formulation.

\subsection{Finite element implementation with \ff}

The algorithms presented below are implemented using the free software \ff \citep{hecht-2012-JNM}.  We illustrate in this section the main principles of programming used in building the toolbox and the numerical settings for the BdG problem.

 One of the main advantages offered by \ff is to  program cumbersome formulae in a compact form, close to the mathematical formulation. For example, the system \eqref{eq-num-GP-stat-Newton} is implemented in a \texttt{Macro} (precisely \texttt{BdG\_1comp/A\_macro/Macro\_problem.edp}) in which integral terms are easy to identify:
{\small\begin{lstlisting}[firstnumber=last]
	NewMacro problemGP
	macro f(ur,ui) (ur^2 + ui^2)//
	macro dfdur(ur,ui) (2.*ur)//
	macro dfdui(ur,ui) (2.*ui)//
	
	varf vGP([q,s],[vr,vi]) = 
	intN(Th,qforder=ord)((Ctrap - mu)*q*vr + .5*grad(q)'*grad(vr)
	+ (Ctrap - mu)*s*vi + .5*grad(s)'*grad(vi)
	+ beta * (f(phir,phii)*q*vr + f(phir,phii)*s*vi)
	+ beta * phir*vr*(dfdur(phir,phii)*q + dfdui(phir,phii)*s)
	+ beta * phii*vi*(dfdur(phir,phii)*q + dfdui(phir,phii)*s))
	+ intN(Th,qforder=ord)((Ctrap - mu)*phir*vr + .5*grad(phir)'*grad(vr)
	+ (Ctrap - mu)*phii*vi + .5*grad(phii)'*grad(vi)
	+ beta * f(phir,phii) * (phir*vr + phii*vi))
	BCGP;
	EndMacro
\end{lstlisting}}
Another advantage of this formulation is that it can be used for any dimension ($d=1, 2$ or $3$) and any  available type of finite elements, by simply declaring these values in the files defining the computational case. For example, for the 1D dark-soliton test case (file \texttt{BdG\_1comp/INIT/1D\_DS.inc}):
{\small\begin{lstlisting}[firstnumber=last]
macro dimension 1//
macro FEchoice P2//		
\end{lstlisting}}		
These choices are then used in the main programs to define the finite-element spaces. For example, in \texttt{FFEM\_GP\_1c\_1D\_2D\_3D.edp}:
{\small\begin{lstlisting}[firstnumber=last]
func Pk = [FEchoice,FEchoice];
...
meshN Th; // Local mesh
fespace Wh(Th,FEchoice);
fespace Whk(Th,Pk);
...
Wh<complex> phi; // Wavefunction
Whk [q,s], [phir,phii];
\end{lstlisting}}		For all programs in this toolbox, we use $P2$ (piece-wise quadratic) finite elements.

\ff also offers a fast mesh generator for 1D, 2D or 3D configurations. The mesh (generically identified as  \texttt{Th}) is made of segments in 1D, triangles in 2D and tetrahedrons in 3D. The initial solution is built specifically for each case as an approximation of the state we want to study (see Sects. \ref{sec-valid1c} and \ref{sec-valid2c}).
Newton iterations are stopped when one of two following criteria is satisfied: 
\begin{equation}\label{eq-bdg-err1}
\left\lVert
\begin{pmatrix}
q\\
s
\end{pmatrix}
\right\rVert_\infty < \epsilon_{\scriptscriptstyle q},
\quad
\left\lVert
\begin{pmatrix}
\mathcal{F}_r\\
\mathcal{F}_i
\end{pmatrix}
\right\rVert_2 < \epsilon_{\scriptscriptstyle F}.
\end{equation}
In practice, we use $\epsilon_{\scriptscriptstyle q} = 10^{-8}$ and $\epsilon_{\scriptscriptstyle F} = 10^{-16}$ and in all considered cases both criteria are satisfied simultaneously. To achieve convergence in the Newton algorithm, the choice of the solver for the linear system resulting from \eqref{eq-num-GP-stat} or \eqref{eq-num-GP2c-stat} is very important. For 1D and 2D problems  we solve the system with a direct LU method using the library \texttt{MUMPS}. 
For 3D problems, we use a  GMRES method, preconditioned by an incomplete LU factorization. 

Branches of stationary solutions are followed by a continuation method on the parameter $\mu_0 \leq \mu \leq \mu_f$. In practice, we start from a value $\mu_0$ for which the initial condition is sufficiently close to the stationary state and use this converged state as an initial guess for the Newton method with chemical potential $\mu_0 + \delta\mu$. The process is repeated until $\mu_f$ is reached. This is especially useful when following states from the linear limit to high values of $\mu$. For the two-component case, the continuation is done first on $\mu_1$ and $\mu_2$ and then on the inter-component interactions $\beta_{12}$ and $\beta_{21}$.

An important tool in \ff is mesh adaptation, that considerably helps in reducing  the computational time while keeping a high degree of accuracy. The mesh is adapted in 2D using the standard \texttt{adaptmesh} command of \ff which creates a new mesh adapted to the Hessian of the solution. In 3D, the adaptation is done through the libraries \texttt{mshmet} and \texttt{mmg} \citep{dapogny2014three} which are directly linked to \ff. When using continuation, we adapt the mesh for different values of $\mu$. Mesh adaptation is mandatory for the  complicated test cases, especially in 3D or for the two-component cases: using a refined mesh for the entire domain would lead to a large memory consumption and an excessively long computational time.

\section{Solving the BdG equations}\label{sec-num-meth-bdg}

The BdG problem \eqref{eq-scal-BdG}  is solved using the \texttt{ARPACK} library \citep{lehoucq1998arpack}. It is directly interfaced with \ff and uses an Arnoldi method to compute the eigenvalues and eigenvectors of a given matrix. We use the following weak formulation corresponding to  \eqref{eq-scal-BdG}:
\begin{equation}
	\label{eq-num-BdG-weak}
\begin{dcases}
\phantom{-}\int_{\mathcal D} \dfrac{1}{2}\nabla A\cdot\nabla v_1 + \int_{\mathcal D} (C_\trap-\mu)A v_1 + \int_{\mathcal D}2\beta |\phi|^2 A v_1 + \int_{\mathcal D}\beta \phi^2B v_1 = \omega \int_{\mathcal D}A v_1,\\
-\int_{\mathcal D}\dfrac{1}{2}\nabla B\cdot\nabla v_2 - \int_{\mathcal D}(C_\trap+\mu)B v_2  - \int_{\mathcal D} 2\beta |\phi|^2 B v_2 - \int_{\mathcal D} \beta \overline{\phi}^2 A v_2 = \omega \int_{\mathcal D}B v_2.
\end{dcases}
\end{equation}
The bilinear terms in the left hand side of this equation form the finite element matrix $M$ that is sent to \texttt{ARPACK}. To check the accuracy of the eigenvalue computation, we compute the residual:
\begin{equation}\label{eq-bdg-residual}
	\left\lVert M \begin{pmatrix} A\\B \end{pmatrix}
	- \omega \begin{pmatrix} A\\B \end{pmatrix}\right\rVert_\infty.
\end{equation}
Numerical tests showed that using a shift leads to an increased accuracy: the residual decreases to $10^{-7}$ and eigenvalues are closer to the expected values when compared to known results. We use a shift $\sigma = 10^{-4}$, which is implemented by adding the following term to the matrix:
\begin{equation}\label{eq-bdg-shift}
	-\int_{\mathcal D} \sigma(A v_1 + B v_2).
\end{equation}
For the two-component case, the numerical method is similar and based on the following weak formulation corresponding to  \eqref{eq-BdG2c-M}-\eqref{eq-BdG2c-M1}:
\begin{equation}\label{eq-num-BdG2c-weak}
\begin{dcases}
\begin{aligned}
\int_{\mathcal D} \frac{1}{2}\nabla A\cdot \nabla v_1 &+ \int_{\mathcal D}(C_\trap - \mu_1) A v_1 + \int_{\mathcal D} \left(2\beta_{11}|\phi_1|^2 + \beta_{12}|\phi_2|^2\right)A v_1\\
& + \int_{\mathcal D} \beta_{11}\phi_1^2B v_1 + \int_{\mathcal D} \beta_{12}\phi_1\overline{\phi_2}C v_1 + \int_{\mathcal D} \beta_{12}\phi_1\phi_2 D v_1 = \omega \int_{\mathcal D} A v_1,\\
\end{aligned}\\
\begin{aligned}
-\int_{\mathcal D}\frac{1}{2}\nabla B\cdot\nabla v_2 &- \int_{\mathcal D}(C_\trap-\mu)B v_2 - \int_{\mathcal D}\left(2\beta_{11}|\phi_1|^2 + \beta_{12}|\phi_2|^2\right)B v_2 \\&- \int_{\mathcal D}\beta_{11}\overline{\phi_1}^2 A v_2  - \int_{\mathcal D}\beta_{12}\overline{\phi_1}\overline{\phi_2}C v_2 -\int_{\mathcal D}\beta_{12}\overline{\phi_1}\phi_2 D v_2 = \omega \int_{\mathcal D}B v_2,\\
\end{aligned}\\
\begin{aligned}
\int_{\mathcal D}\frac{1}{2}\nabla C\cdot \nabla v_3 &+ \int_{\mathcal D}(C_\trap-\mu)C v_3 + \int_{\mathcal D}\left(2\beta_{22}|\phi_2|^2 + \beta_{21}|\phi_1|^2 \right)C v_3 \\&+ \int_{\mathcal D}\beta_{21}\overline{\phi_1}\phi_2 A v_3 + \int_{\mathcal D}\beta_{21}\phi_1\phi_2 B v_3 + \int_{\mathcal D}\beta_{22}\phi_2^2 D v_3 = \omega \int_{\mathcal D}C v_3,\\
\end{aligned}\\
\begin{aligned}
-\int_{\mathcal D}\frac{1}{2}\nabla D\cdot \nabla v_4 &- \int_{\mathcal D}(C_\trap-\mu)D v_4 - \int_{\mathcal D}\left(2\beta_{22}|\phi_2|^2 + \beta_{21}|\phi_1|^2\right)D v_4 \\&-\int_{\mathcal D}\beta_{21}\overline{\phi_1}\overline{\phi_2}A v_4 - \int_{\mathcal D}\beta_{21}\phi_1\overline{\phi_2}B v_4 - \int_{\mathcal D}\beta_{22}\overline{\phi_2}^2C v_4 = \omega \int_{\mathcal D}D v_4.\\
\end{aligned}\\
\end{dcases}
\end{equation}

\section{Validation test cases for the one-component BEC}\label{sec-valid1c}

We start by validating the codes for the one-component BEC against well known benchmarks. A summary of the considered cases, together with typical computational times and mesh sizes, is provided in Tab. \ref{tab-cputime-1c}. When mesh adaptation is used, we indicate the size of the mesh for the last step of the continuation procedure. Note that we considered for all cases the non-dimensional equations in the setting for which $\beta=1$.
\begin{table}[h!]
	\resizebox{\textwidth}{!}{%
		\begin{tabular}{l|rrr|rrr|}
			\cline{2-7}
			& \multicolumn{3}{c|}{Without mesh adaptation} & \multicolumn{3}{c|}{With mesh adaptation}   \\
			& CPU time GP  & CPU time BdG & Mesh size & CPU time GP & CPU time BdG & Mesh size \\ \hline
	           \multicolumn{1}{|l|}{1D ground state}          &  00:00:01          & 00:00:05           & 3602      &             &              &           \\
	\multicolumn{1}{|l|}{1D dark soliton}          & 00:00:01           & 00:00:02           & 1356      &             &              &           \\
	\multicolumn{1}{|l|}{2D ground state}          & 00:00:02    & 00:00:24          & 11552     & 00:00:05          & 00:00:20          & 9942     \\
	\multicolumn{1}{|l|}{2D dark soliton}          &   00:20:49   & 05:34:28        & 45000     & 00:09:32        & 00:58:07        & 14498     \\
	\multicolumn{1}{|l|}{2D central vortex}        & 00:07:26         & 00:58:07         & 14200     & 00:09:30        & 01:03:40        & 18775     \\
	\multicolumn{1}{|l|}{3D ground state}          & 00:04:22         & 01:09:00        & 24576     & 00:07:45        & 01:20:38        & 30317     \\
	 \hline
		\end{tabular}%
	}
	\caption{Test cases for the one-component BEC. Computational time  and mesh size (number of elements). All computation were performed on a Macbook pro M1, 16GB of DDR4 2400 MHz RAM.}
	\label{tab-cputime-1c}
\end{table}

\subsection{1D case: ground state}

The first test case is the computation of eigenvalues of  the ground state of a one-dimensional BEC with trapping potential $V_\trap=\frac{1}{2} m \omega_z^2 z^2$. In the Thomas-Fermi limit, the explicit expressions for eigenvalues are known \citep{kevrekidis2010distribution}:
\begin{equation}\label{eq-bdg-1D-TF}
	\omega_n^\TF = \omega_z \sqrt{\frac{n(n+1)}{2}},\quad n\in\N.
\end{equation}

We compare in Tab. \ref{tab-1D-TF} numerical and theoretical values of eigenvalues  $\omega$ for $\mu = 6$ and $\omega_z = 0.025$. 
The Thomas-Fermi solution \eqref{eq-TF} was used to initialize the Newton algorithm. We could check from Tab. \ref{tab-1D-TF} that the computed eigenvalues verify the following expected properties (see also Sect. \ref{sec-BdG}): \\(i)  all eigenvalues are real (the stationary state is dynamically stable) and form pairs $(+\omega,-\omega)$, \\(ii) the first eigenvalue is $\omega=0$,\\ (iii) the other eigenvalues correspond to theoretical predictions  \eqref{eq-bdg-1D-TF},\\ (iv) all Krein signatures are positive (the stationary state is energetically stable).

\begin{table}[ht!]
	\centering
	\resizebox{.7\textwidth}{!}{%
		\begin{tabular}{c|rrr|l|}
			\cline{2-5}
			& $Re(\omega)$ & $Im(\omega)$ & K & $\omega_n^\TF$ from \eqref{eq-bdg-1D-TF}                                      \\ \hline
			\multicolumn{1}{|l|}{$\omega_1$}    & -2.89857e-15  & 2.16087e-07                & 1 & \multirow{2}{*}{$\omega_0^\TF = 0$}                               \\
			\multicolumn{1}{|l|}{$\omega_2$}    & 6.18933e-15& -2.16087e-07                 & 1 &                                                 \\ \hline
			\multicolumn{1}{|l|}{$\omega_3$}    & -0.025       & -8.80682e-11                & 1 & \multirow{2}{*}{$\omega_1^\TF = \omega_z = 0.025$}  \\
			\multicolumn{1}{|l|}{$\omega_4$}    & 0.025        & 2.76512e-11                 & 1 &                                                 \\ \hline
			\multicolumn{1}{|l|}{$\omega_5$}    & -0.0433018    & -4.41549e-11                & 1 & \multirow{2}{*}{$\omega_2^\TF \approx 0.043301270$} \\
			\multicolumn{1}{|l|}{$\omega_6$}    & 0.0433018     & -1.21387e-11                & 1 &                                                 \\ \hline
			\multicolumn{1}{|l|}{$\omega_7$}    & -0.0612394   & -2.87955e-10                & 1 & \multirow{2}{*}{$\omega_3^\TF \approx 0.061237243$} \\
			\multicolumn{1}{|l|}{$\omega_8$}    & 0.0612394    & 1.64467e-10                 & 1 &                                                 \\ \hline
			\multicolumn{1}{|l|}{$\omega_9$}    & -0.0790624    & -1.09235e-10               & 1 & \multirow{2}{*}{$\omega_4^\TF \approx 0.07905694$}  \\
			\multicolumn{1}{|l|}{$\omega_{10}$} & 0.0790624   & 8.67993e-11                & 1 &                                                 \\ \hline
		\end{tabular}%
	}
	\caption{1D  ground state: eigenvalues and Krein signatures.}
	\label{tab-1D-TF}
\end{table}

\subsection{1D case: dark soliton}

\begin{figure}[!ht]
	\centering
	\begin{subfigure}[b]{0.45\textwidth}
		\centering
		\includegraphics[width=\textwidth]{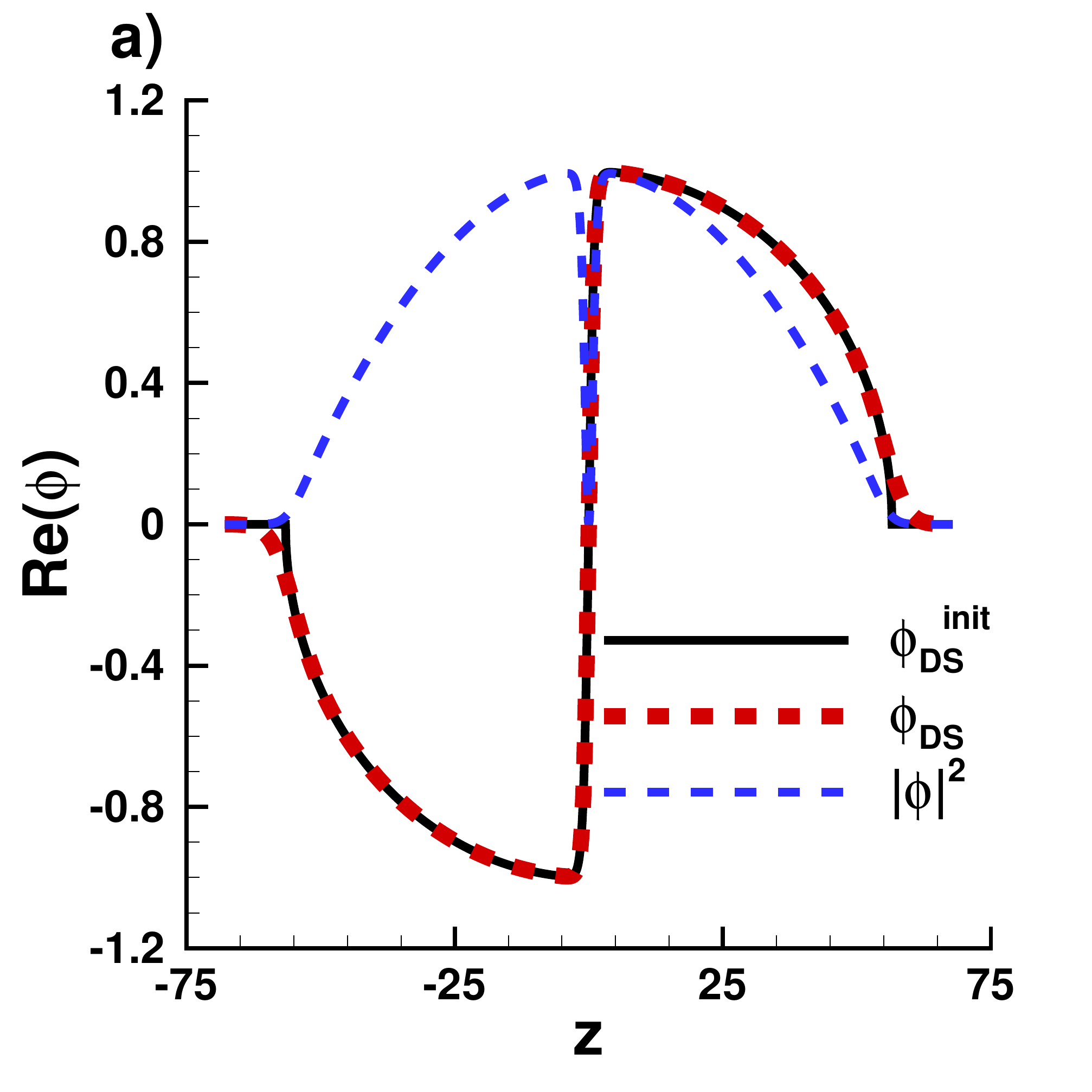}
		\label{fig:1D_DS_mu_1_a}
	\end{subfigure}
	\hfill
	\begin{subfigure}[b]{0.45\textwidth}
		\begin{subfigure}[]{\textwidth}
			\centering
			\includegraphics[width=\textwidth]{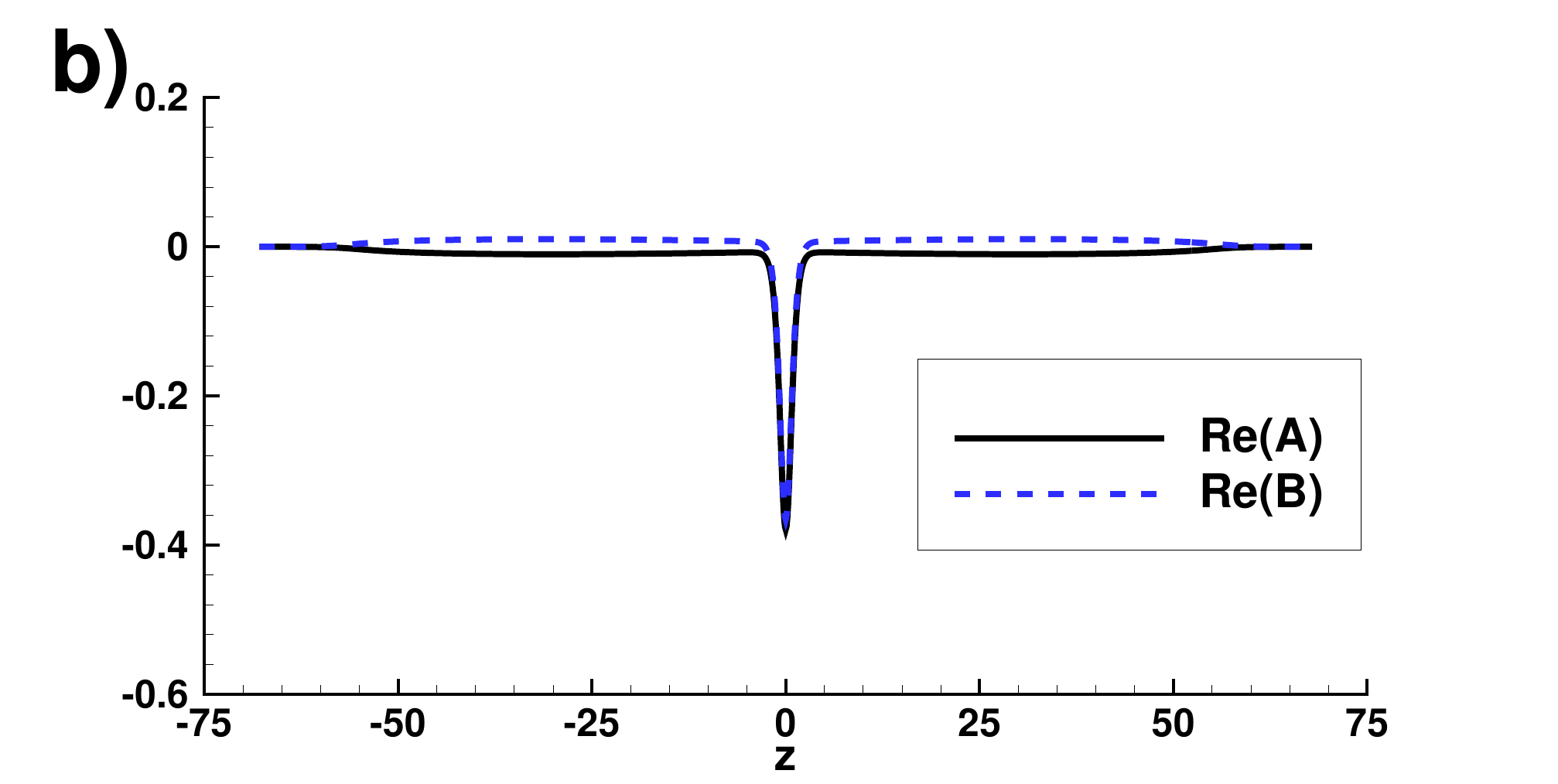}
			\label{fig:1D_DS_mu_1_b}
		\end{subfigure}
		\begin{subfigure}[b]{\textwidth}
			\centering
			\includegraphics[width=\textwidth]{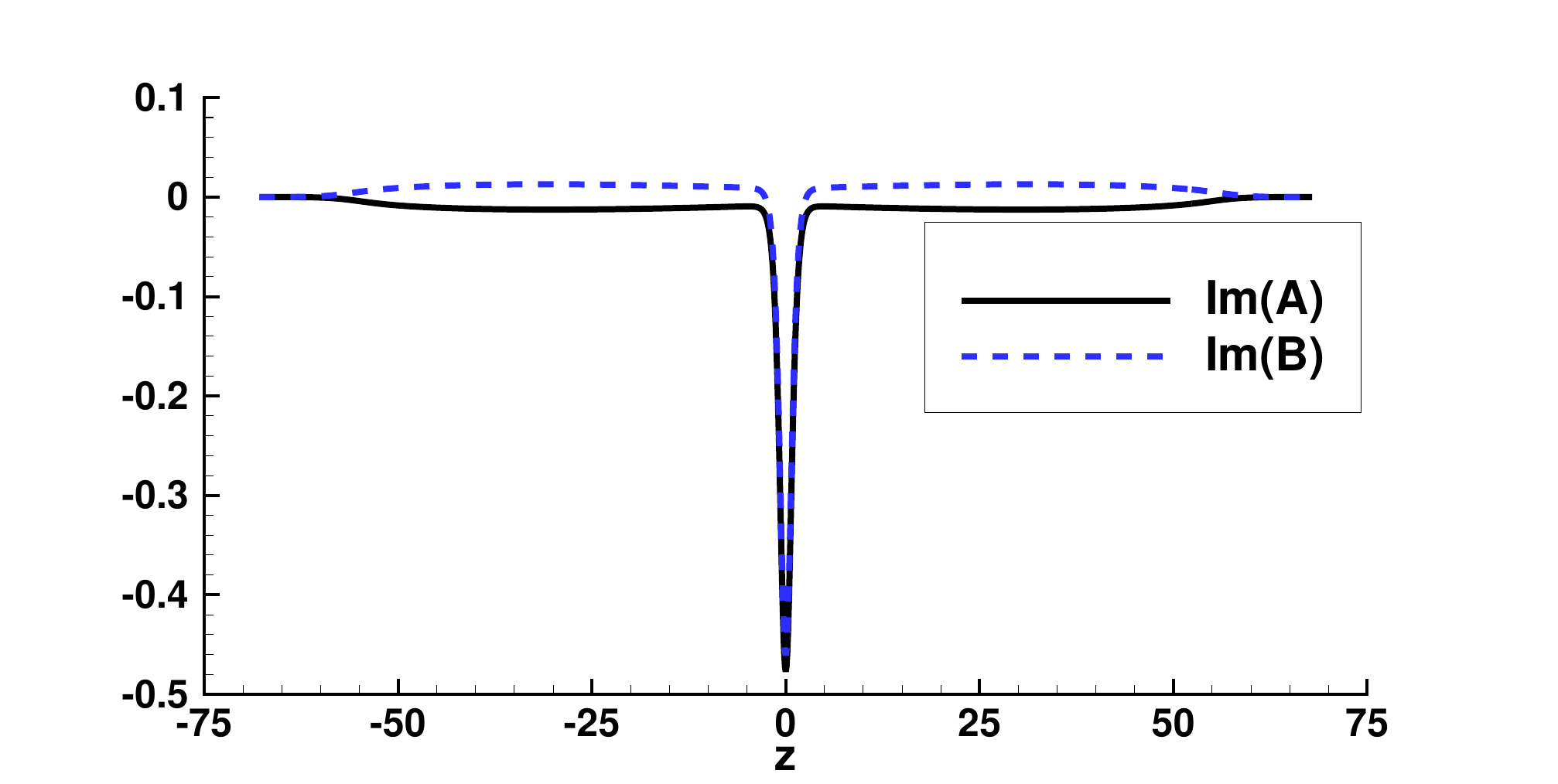}
			\label{fig:1D_DS_mu_1_c}
		\end{subfigure}
	\end{subfigure}
	\caption{1D dark soliton: a) initial state and stationary solution, b) anomalous mode with $\omega_4 \approx \frac{\omega_z}{\sqrt{2}}$.}
	\label{fig-1D-DS}
\end{figure}

We analyse for the second 1D  test case an excited state, obtained by adding a dark solition to the Thomas-Fermi density previously computed. The initial condition for the Newton algorithm is thus built as:
\begin{equation}\label{eq-DS}
\phi_{DS}^{init} = \sqrt{n_\TF}\tanh(\sqrt{\mu}z).
\end{equation}
We plot in  Fig. \ref{fig-1D-DS}(a) the initial condition and the converged stationary state. Eigenvalues are displayed in Tab. \ref{tab-bdg-1D-DS}. As expected, all eigenvalues are real, as the dark soliton is dynamically stable in 1D. A complete characterization of the BdG modes is offered in 
\cite{frantzeskakis2010dark}.
\begin{itemize}
	\item The mode with $\omega_4 \approx \frac{\omega_z}{\sqrt{2}} \approx 0.017677669$ is the anomalous mode; it is the only mode with a negative Krein signature. It is represented in Fig. \ref{fig-1D-DS}(b) and we retrieve the profile obtained in \cite{law2002quantum}.
	
	\item The dipole or Kohn mode at $\omega_6 \approx \omega_z = 0.025$ corresponds to oscillations of the center of mass of the condensate.
	
	\item The quadrupole mode (or the breathing mode) is obtained for $\omega_8 \approx \omega_z\sqrt{3} \approx 0.04330127$. This mode is particular to the one-dimensionality of the system.
\end{itemize}

\begin{table}[ht!]
	\centering
	\begin{tabular}{c|rrr|}
		\cline{2-4}
		& $Re(\omega)$ & $Im(\omega)$ & K  \\ \hline
		\multicolumn{1}{|c|}{$\omega_1$}    & -2.57971e-07  & -3.29591e-15                 & 1  \\
		\multicolumn{1}{|c|}{$\omega_2$}    & 2.57971e-07 & 3.25101e-15                & 1  \\
		\multicolumn{1}{|c|}{$\omega_3$}    & -0.0178197    & -1.09577e-12                 & -1 \\
		\multicolumn{1}{|c|}{$\omega_4$}    & 0.0178197   & 1.00791e-12                  & -1 \\
		\multicolumn{1}{|c|}{$\omega_5$}    & -0.025       & -5.42312e-12                 & 1  \\
		\multicolumn{1}{|c|}{$\omega_6$}    & 0.025        & 6.69980e-12                 & 1  \\
		\multicolumn{1}{|c|}{$\omega_7$}    & -0.0435553     & -8.24994e-12                & 1  \\
		\multicolumn{1}{|c|}{$\omega_8$}    & 0.0435553    & 9.60204e-12                 & 1  \\
		\multicolumn{1}{|c|}{$\omega_9$}    & -0.0616151   & 4.28088e-13                 & 1  \\
		\multicolumn{1}{|c|}{$\omega_{10}$} & 0.0616151    & 2.16934e-13                & 1  \\ \hline
	\end{tabular}
	\caption{1D dark soliton: eigenvalues and Krein signatures.}
	\label{tab-bdg-1D-DS}
\end{table}

\subsection{2D case: ground state}

We switch now to 2D BEC configurations with trapping potential $V_\trap=\frac{1}{2} m \omega_\perp^2 r^2$, where $r^2=x^2+y^2$. For this case, the eigenvalues in the Thomas-Fermi limit are also known \citep{kevrekidis2010distribution}:
\begin{equation}
	\label{eq-2DS-pred}
	\omega_{m,k}^\TF = \omega_\perp \sqrt{m + 2k^2 + 2k(1+m)},
\end{equation}
where $m, k \geq 0$ are integers. We present in Tab. \ref{tab-2D-TF} the first 20 eigenvalues computed for $\mu=6$  and  $\omega_\perp = 0.2$, with and without mesh adaptation. We find all Krein signatures to be $1$, which is the expected result in the absence of topological excitations. This is a perfect case to check that computations using mesh adaptation provide the same results as computations with a refined fixed mesh. Results in Tab. \ref{tab-2D-TF} show that this is indeed the case for our BdG solvers.

\begin{table}[h!]
	\resizebox{\textwidth}{!}{%
		\begin{tabular}{c|rrr|rrr|l}
			\cline{2-7}
			& \multicolumn{3}{c|}{No mesh adaptation}             & \multicolumn{3}{c|}{With mesh adaptation}           &                                                                           \\ \cline{2-8} 
			& $Re(\omega)$ & $Im(\omega)$ & K & $Re(\omega)$ & $Im(\omega)$ & K & \multicolumn{1}{l|}{$\omega_{m,k}$ from \eqref{eq-2DS-pred}}                                       \\ \hline
			\multicolumn{1}{|l|}{$\omega_1$}    & -2.07687e-06 & 5.60174e-16                & 1 & -6.24135e-15 & 1.37474e-07                 & 1 & \multicolumn{1}{l|}{\multirow{2}{*}{$\omega_{0,0}^\TF = 0$}}                  \\
			\multicolumn{1}{|l|}{$\omega_2$}    & 2.07687e-06 & -5.08061e-16                 & 1 & 6.23144e-15   & -1.37474e-07                  & 1 & \multicolumn{1}{c|}{}                                                     \\ \hline
			\multicolumn{1}{|l|}{$\omega_3$}    & -0.2         & -4.04752e-11                 & 1 & -0.2         & 1.16809e-11                 & 1 & \multicolumn{1}{l|}{\multirow{4}{*}{$\omega_{1,0}^\TF = 0.2$}}                \\
			\multicolumn{1}{|l|}{$\omega_4$}    & 0.2          & 1.28499e-11                & 1 & 0.2          & -2.51572e-11                 & 1 & \multicolumn{1}{l|}{}                                                     \\
			\multicolumn{1}{|l|}{$\omega_5$}    & -0.2         & -9.72650e-12                 & 1 & -0.2         & -5.27780e-12                & 1 & \multicolumn{1}{c|}{}                                                     \\
			\multicolumn{1}{|l|}{$\omega_6$}    & 0.2          & -1.80613e-11                  & 1 & 0.2          & 4.37562e-11                & 1 & \multicolumn{1}{c|}{}                                                     \\ \hline
			\multicolumn{1}{|l|}{$\omega_7$}    & -0.283446     & 5.84768e-11                & 1 & -0.283447     & 4.45534e-12                & 1 & \multicolumn{1}{l|}{\multirow{4}{*}{$\omega_{2,0}^\TF = 0.28284271$}}         \\
			\multicolumn{1}{|l|}{$\omega_8$}    & 0.283446    & 6.54561e-11                  & 1 & 0.283447    & 3.70927e-12                 & 1 & \multicolumn{1}{c|}{}                                                     \\
			\multicolumn{1}{|l|}{$\omega_9$}    & -0.283447    & 2.32827e-11                 & 1 & -0.283447     & 1.65992e-12                & 1 & \multicolumn{1}{c|}{}                                                     \\
			\multicolumn{1}{|l|}{$\omega_{10}$} & 0.283447     & 2.88143e-11                & 1 & 0.283447    & 6.51049e-12                   & 1 & \multicolumn{1}{c|}{}                                                     \\ \hline
			\multicolumn{1}{|l|}{$\omega_{11}$} & -0.348749    & -3.21680e-12                 & 1 & -0.348750     & 1.02905e-11                 & 1 & \multicolumn{1}{l|}{\multirow{4}{*}{$\omega_{3,0}^\TF = 0.34641016$}}         \\
			\multicolumn{1}{|l|}{$\omega_{12}$} & 0.348749     & 2.38853e-11                 & 1 & 0.348750      & 1.33981e-11                 & 1 & \multicolumn{1}{c|}{}                                                     \\
			\multicolumn{1}{|l|}{$\omega_{13}$} & -0.348749     & -4.01642e-11                & 1 & -0.348751    & -5.37018e-12                 & 1 & \multicolumn{1}{c|}{}                                                     \\
			\multicolumn{1}{|l|}{$\omega_{14}$} & 0.348749    & -9.62656e-12                & 1 & 0.348751     & 6.96459e-11                 & 1 & \multicolumn{1}{c|}{}                                                     \\ \hline
			\multicolumn{1}{|l|}{$\omega_{15}$} & -0.399998    & -6.47856e-11                & 1 & -0.399999     & 4.66425e-11                & 1 & \multicolumn{1}{l|}{\multirow{6}{*}{$\omega_{4,0}^\TF = \omega_{0,1}^\TF = 0.4$}} \\
			\multicolumn{1}{|l|}{$\omega_{16}$} & 0.399998     & 7.43875e-11                & 1 & 0.399999    & -2.73013e-12                 & 1 & \multicolumn{1}{c|}{}                                                     \\
			\multicolumn{1}{|l|}{$\omega_{17}$} & -0.405630     & 1.00037e-11                & 1 & -0.405633    & 4.30066e-11                & 1 & \multicolumn{1}{c|}{}                                                     \\
			\multicolumn{1}{|l|}{$\omega_{18}$} & 0.405630      & 2.17681e-11                 & 1 & 0.405633     & 5.11857e-11                & 1 & \multicolumn{1}{c|}{}                                                     \\
			\multicolumn{1}{|l|}{$\omega_{19}$} & -0.405630     & -2.72905e-11                & 1 & -0.405633     & 2.77070e-11                & 1 & \multicolumn{1}{c|}{}                                                     \\
			\multicolumn{1}{|l|}{$\omega_{20}$} & 0.405630      & -1.36630e-11                 & 1 & 0.405633    & -1.52304e-11                & 1 & \multicolumn{1}{c|}{}                                                     \\ \hline
		\end{tabular}%
	}
	\caption{2D ground state: eigenvalues and Krein signatures.}
	\label{tab-2D-TF}
\end{table}

\subsection{2D case: dark soliton}

\begin{figure}[h!]
	\centering
	\begin{subfigure}[t]{0.45\textwidth}
		\centering
		\includegraphics[width=\textwidth]{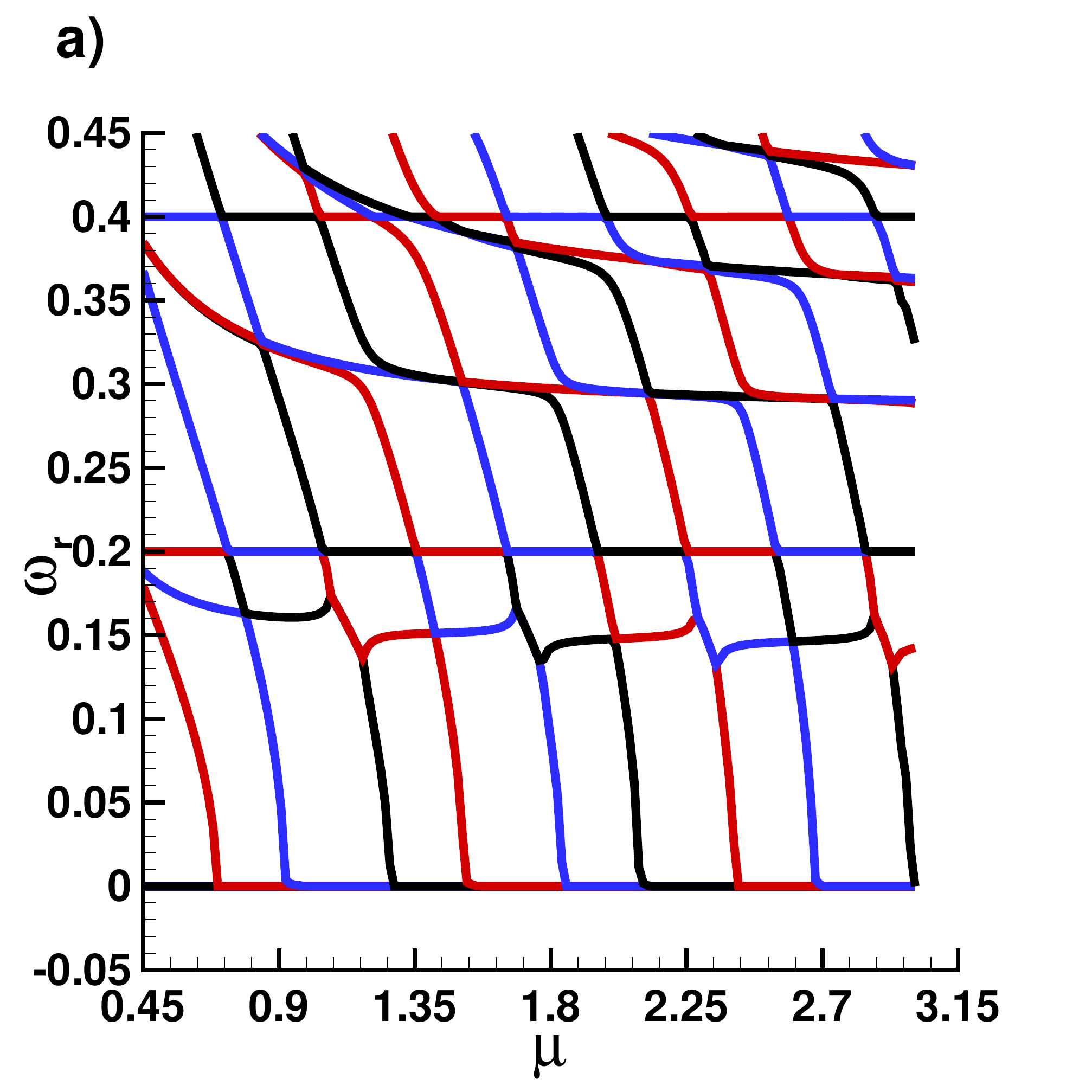}
		\label{fig:2D_DSH_a}
	\end{subfigure}
	\begin{subfigure}[t]{0.45\textwidth}
		\centering
		\includegraphics[width=\textwidth]{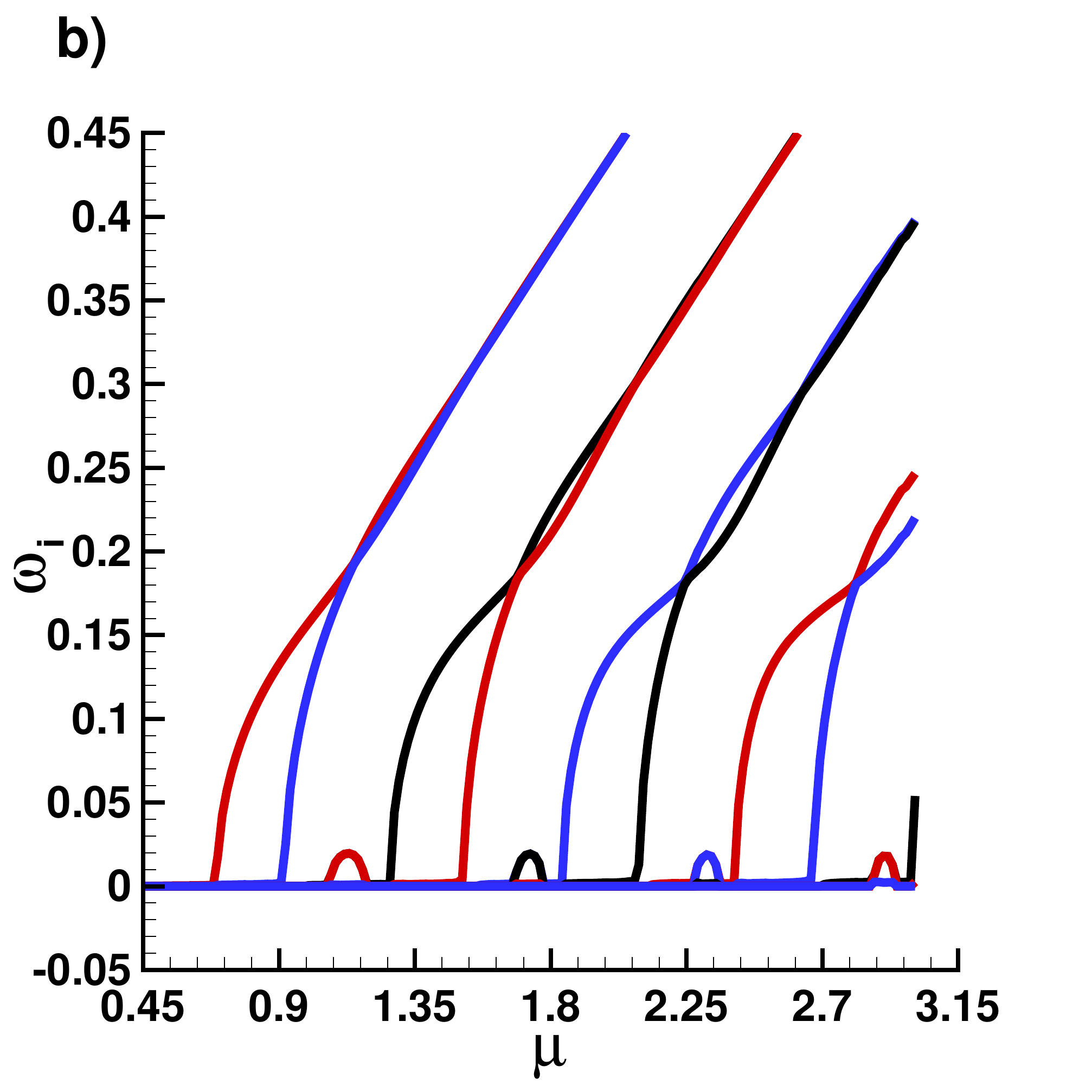}
		\label{fig:2D_DSH_b}
	\end{subfigure}
	\begin{subfigure}[t]{0.45\textwidth}
		\centering
		\includegraphics[width=\textwidth]{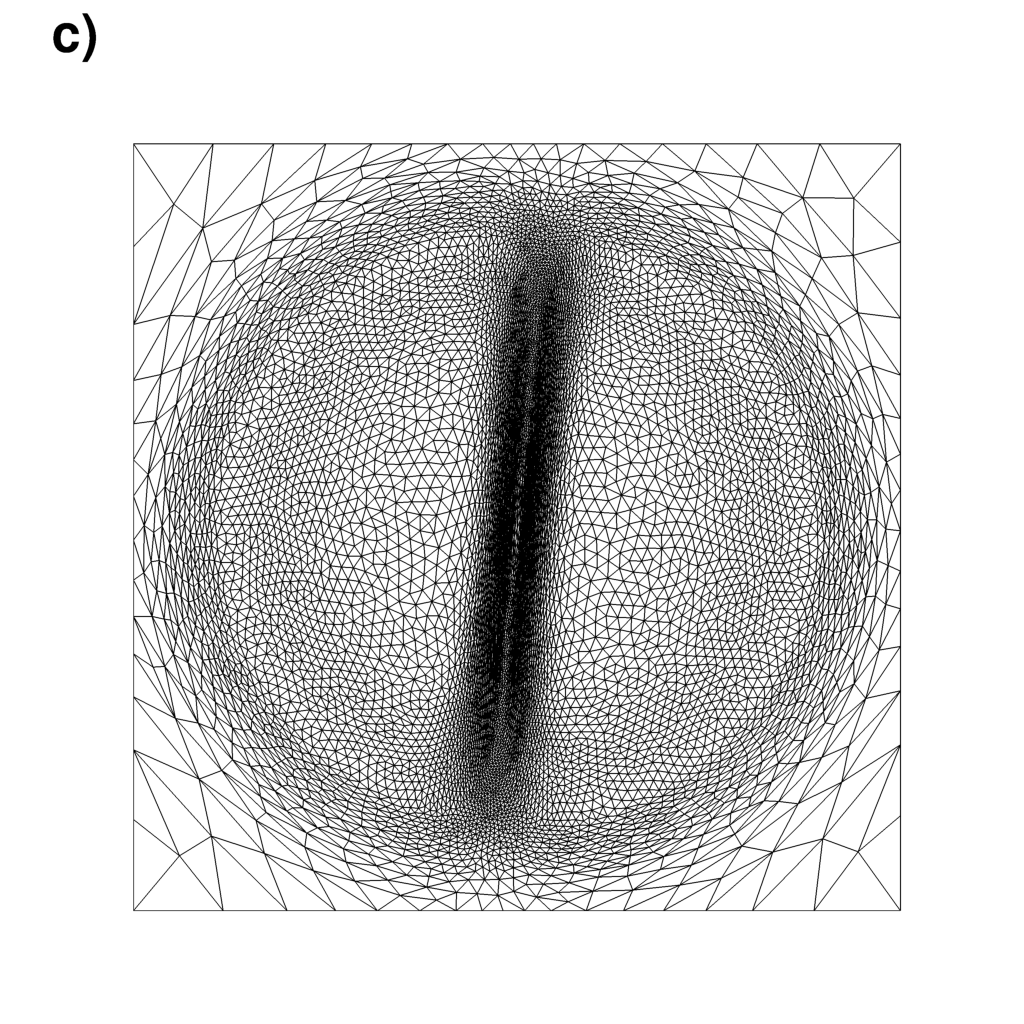}
		\label{fig:2D_DSH_c}
	\end{subfigure}
	\begin{subfigure}[t]{0.45\textwidth}
		\centering
		\includegraphics[width=\textwidth]{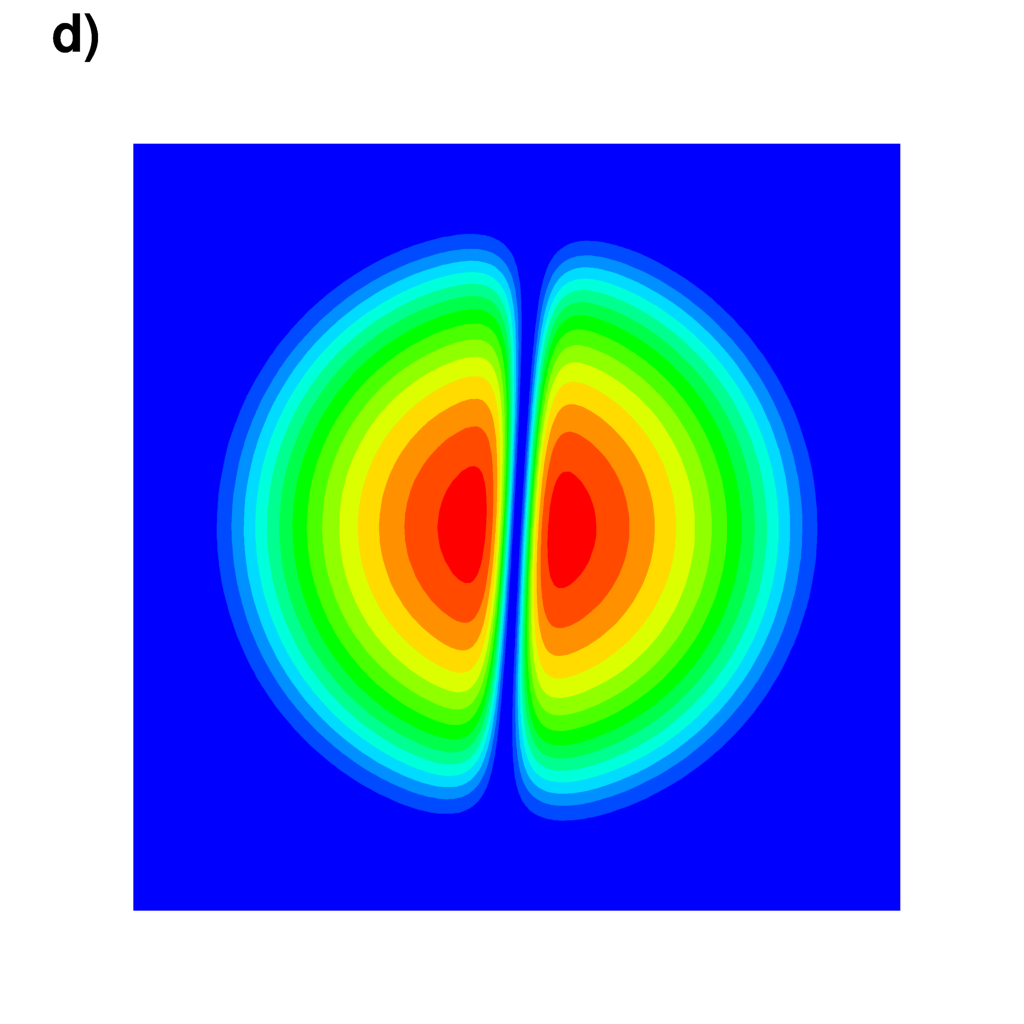}
		\label{fig:2D_DSH_d}
	\end{subfigure}
	\caption{2D dark soliton: a) real part $\omega_r$ and b) imaginary part $\omega_i$ of eigenvalues as a function of $\mu$. Solution for  $\mu=3.007$: c) adapted mesh and d) atomic density $|\phi|^2$.}
	\label{fig-2D-DSH}
\end{figure}

Following the same procedure as in the 1D case, we add to the previously computed 2D ground state a centered dark soliton (Fig. \ref{fig-2D-DSH}). This is an interesting case to test the continuation procedure in following a branch of stationary solutions. The initial condition is given by the $\ket{1,0}$ state in the linear limit:
\begin{equation}
	\phi_{DS} = \sqrt{\frac{\omega_\perp}{2\pi}} H_0(\sqrt{\omega_\perp}x)H_1(\sqrt{\omega_\perp}y)e^{-\frac{1}{2}\omega_\perp(x^2+y^2)},
\end{equation}
where $H_n$ are Hermite polynomials. We set  $\omega_\perp = 0.2$ and follow this solution up to $\mu = 3$.
Real and imaginary parts of eigenvalues are presented in Fig. \ref{fig-2D-DSH}(a, b) and are identical to the results published in \cite{middelkamp2010bifurcations}. 
This state does not have an azimuthal symmetry. 
Due to the space discretization, there exists a preferred  direction along which the soliton will tend to align itself. When adapting the mesh, this direction changes and the wave function will then rotate. To avoid this phenomenon, we only adapt the mesh every 5 iterations during the continuation procedure. This permits to optimally adapt the size of the mesh  while reducing the effects of the rotation. 

Table \ref{tab-cputime-1c} shows that this is an efficient approach to reduce the computational time. The final adapted mesh is presented in Fig. \ref{fig-2D-DSH}(c), with the corresponding atomic density $|\phi|^2$ in Fig. \ref{fig-2D-DSH}(d).

\newpage

\subsection{2D case: central vortex}

\begin{figure}[h!]
	\centering
	\begin{subfigure}[t]{0.4\textwidth}
		\centering
		\includegraphics[width=\textwidth]{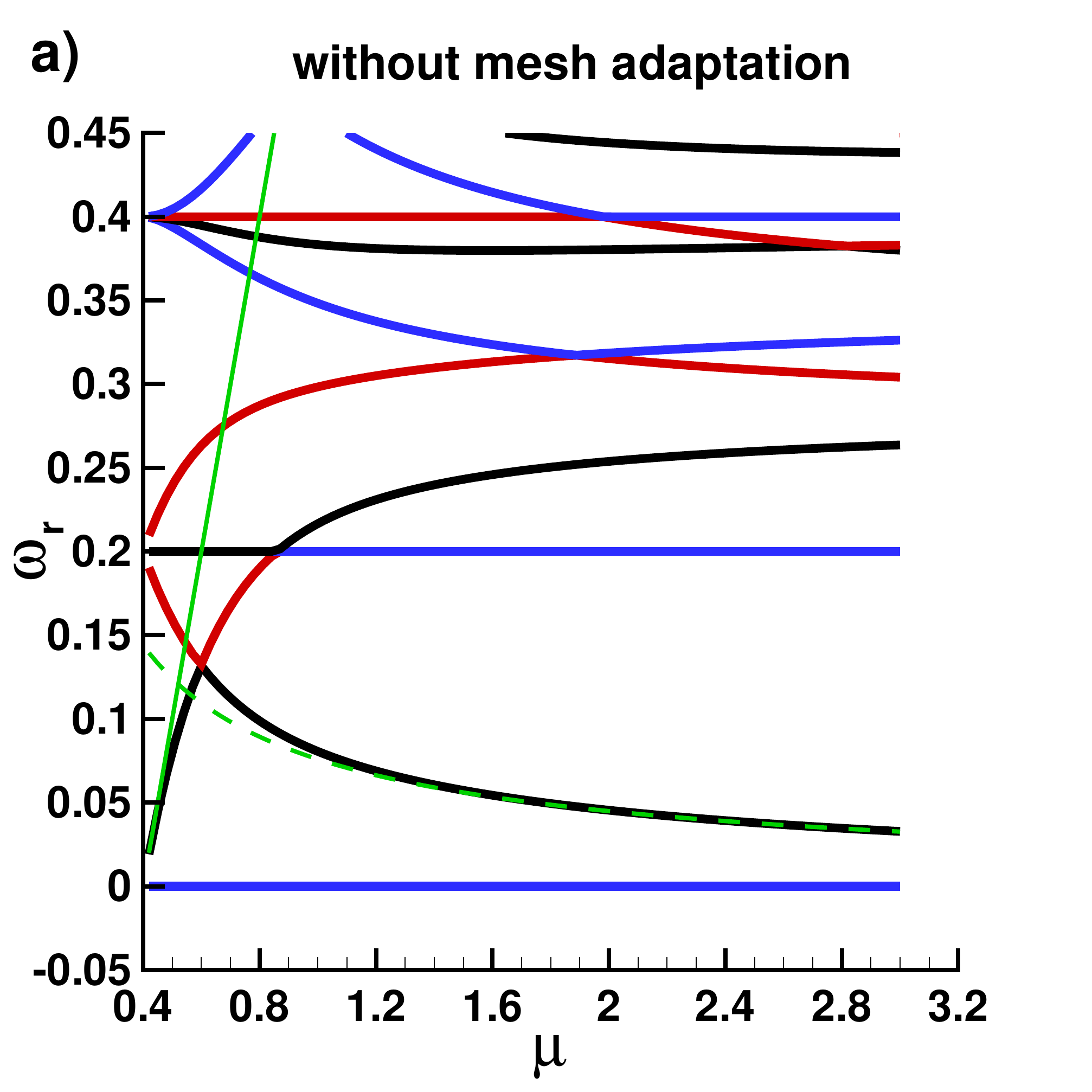}
		\label{fig:2D_VSH_a}
	\end{subfigure}
	\begin{subfigure}[t]{0.4\textwidth}
		\centering
		\includegraphics[width=\textwidth]{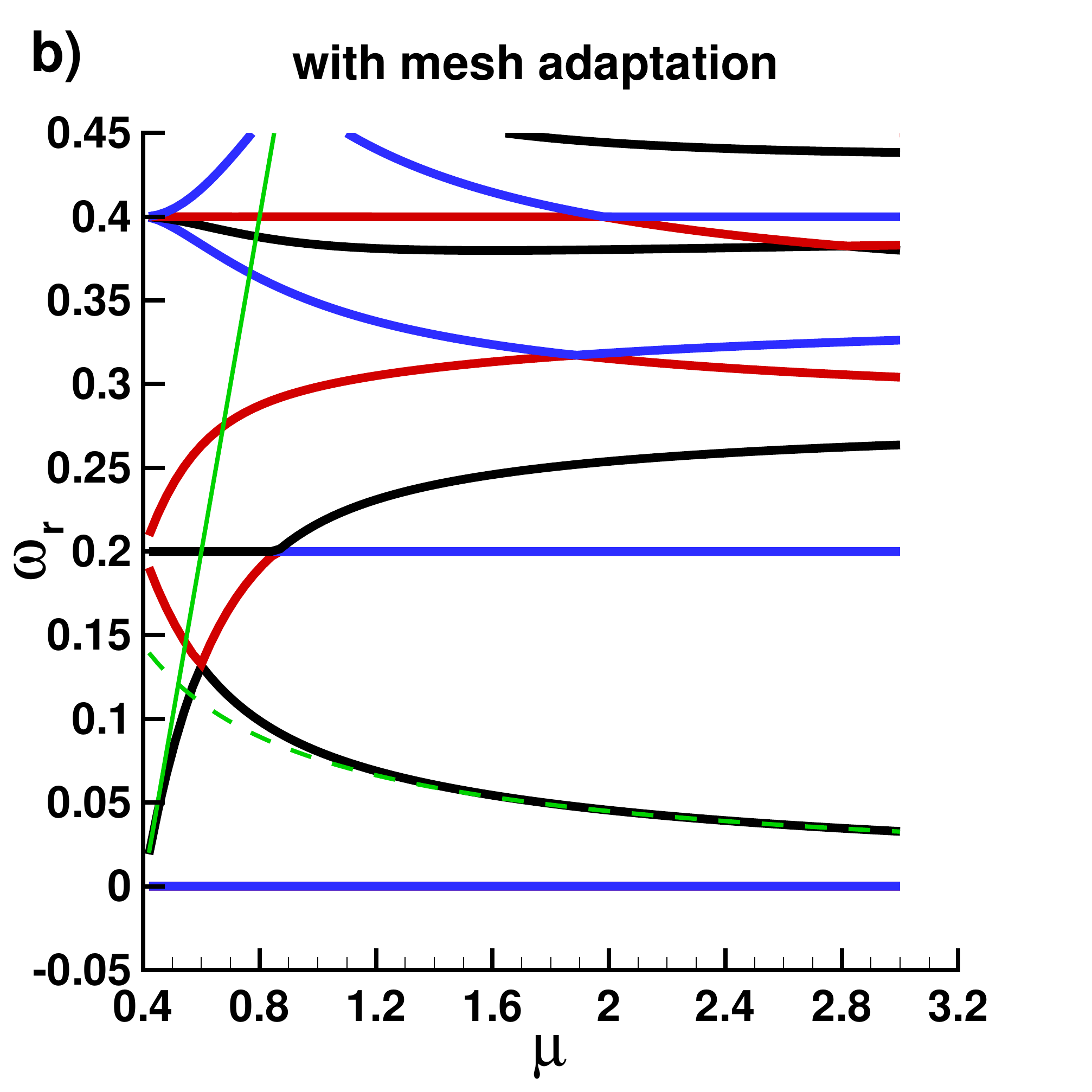}
		\label{fig:2D_VSH_b}
	\end{subfigure}
	\begin{subfigure}[t]{0.4\textwidth}
		\centering
		\includegraphics[width=\textwidth]{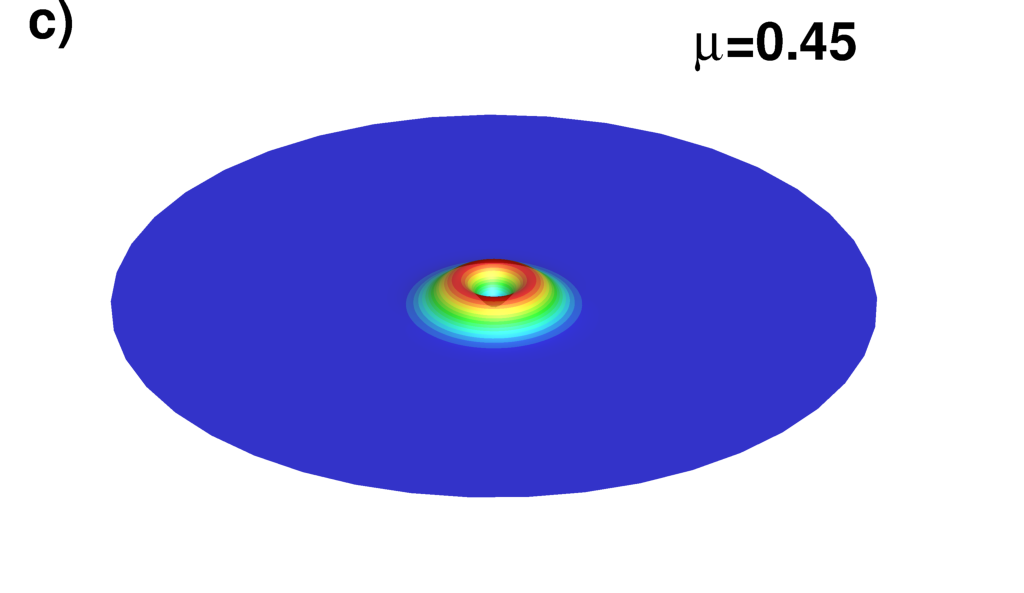}
		\label{fig:2D_VSH_c}
	\end{subfigure}
	\begin{subfigure}[t]{0.4\textwidth}
		\centering
		\includegraphics[width=\textwidth]{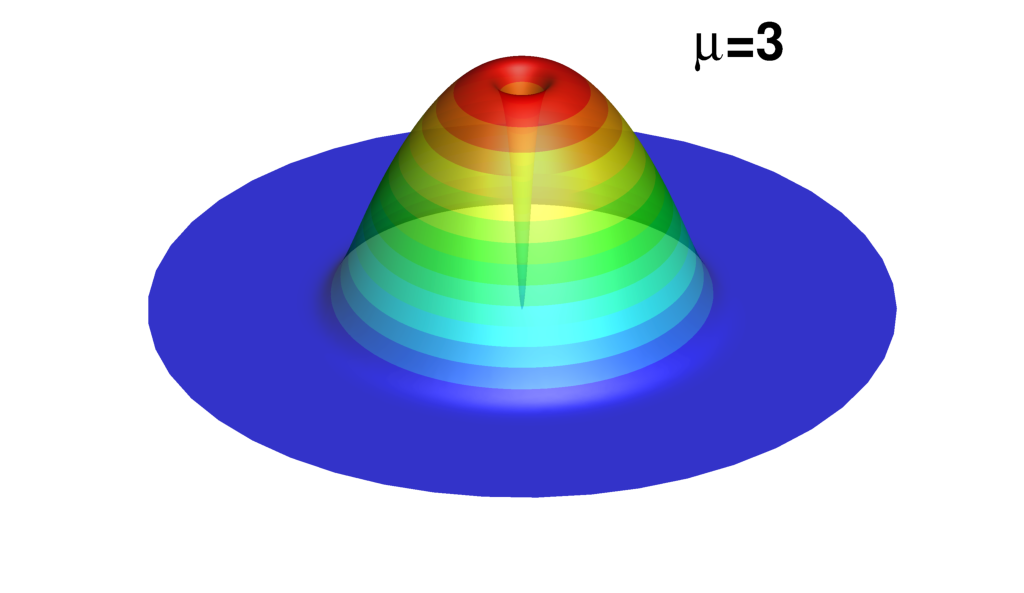}
		\label{fig:2D_VSH_d}
	\end{subfigure}
	\caption{2D BEC with a centered vortex. Real part $\omega_r$ of  eigenvalues as a function of $\mu$ computed a) without and b) with mesh adaptation. c) Atomic density $|\phi|^2$ for $\mu=0.45$ and $\mu=3$.}
	\label{fig-2D-VSH}
\end{figure}

We compute another solution studied in \cite{middelkamp2010bifurcations,middelkamp2010stability}. It consists of a disk-shaped BEC with a centered vortex. 
The initial condition is given by the $\ket{0,1}$ state in cylindrical coordinates $(r, \theta)$:
\begin{equation}
\phi_{VS} \propto r L_0^1(\omega_\perp r^2) e^{i\theta}e^{-\frac{1}{2}\omega_\perp r^2},
\end{equation}
where $L_0^1$  is the Laguerre polynomial.  We set, as in the previous case, $\omega_\perp = 0.2$. Eigenvalues computed with and without mesh adaptation are displayed in Fig. \ref{fig-2D-VSH}(a) and (b). We checked that both methods give the same results as those obtained in \cite{middelkamp2010bifurcations,middelkamp2010stability}. The atomic density is presented in Fig. \ref{fig-2D-VSH}(c) and (d) for two values of $\mu$.

\newpage

\begin{figure}[h!]
	\centering
	\begin{subfigure}[t]{0.45\textwidth}
		\centering
		\includegraphics[width=\textwidth]{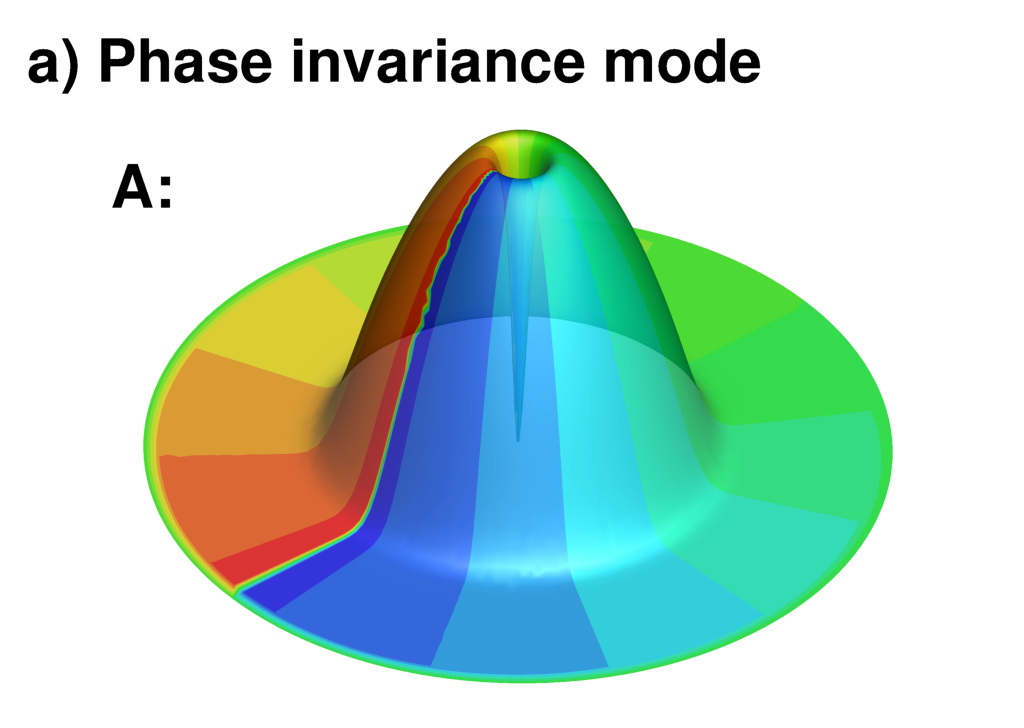}
		\label{fig:2D_VSH_eig_a}
	\end{subfigure}
	\begin{subfigure}[t]{0.45\textwidth}
		\centering
		\includegraphics[width=\textwidth]{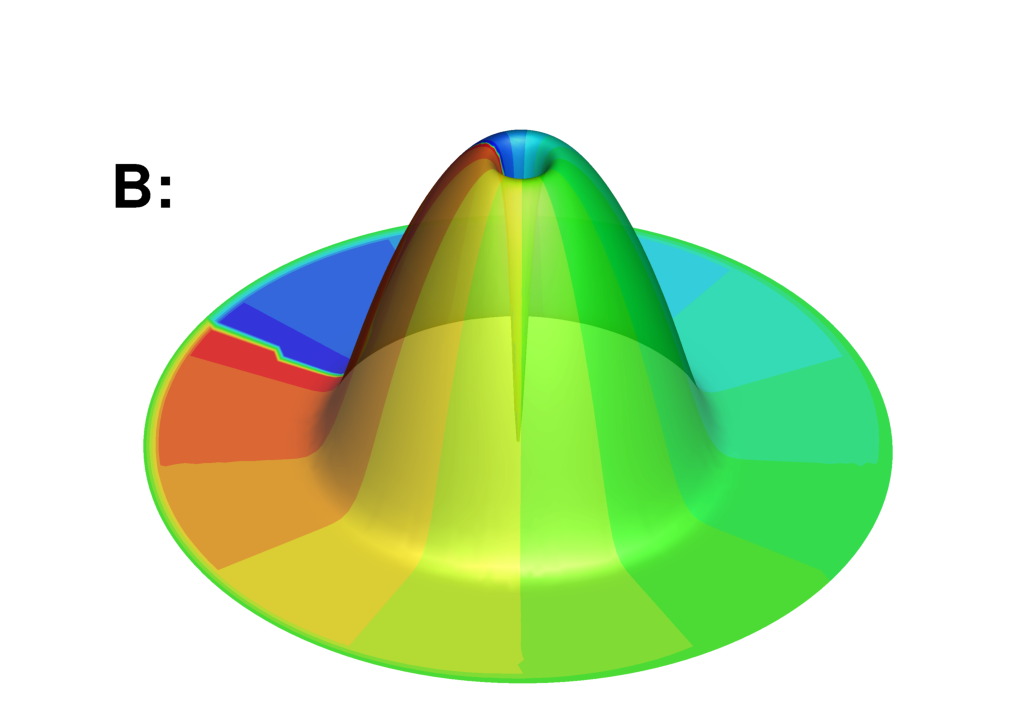}
		\label{fig:2D_VSH_eig_b}
	\end{subfigure}
	\begin{subfigure}[t]{0.45\textwidth}
		\centering
		\includegraphics[width=\textwidth]{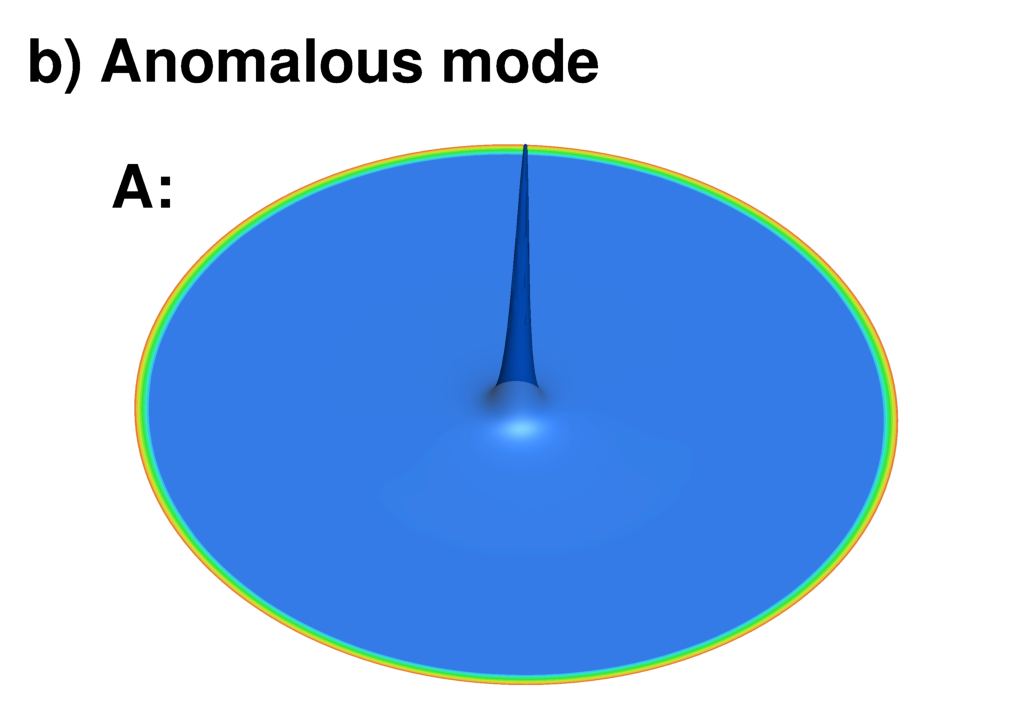}
		\label{fig:2D_VSH_eig_c}
	\end{subfigure}
	\begin{subfigure}[t]{0.45\textwidth}
		\centering
		\includegraphics[width=\textwidth]{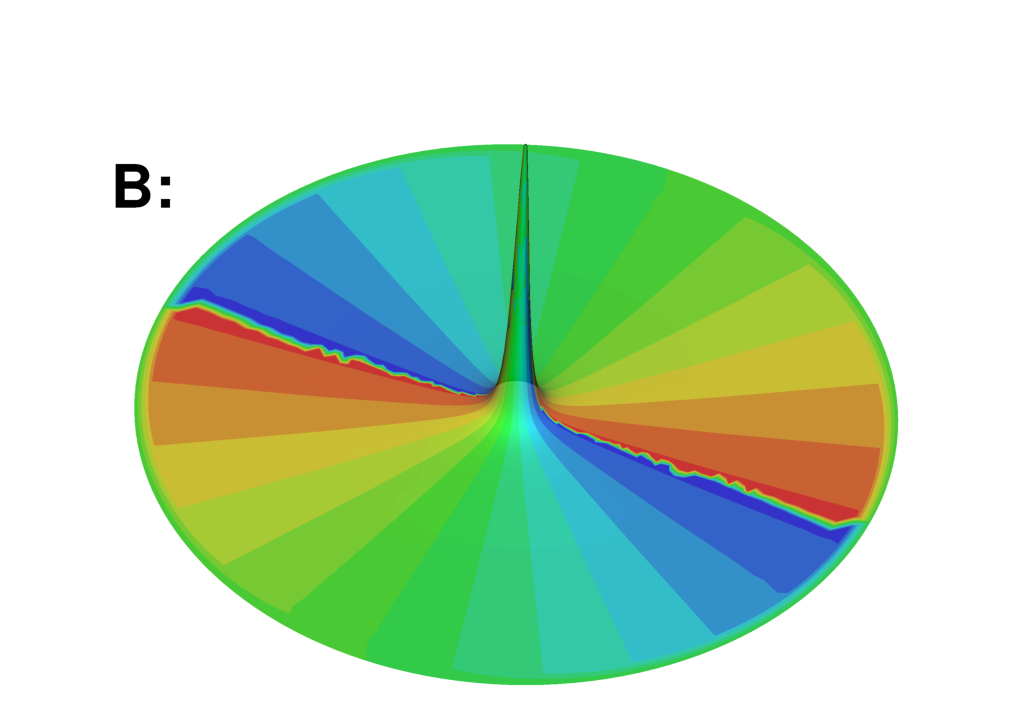}
		\label{fig:2D_VSH_eig_d}
	\end{subfigure}
	\begin{subfigure}[t]{0.45\textwidth}
		\centering
		\includegraphics[width=\textwidth]{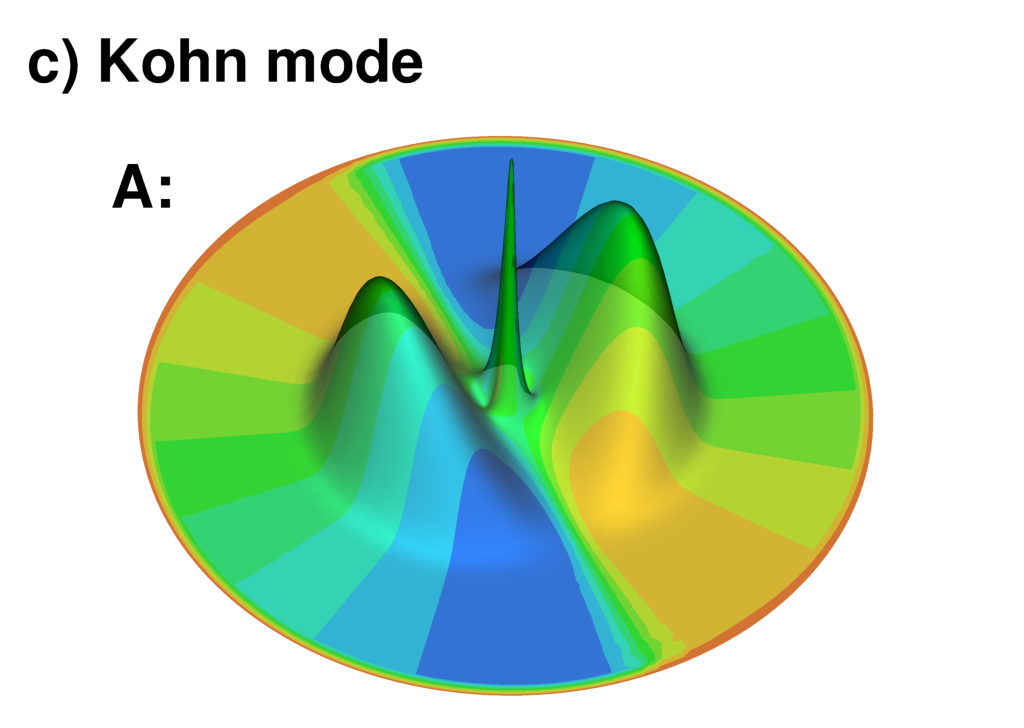}
		\label{fig:2D_VSH_eig_e}
	\end{subfigure}
	\begin{subfigure}[t]{0.45\textwidth}
		\centering
		\includegraphics[width=\textwidth]{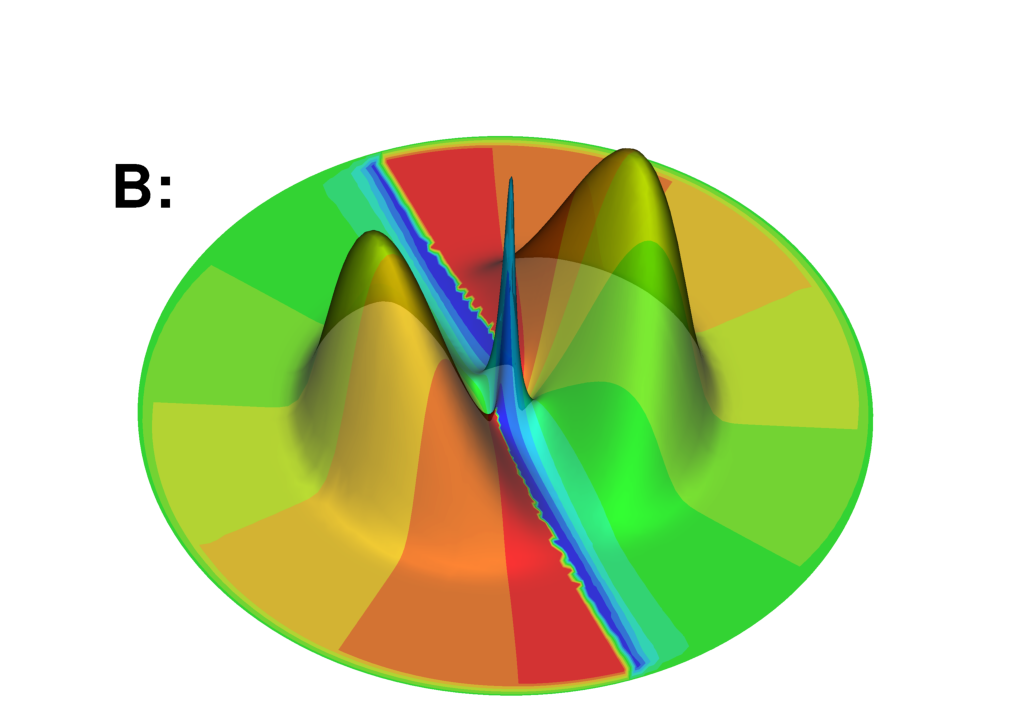}
		\label{fig:2D_VSH_eig_f}
	\end{subfigure}
	\begin{subfigure}[t]{0.45\textwidth}
		\centering
		\includegraphics[width=\textwidth]{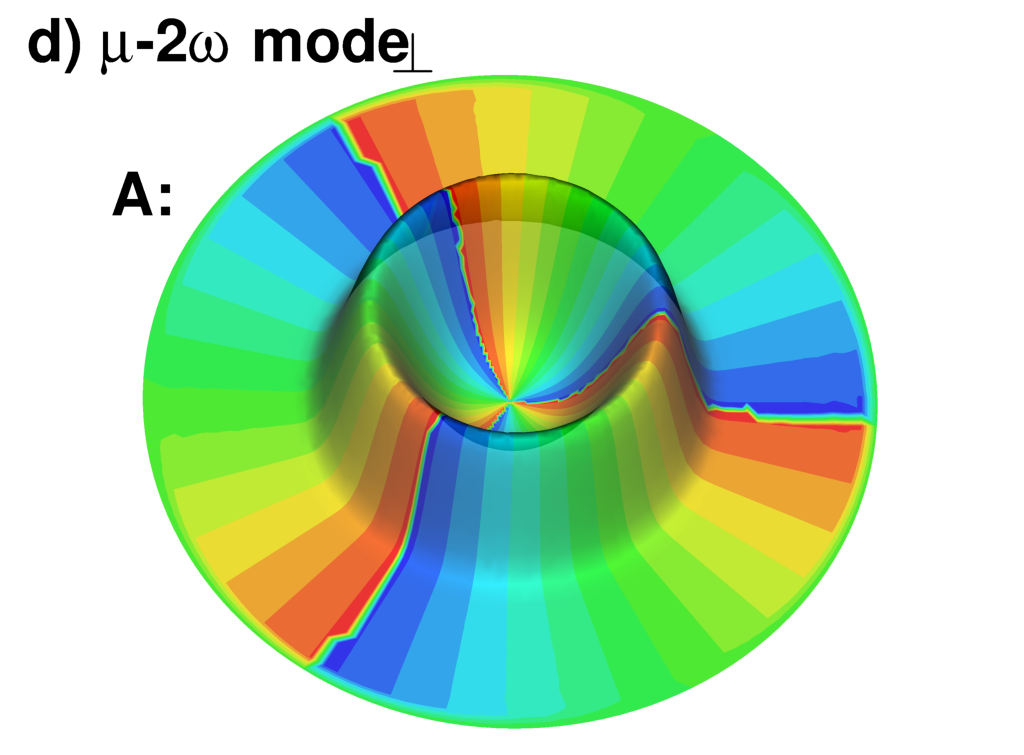}
		\label{fig:2D_VSH_eig_g}
	\end{subfigure}
	\begin{subfigure}[t]{0.45\textwidth}
		\centering
		\includegraphics[width=\textwidth]{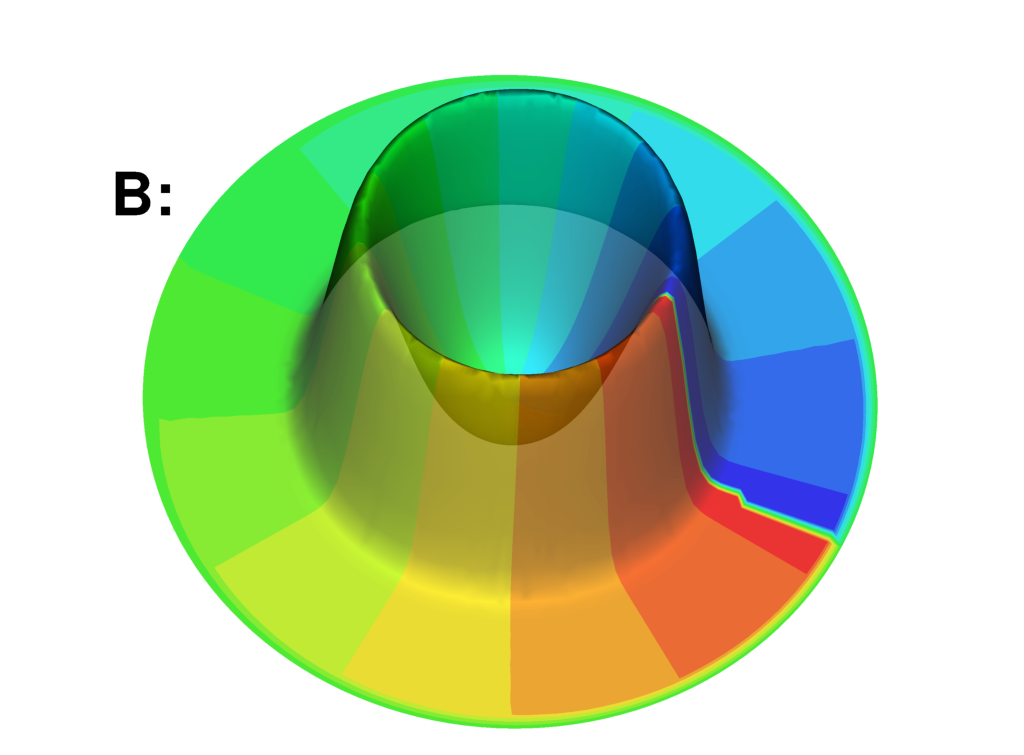}
		\label{fig:2D_VSH_eig_h}
	\end{subfigure}
	\caption{2D BEC with a centered vortex: first four BdG modes $A$ and $B$. Surfaces of modulus coloured by the phase.}
	\label{fig-2D-VSH-eig}
\end{figure}

The first BdG  modes ($A, B$) for this state are displayed in Fig. \ref{fig-2D-VSH-eig} by plotting their modulus coloured by the phase. We can distinguish:
\begin{itemize}
	\item The zero-energy mode $\omega = 0$ (Fig. \ref{fig-2D-VSH-eig}(a)), associated to the phase invariance of the GP equation.
	\item The anomalous mode (Fig. \ref{fig-2D-VSH-eig}(b)) corresponds to the following approximation of its eigenvalue in the Thomas-Fermi limit  
	 \citep{middelkamp2010bifurcations}: 
	\begin{equation}
	\omega = \frac{\omega_\perp^2}{2\mu}\ln(A\frac{\mu}{\omega_\perp}) \approx 0.03261667238,\quad A\approx 2\sqrt{2}\pi \approx 8.886.
	\end{equation}
	This value  is represented by a dashed green line in Figs. \ref{fig-2D-VSH} (a) and (b).
	\item The dipole or Kohn mode (Fig. \ref{fig-2D-VSH-eig}(c)) corresponds to $\omega = \omega_\perp$ and  is independent of $\mu$. 
	\item The 4-th mode (Fig. \ref{fig-2D-VSH-eig}(d)) corresponding to $\omega = \mu - 2\omega_\perp$ in the linear limit \citep{middelkamp2010stability}. This value is
      represented by a continuous green line in Figs. \ref{fig-2D-VSH} (a) and (b).
\end{itemize}

\newpage

\clearpage

\subsection{3D case: ground state}

\begin{figure}[!h]
	\centering
	\begin{subfigure}[t]{0.45\textwidth}
		\centering
		\includegraphics[width=\textwidth]{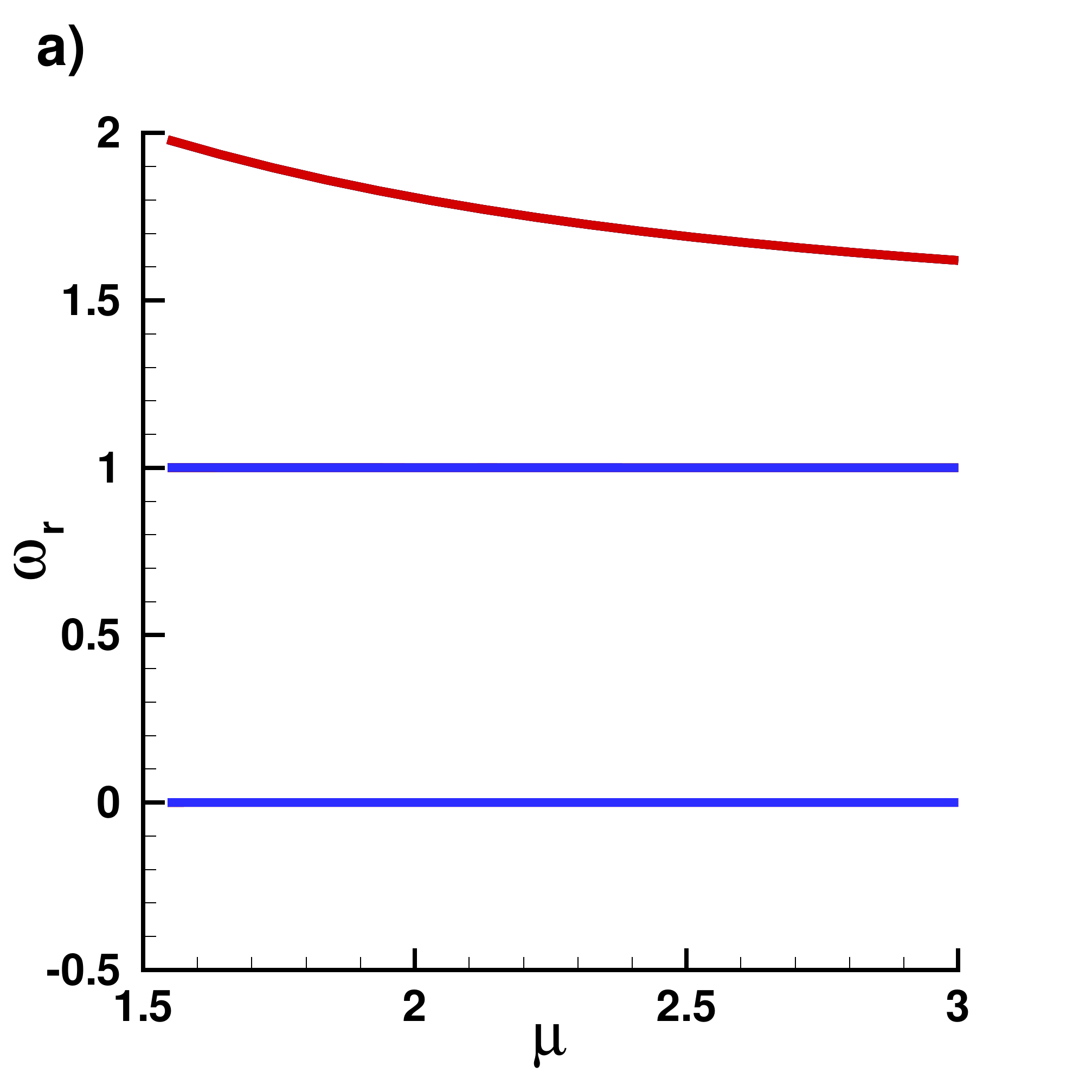}
		\label{fig:3D_TF_a}
	\end{subfigure}
	\hfill
		\begin{subfigure}[t]{0.45\textwidth}
			\centering
			\includegraphics[width=\textwidth]{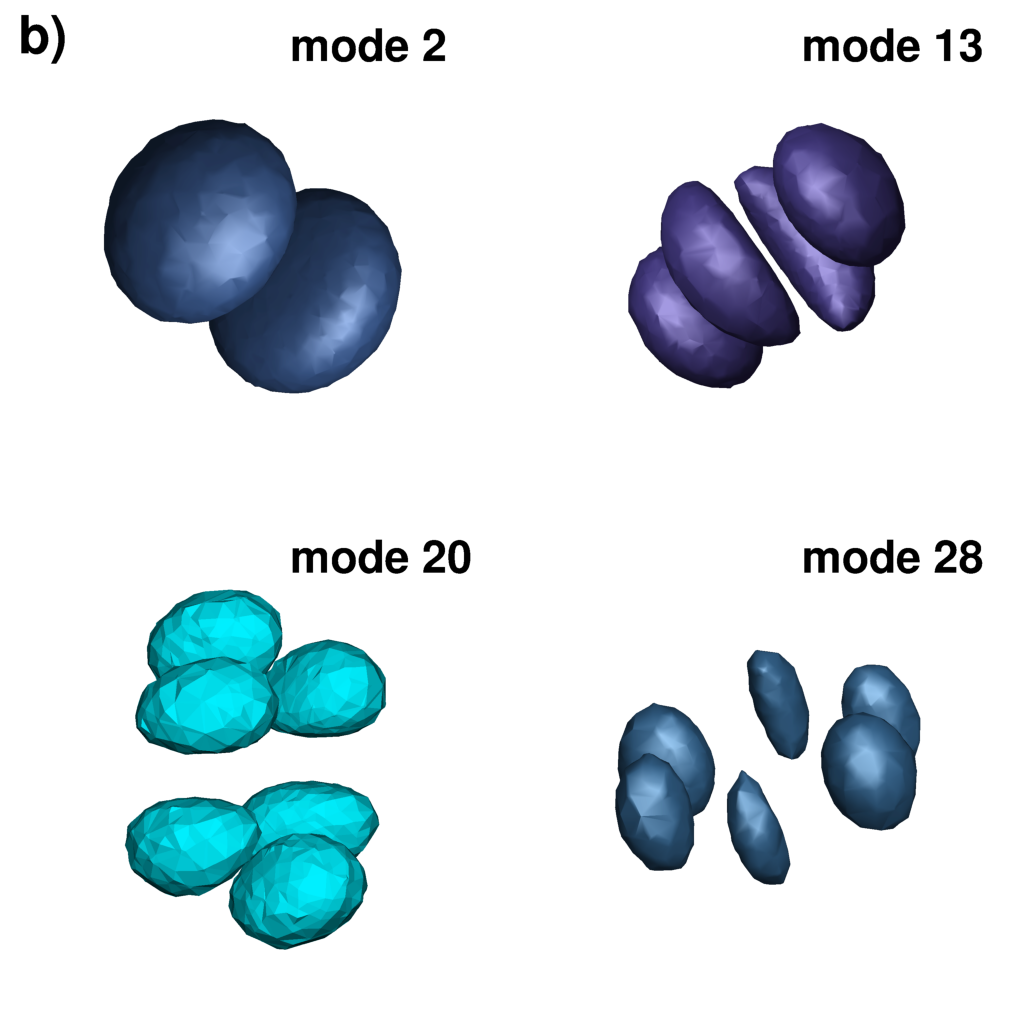}
			\label{fig:3D_TF_b}
		\end{subfigure}

	\caption{3D case: ground state. a) Real part $\omega_r$ of the eigenvalues as a function of $\mu$, b) illustration of four BdG modes (iso-surfaces of the modulus).}
\label{fig-3D-TF}
\end{figure}

Computing the BdG modes for a  3D BEC is a challenging numerical problem. Even with mesh adaptation, the number of degrees of freedom is high and increases with the size of condensate (\ie with $\mu$). For this test case, we computed the spectrum of the ground state of a spherical BEC with trapping potential $V_\trap=\frac{1}{2} m \omega_\perp^2 r^2$, where $r^2=x^2+y^2+z^2$. We set $\omega_\perp=1$. The eigenvalues presented in Fig. \ref{fig-3D-TF} are in very good agreement with  numerical results obtained in \cite{bisset2015robust}. This case shows that our finite-element toolbox can be used to study simple 3D configurations. For more complicated states, the use of parallelization is mandatory to reduce the computational time and memory requirements.

\pagebreak

\section{Validation test cases for the two-component BEC}\label{sec-valid2c}

For the two-component BEC, we compute BdG modes for the dark-antidark solitary 1D or 2D waves studied in \cite{dan-2016-PRA}. Antidark solitary waves
are bright solitary waves on top of a finite background. Such states appear in a two-component system with inter-component repulsion: a dark soliton or a vortex (or a ring) in one component will induce an effective potential,
through the inter-component nonlinearity, on the second component. The result is that atoms of the second component are attracted into the dip of the first
one. We consider the system \eqref{eq-scal-GP2c-stat} in the case of repulsive inter-component interactions with miscibility condition $0 \leq \beta_{12} < \sqrt{\beta_{11} \beta_{22}}$ that ensures that the two components co-exist outside the dark-antidark state. To simplify the case study, since only the ratio between non-linear interaction constants matters, we set $\beta_{11}=\beta_{22}=\beta=1$, $\beta_{12}=\beta_{21}$ and $0<\beta_{12} < \beta$. 

The considered two-component cases are summarized in Tab. \ref{tab-cputime-2c} displaying the necessary computational times and mesh sizes. Note that the codes consider independent values for coefficients $\beta_{ij}, 1 \leq i,j \leq 2$, and thus can be used to study configurations different from those analysed in  \cite{dan-2016-PRA}.

\begin{table}[h!]
	\resizebox{\textwidth}{!}{%
		\begin{tabular}{l|rrr|rrr|}
			\cline{2-7}
			& \multicolumn{3}{c|}{Without mesh adaptation} & \multicolumn{3}{c|}{With mesh adaptation}   \\
			& CPU time GP  & CPU time BdG & mesh size & CPU time GP & CPU time BdG & mesh size \\ \hline
   
\multicolumn{1}{|l|}{1D dark-antidark state}   & 00:00:30          & 00:08:48         & 2714      &             &              &           \\
\multicolumn{1}{|l|}{2D vortex-antidark state} & 00:07:46         & 01:02:21         & 10469     & 00:14:34        & 01:27:09         & 7874      \\
\multicolumn{1}{|l|}{2D ring-antidark state}   & 00:09:22         & 01:48:45        & 10469     & 00:11:05        & 01:59:30        & 9533     \\ \hline			
		\end{tabular}%
	}
	\caption{Test cases for the two-component BEC. Computational time  and mesh size (number of elements). All computation were performed on a Macbook pro M1, 16GB of DDR4 2400 MHz RAM.}
	\label{tab-cputime-2c}
\end{table}

\subsection{1D case: dark-antidark soliton}

\begin{figure}[ht!]
	\centering
	\begin{subfigure}[t]{0.35\textwidth}
		\centering
		\includegraphics[width=\textwidth]{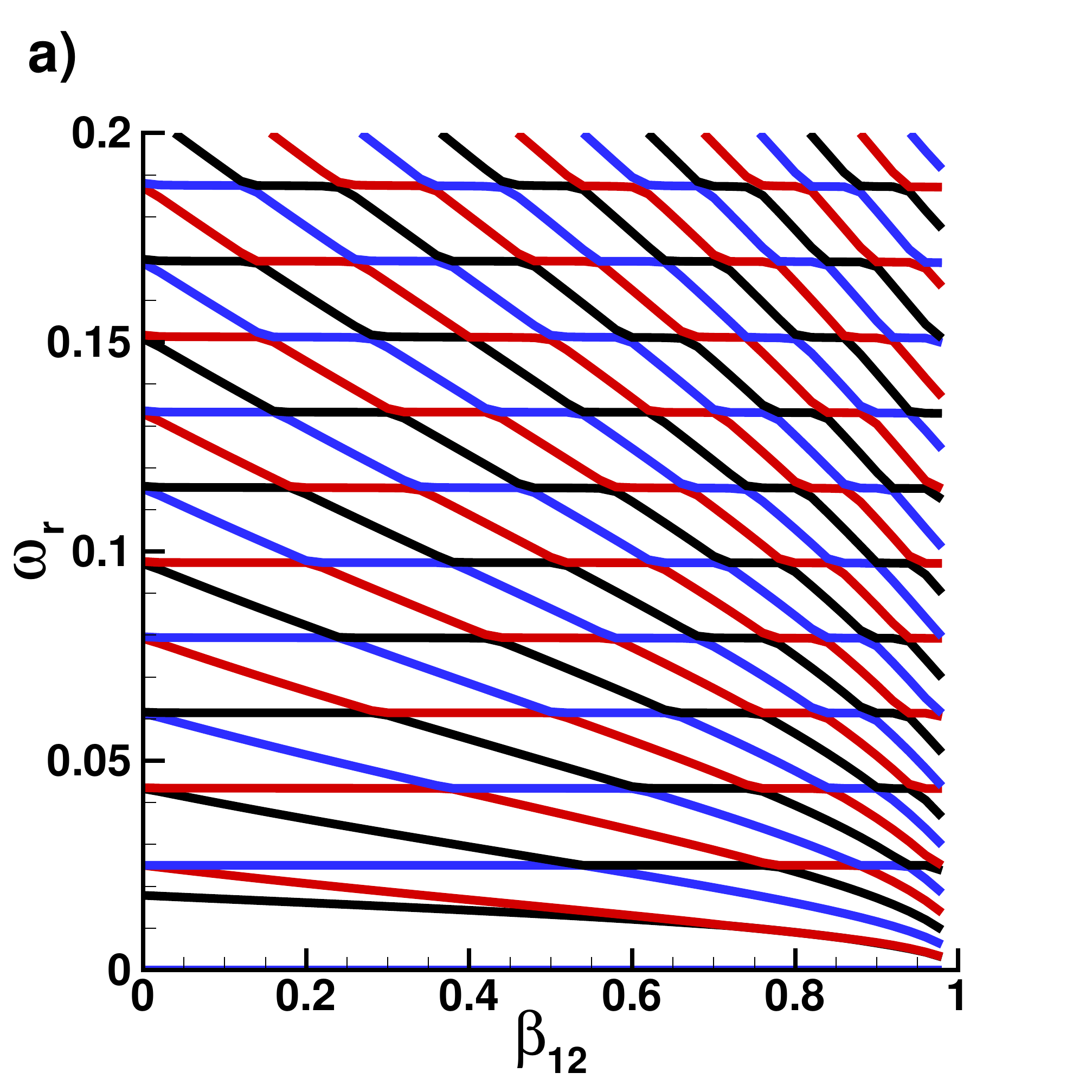}
		\label{fig:1D_DS_TF_a}
	\end{subfigure}
	\begin{subfigure}[t]{0.35\textwidth}
		\centering
		\includegraphics[width=\textwidth]{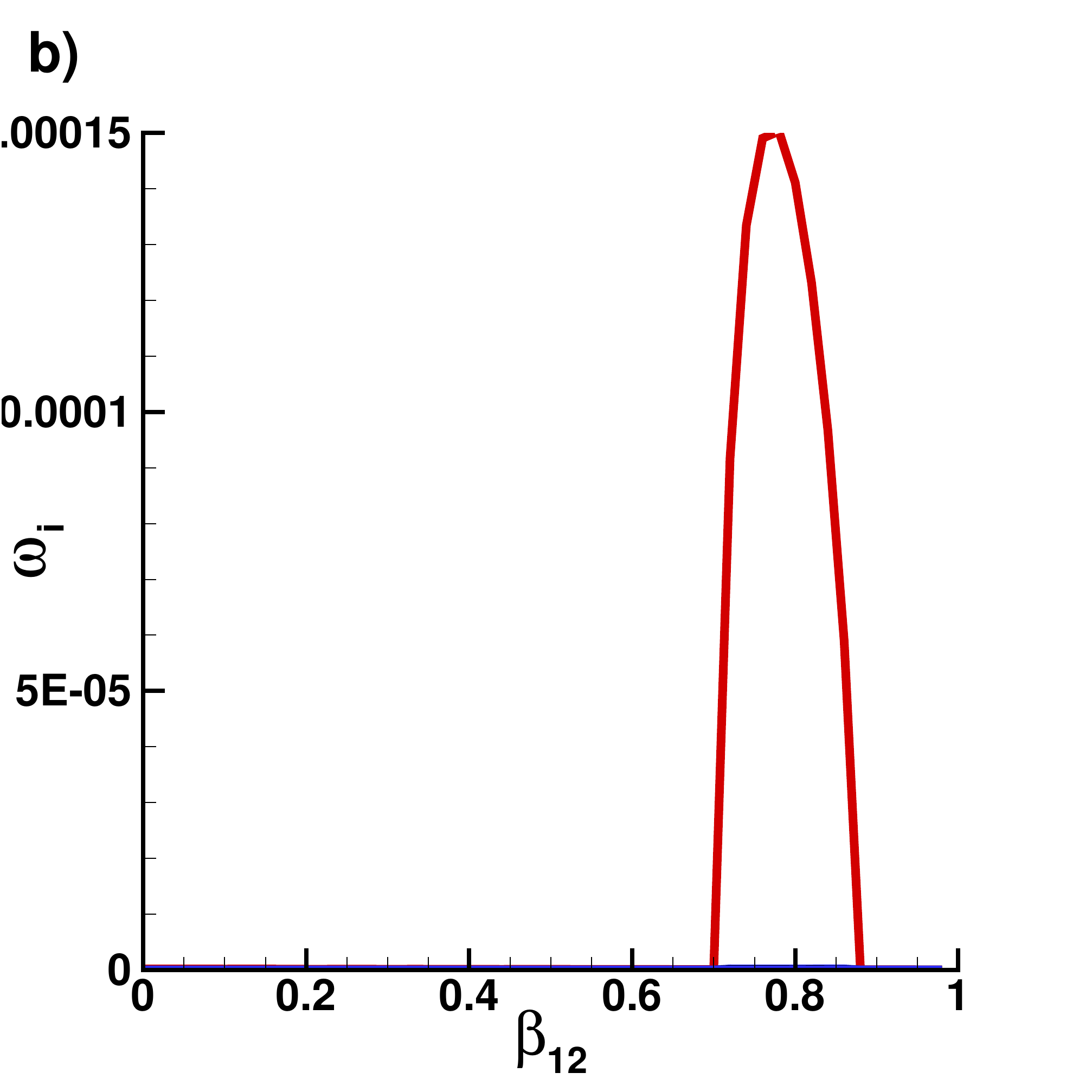}
		\label{fig:1D_DS_TF_b}
	\end{subfigure}\\
	\begin{subfigure}[t]{0.3\textwidth}
		\centering
		\includegraphics[width=\textwidth]{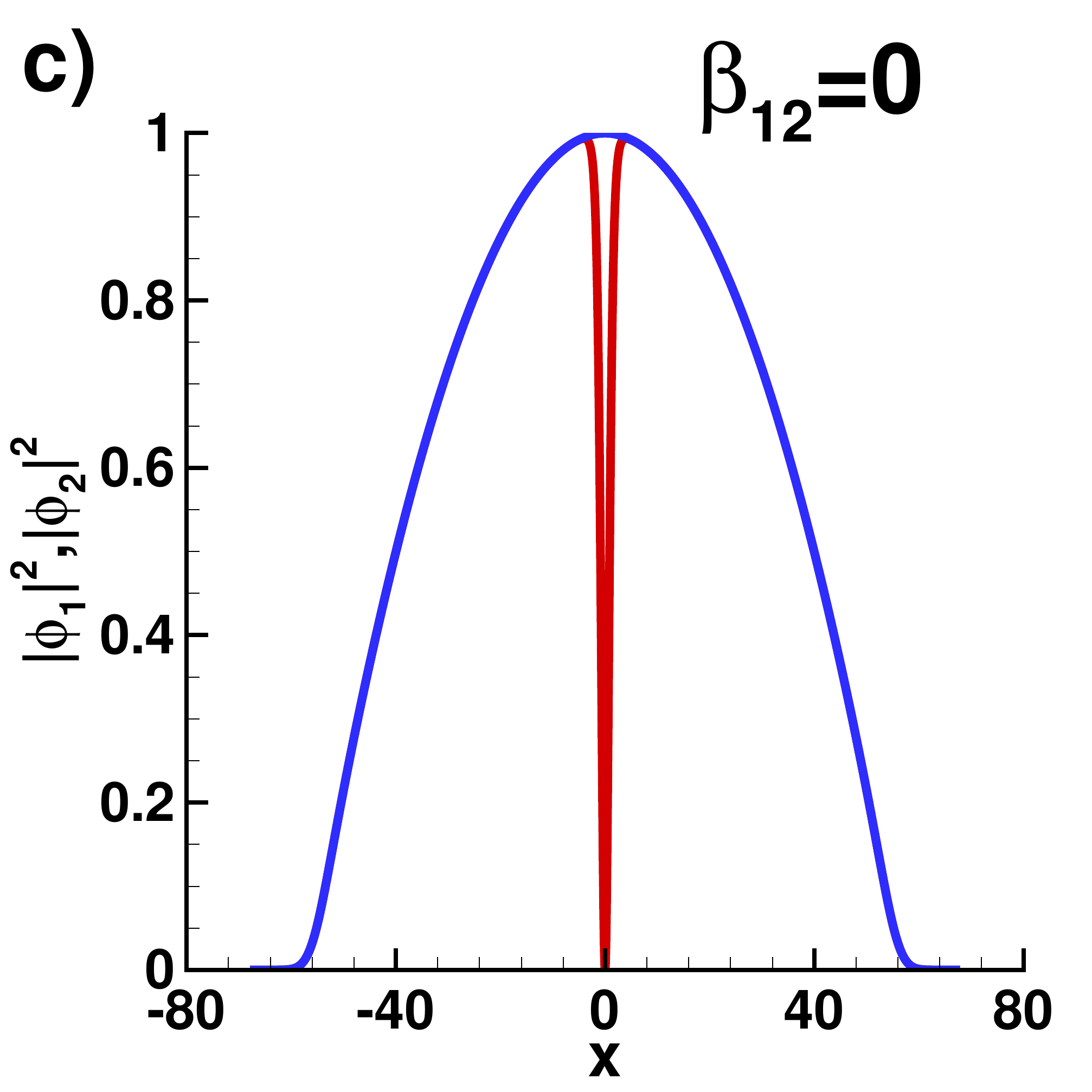}
		\label{fig:1D_DS_TF_c}
	\end{subfigure}
	\begin{subfigure}[t]{0.3\textwidth}
		\centering
		\includegraphics[width=\textwidth]{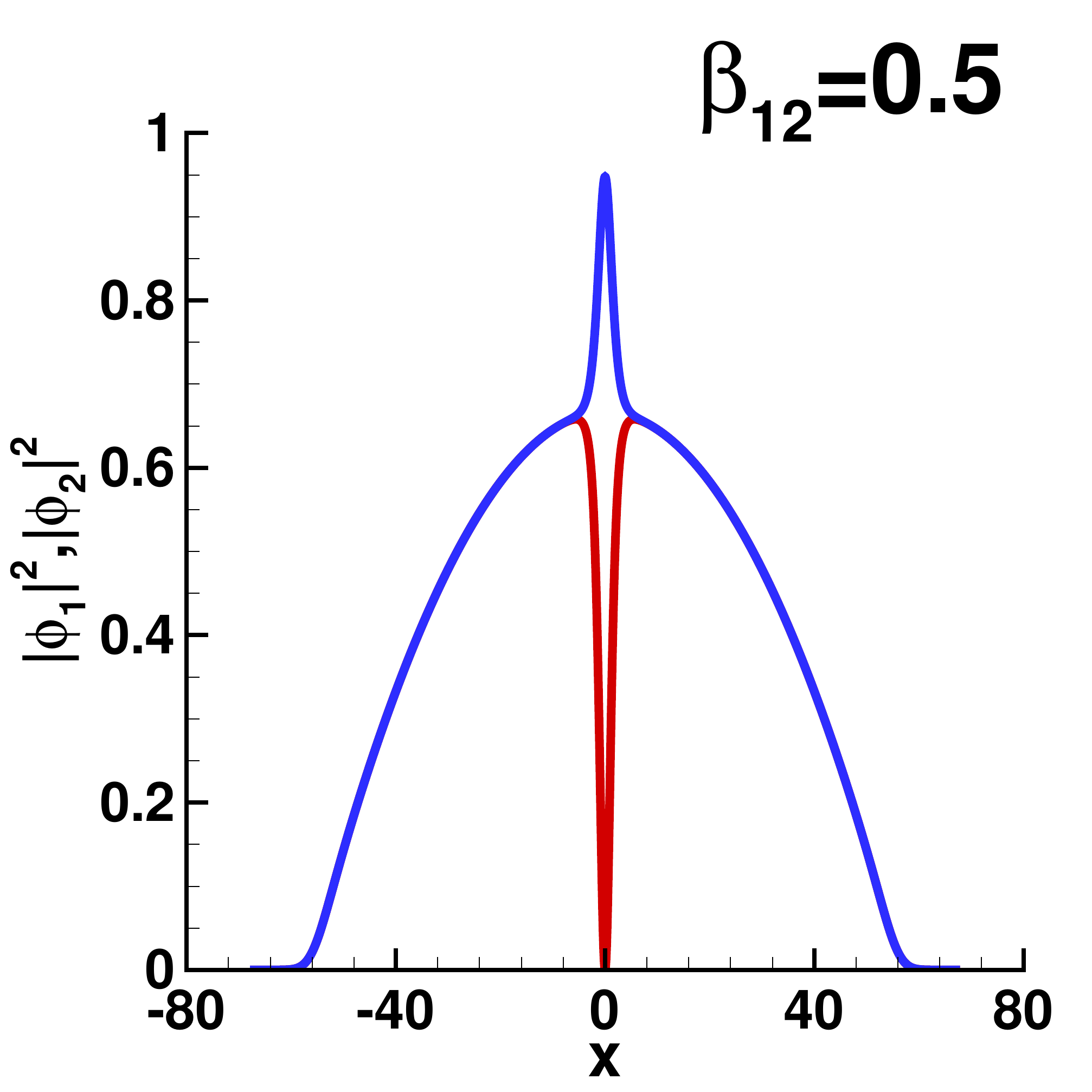}
		\label{fig:1D_DS_TF_d}
	\end{subfigure}
	\begin{subfigure}[t]{0.3\textwidth}
		\centering
		\includegraphics[width=\textwidth]{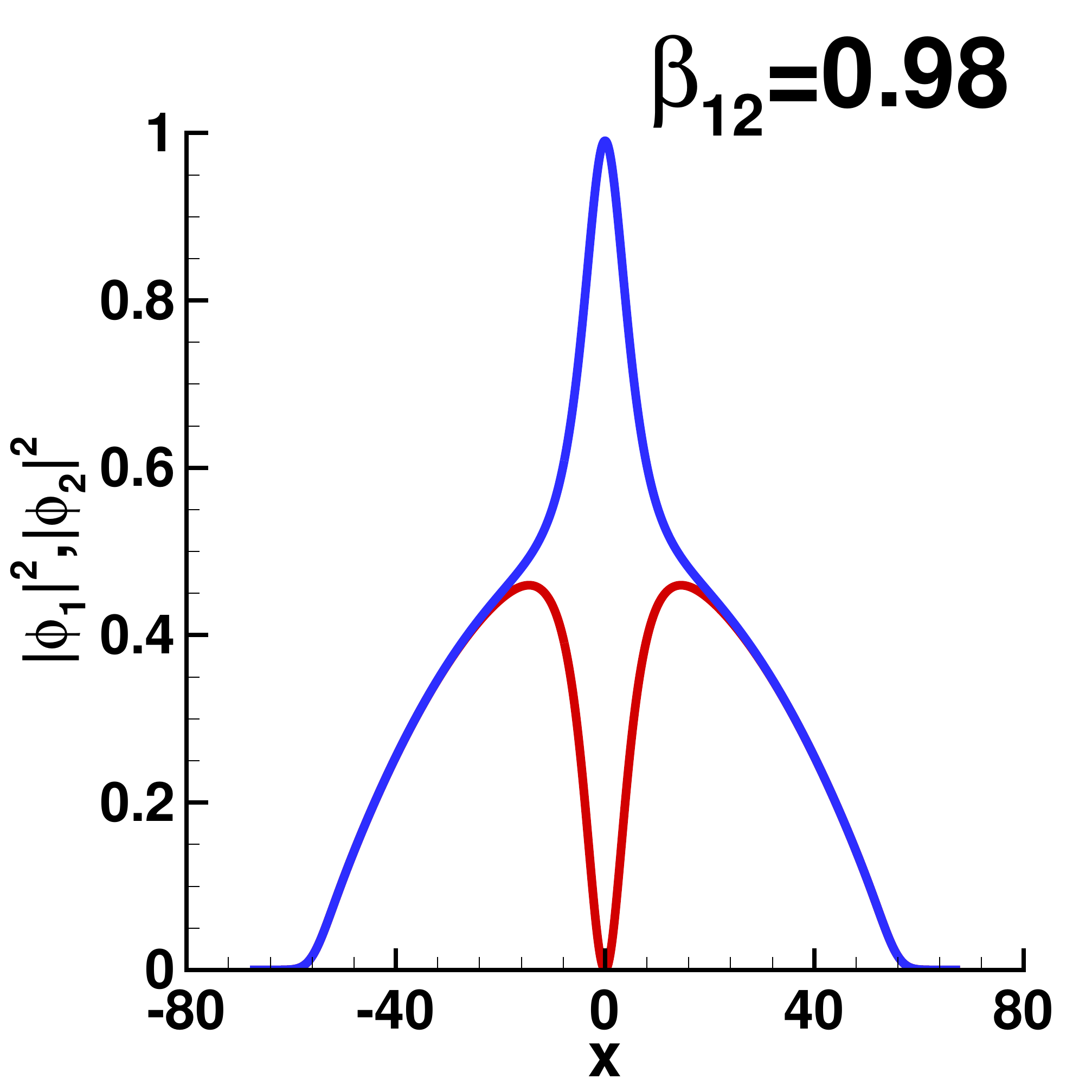}
		\label{fig:1D_DS_TF_e}
	\end{subfigure}
	\caption{1D two-component case: dark-antidark solitary wave. a) Real part $\omega_r$ and b) imaginary part $\omega_i$ of BdG eigenvalues, c) atomic density profiles for three values of $\beta_{12}$.}
	\label{fig-1D-DS-TF}
\end{figure}

The first state  is a dark-antidark solitary wave in 1D. We set a soliton solution (constructed as in Eq. \eqref{eq-DS}) in the first component, while the second component is in the Thomas-Fermi ground state. Obtained eigenvalues are shown in Fig. \ref{fig-1D-DS-TF}(a) and (b) and correspond to the results of \cite{dan-2016-PRA}. The small imaginary instability around $\beta_{12} = \beta_{21} = 0.8$ is well resolved. Profiles of the atomic density for different values of the interaction coefficient are presented in Fig. \ref{fig-1D-DS-TF}(c).

\subsection{2D two-component case: ring-antidark-ring state}

\begin{figure}[h!]
	\centering
	\begin{subfigure}[t]{0.45\textwidth}
		\centering
		\includegraphics[width=\textwidth]{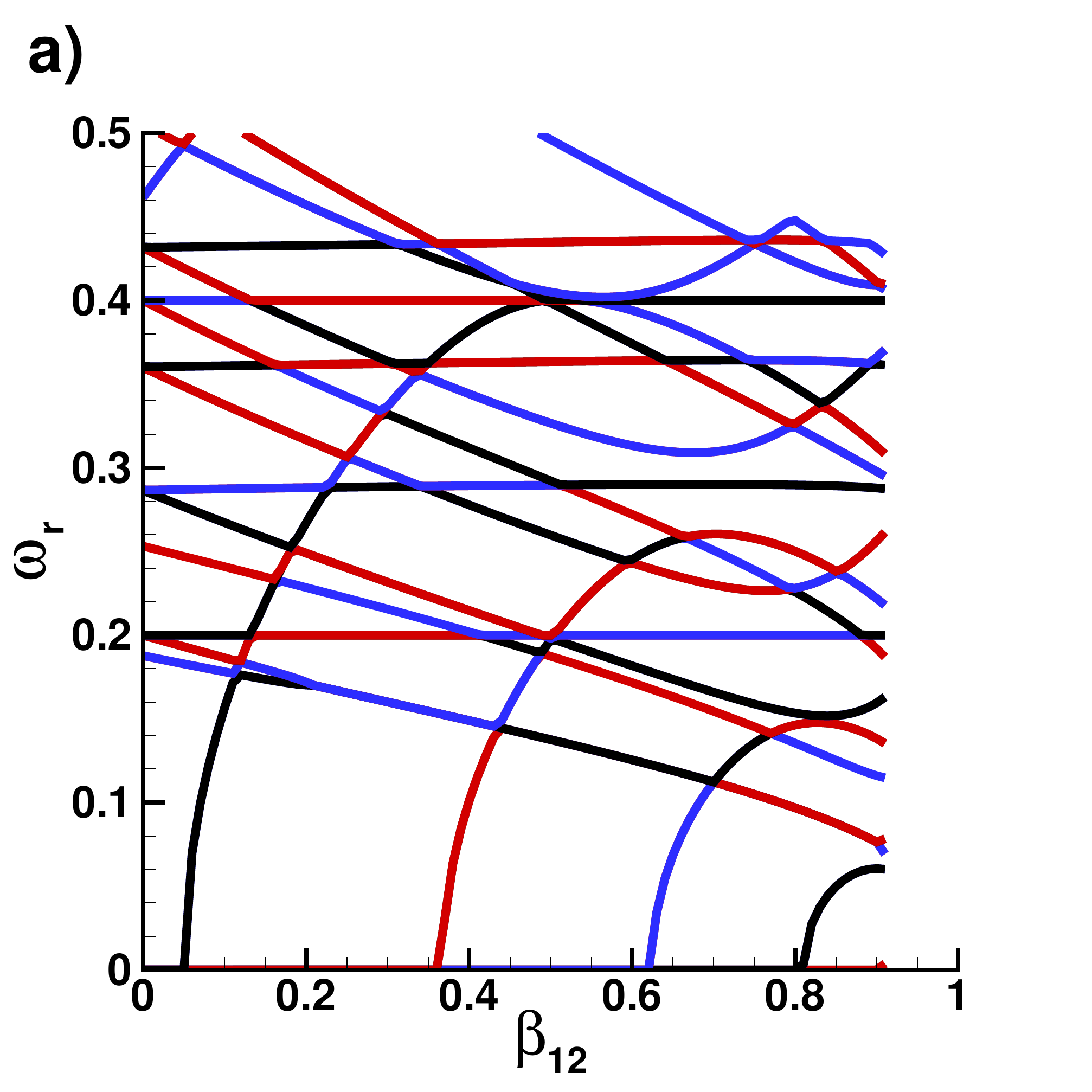}
		\label{fig:2D_RS_TF_a}
	\end{subfigure}
	\begin{subfigure}[t]{0.45\textwidth}
		\centering
		\includegraphics[width=\textwidth]{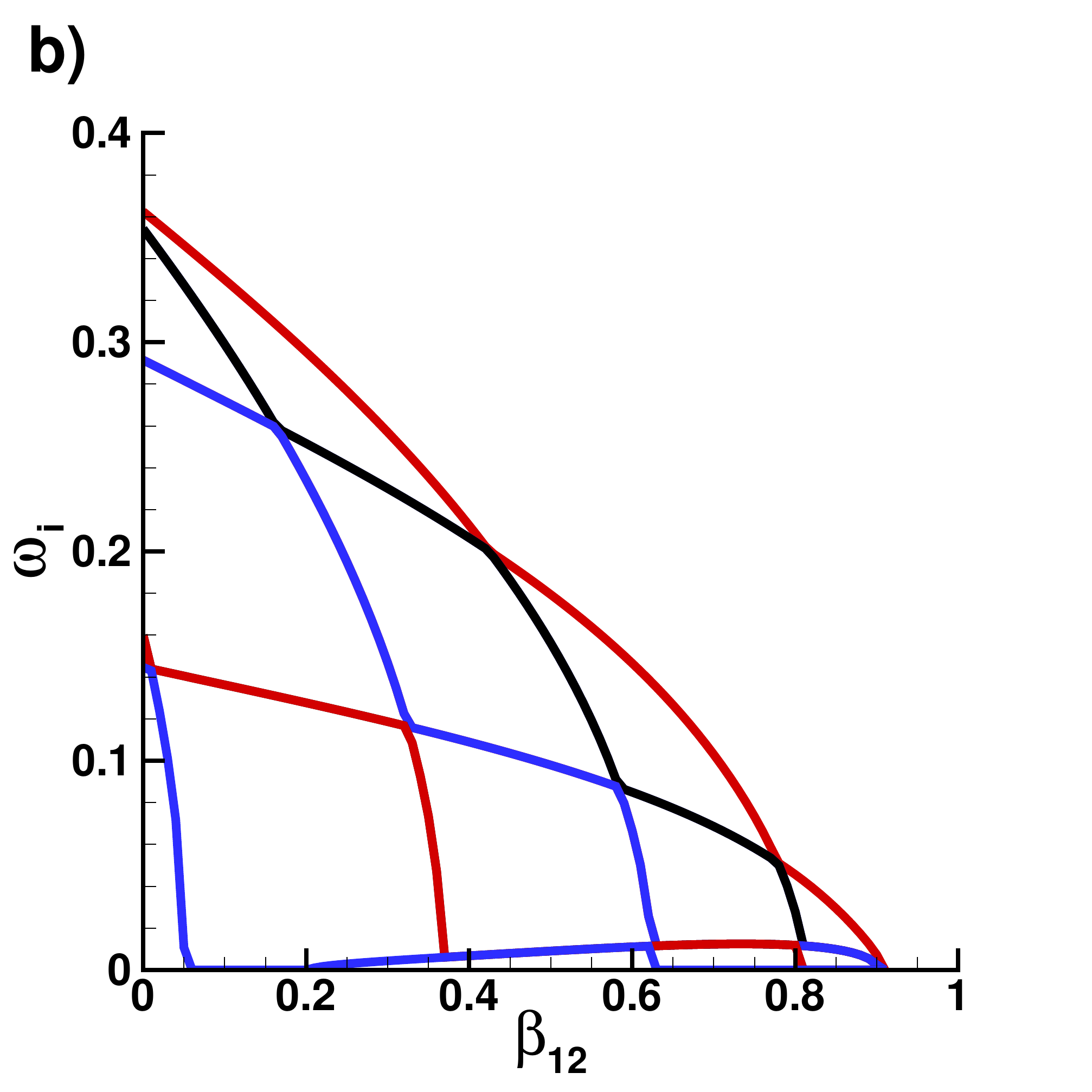}
		\label{fig:2D_RS_TF_b}
	\end{subfigure}
	\begin{subfigure}[t]{0.3\textwidth}
		\centering
		\includegraphics[width=\textwidth]{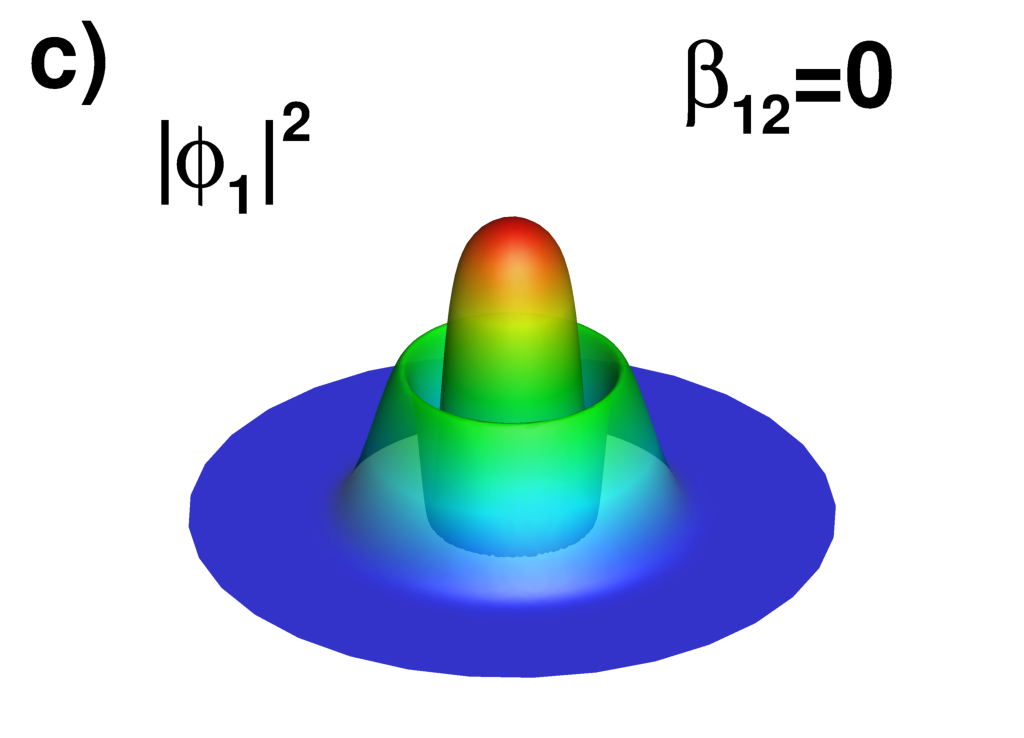}
		\label{fig:2D_RS_TF_c}
	\end{subfigure}
	\begin{subfigure}[t]{0.3\textwidth}
		\centering
		\includegraphics[width=\textwidth]{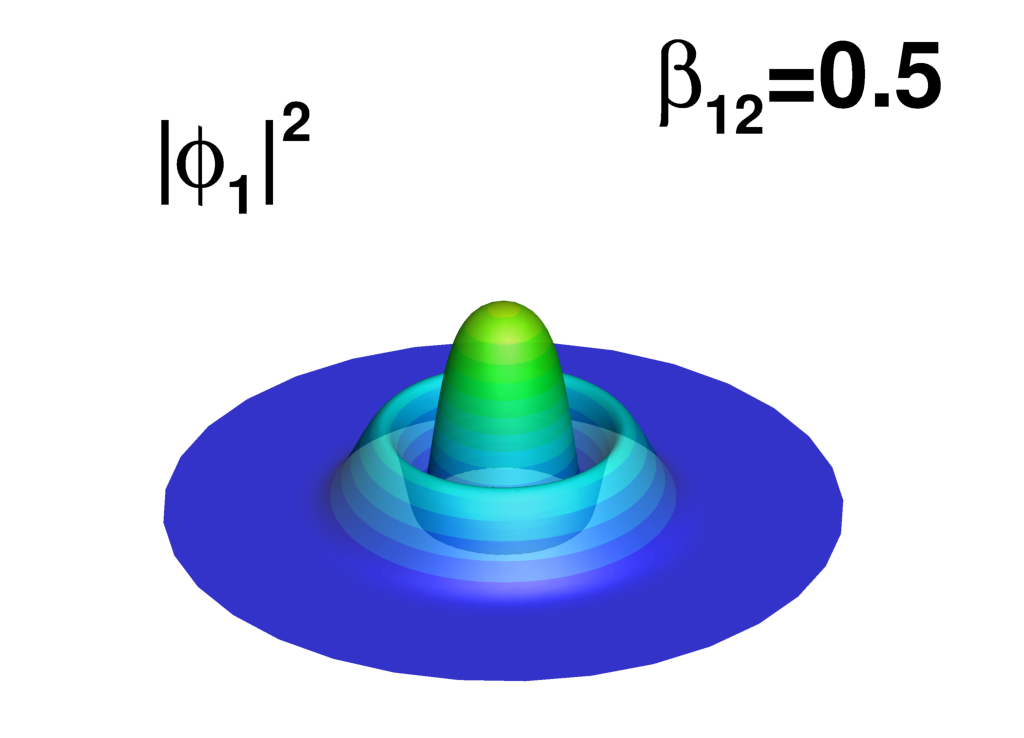}
		\label{fig:2D_RS_TF_d}
	\end{subfigure}
	\begin{subfigure}[t]{0.3\textwidth}
		\centering
		\includegraphics[width=\textwidth]{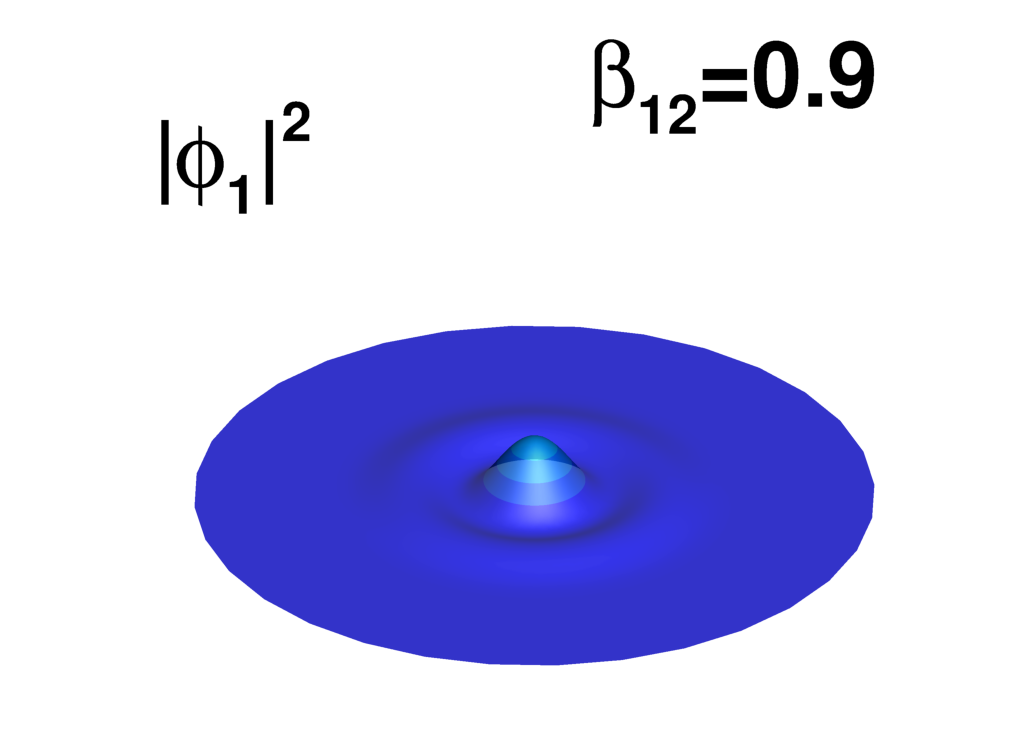}
		\label{fig:2D_RS_TF_e}
	\end{subfigure}
	\begin{subfigure}[t]{0.3\textwidth}
		\centering
		\includegraphics[width=\textwidth]{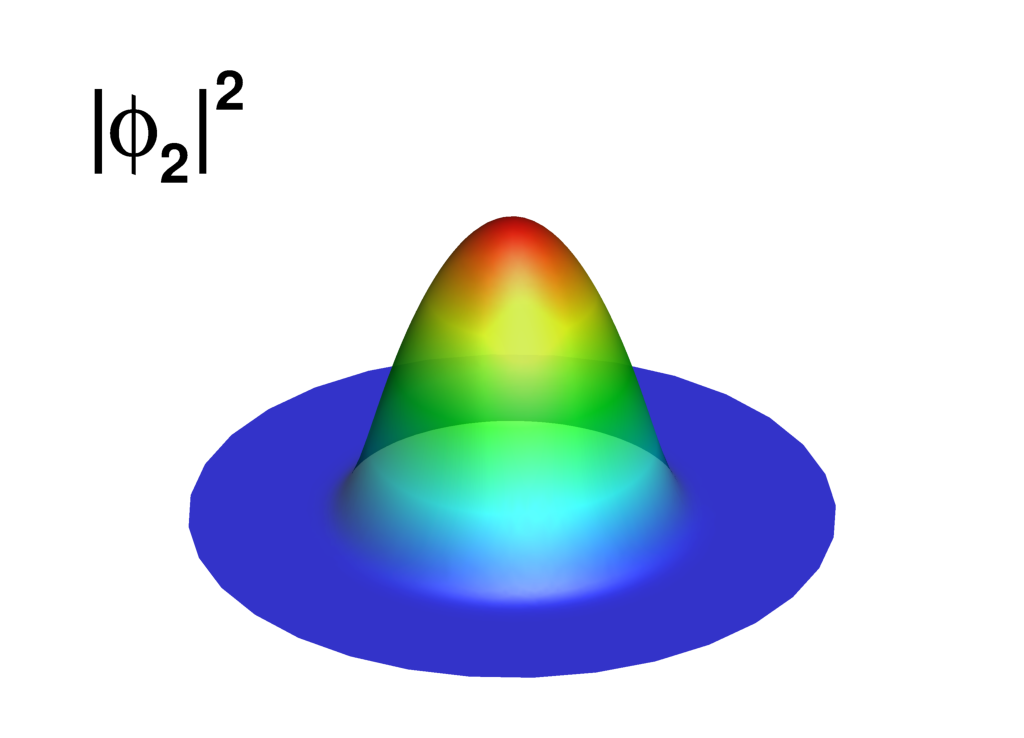}
		\label{fig:2D_RS_TF_f}
	\end{subfigure}
	\begin{subfigure}[t]{0.3\textwidth}
		\centering
		\includegraphics[width=\textwidth]{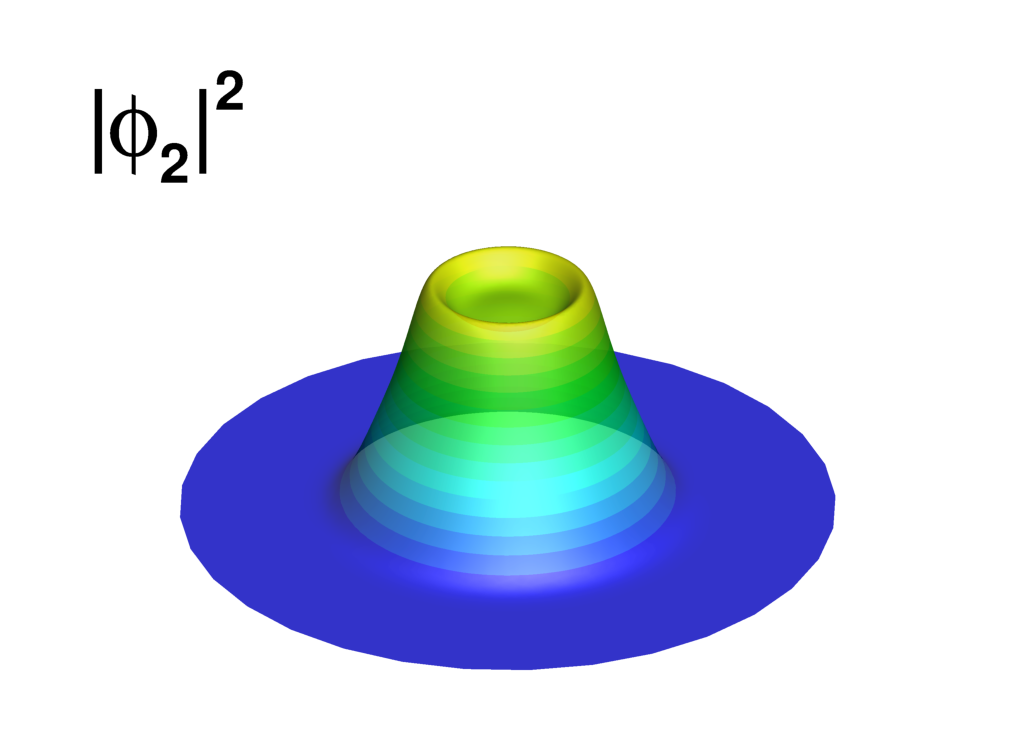}
		\label{fig:2D_RS_TF_g}
	\end{subfigure}
	\begin{subfigure}[t]{0.3\textwidth}
		\centering
		\includegraphics[width=\textwidth]{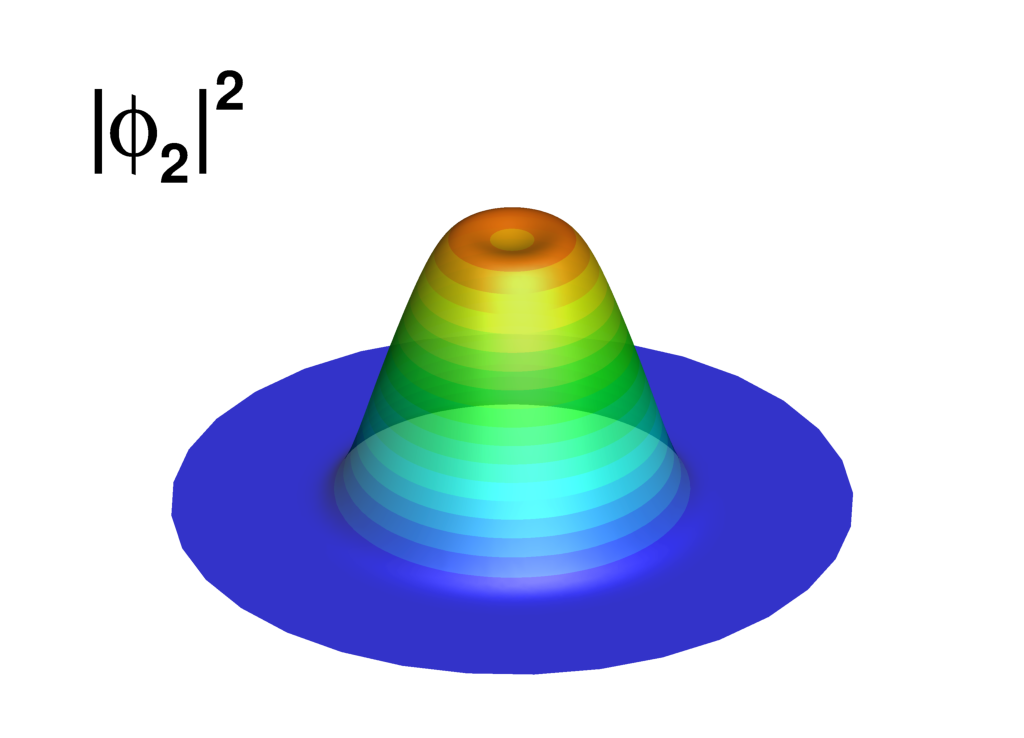}
		\label{fig:2D_RS_TF_h}
	\end{subfigure}
	\caption{2D two-component case: ring-antidark-ring state. a) Real part $\omega_r$ and b) imaginary part of eigenvalues, c) density profiles for three values of $\beta_{12}$.}
	\label{fig-2D-RS-TF}
\end{figure}

With the toolbox, we provide two test cases for 2D configurations: the vortex-antidark  and ring-antidark solitary waves. We show here only the case with ring-antidark solitary waves. The first component contains a ring soliton and the second is in the ground state. Results are shown in Fig. \ref{fig-2D-RS-TF}(a) and (b) for the real and imaginary parts of the eigenvalues. Figure \ref{fig-2D-RS-TF}(c) shows for the atomic density profiles which correspond to the figures presented in \cite{dan-2016-PRA}.

\pagebreak
\clearpage

\section{Description of the programs }\label{sec-desc-prog}

In this section, we first describe the architecture of the programs and the organisation of the provided files. We then present the input parameters and the structure of the output files.

\subsection{Program architecture}

Codes and data files forming the BdG toolbox are stored in the \texttt{FFEM$\_$BdG$\_$toolbox} directory, which is organized around two main subdirectories: \texttt{BdG\_1comp} and \texttt{BdG\_2comp}, corresponding to the one- and two-component codes. Each subdirectory contains two main files: {\em FFEM\_GP\_\$case.edp}, which is the main \ff script file for the computation of the stationary state, and {\em FFEM\_BdG\_\$case.edp} which is the main \ff\ script file for the computation of the BdG eigenvalues. To run the computation of the Gross-Pitaevskii stationary state, the user can use the command \texttt{FreeFem++ FFEM$\_$GP$\_$\$case.edp}. BdG eigenvalues can then be computed with the command \texttt{FreeFem++ FFEM$\_$BdG$\_$\$case.edp}. Parameter files for the examples presented in this paper are stored in the \texttt{INIT} folder.

The obtained solutions are saved in the \texttt{dircase} directory. Depending on the output format selected by the user, data files are generated in specific folders for visualization with Tecplot, Paraview or Gnuplot. We also provide in the folder \texttt{Figures} ready-made layouts for Tecplot. The user can thus obtain the figures from this paper using newly generated data. More details about the output structure are given in Sect. \ref{sec-outputs-bdg}.

The complete architecture of the \texttt{BdG\_1comp} directory is the following (the architecture of the \texttt{BdG\_2comp} directory is almost identical):
\begin{enumerate}
	\item {\em FFEM\_GP\_\$case.edp}: the main script for the computation of the GP stationary states.
	\item {\em FFEM\_BdG\_\$case.edp}: the main script for computing eigenvalues.
	\item {\em param\_num\_common.inc}: a parameter file for the main numerical parameters.
	\item \texttt{INIT}: directory storing the parameter files for the examples presented in Sect. \ref{sec-valid1c}.
	\item \texttt{Figures}:  directory containing Tecplot layouts used to replot the figures shown in Sect. \ref{sec-valid1c}. The main code must be run with the associated example before opening the layout to replot the figure. For some examples, it is necessary to run the case with different parameters (e.g. with and without mesh adaptation) before opening the layout.
	\item \texttt{A\_macro}: directory containing macros used in the main scripts.
\end{enumerate}

\subsection{Macros and functions}

The different macros and functions used in the toolbox for the sequential code are stored in the \texttt{A\_macro} folders:
\begin{itemize}
	\item {\em Macro\_BdGsolve.edp}: macro computing the BdG eigenvalues corresponding to  matrices in Eqs. \eqref{eq-num-BdG-weak} and \eqref{eq-num-BdG2c-weak}.
	\item {\em Macro\_createdir.edp}: macro creating the file structure of the \texttt{dircase} folder.
	\item {\em Macro\_GPsolve.edp}: macro computing the GP stationary state with a Newton method (see Eqs. \eqref{eq-num-GP-stat-Newton} and \eqref{eq-num-GP2c-stat-Newton-1}-\eqref{eq-num-GP2c-stat-Newton-4}).
	\item {\em Macro\_meshAdapt.edp}: macro adapting the mesh to the wave function.
	\item {\em Macro\_operator.edp}: definitions of useful macros and functions: gradients, energy \eqref{eq-NRJ}, chemical potential \eqref{eq-GP-mu}, Hermite polynomials, etc. Also contains a macro creating a spherical mesh for  3D problems.
	\item {\em Macro\_output.edp}: macros used to save data in Tecplot and Paraview formats.
	\item {\em Macro\_plotEigenvector.edp}: macro plotting the real and imaginary parts of a BdG eigenvector.
	\item {\em Macro\_plotphi.edp}: macro plotting the complex wave function. The user can press "k"  to alternate between plots of the density, phase and real and imaginary parts of the wave function.
	\item {\em Macro\_problem.edp}: definitions of the weak formulations for the GP (Eqs. \eqref{eq-num-GP-stat-Newton} or \eqref{eq-num-BdG2c-weak}) and BdG problems  (Eqs. \eqref{eq-num-BdG-weak} or  \eqref{eq-num-GP2c-stat-Newton-1}-\eqref{eq-num-GP2c-stat-Newton-4}).
	\item {\em Macro\_restart.edp}: macros used to save and load the wave function to or from \ff files.
	\item {\em Macro\_saveData.edp}: macro saving the stationary wave function.
	\item {\em Macro\_saveEigenvalues.edp}: macro saving the BdG eigenvalues and eigenvectors.
\end{itemize}

\subsection{Input parameters}

Parameters are separated in two files. Numerical parameters used in all computations are in {\em param\_num\_common.inc}. Files in the \texttt{INIT} directory specify physical parameters describing the state that will be studied during a computation and numerical parameters specific to this problem. The files distributed with the toolbox provide a variety of examples that can be used as a starting point when selecting parameters for the study of new states. \\
\textbf{(1)} In the file \texttt{param\_num\_common.inc}, the parameters are:
\begin{itemize}
	\item \textbf{displayplot}: controls the output information to plot. Possible values range from $0$ (no plots), to {$2$} (plots data at all iterations of the Newton method and all eigenvectors computed by the BdG code).
	\item \textbf{iwait}: a Boolean indicating if the code must wait for user input when a plot is shown (\texttt{true}) or it can continue (\texttt{false}) with the next plot.
	\item \textbf{cutXY}, \textbf{cutXZ}, \textbf{cutYZ}: (only for 3D cases) Booleans indicating whether to plot cuts of the wave function along the different axis at $x=0$, $y=0$ or $z=0$.
	\item \textbf{Tecplot}: a Boolean indicating whether to save data in the Tecplot format.
	\item \textbf{Paraview}: a Boolean indicating whether to save data in the Paraview format (only in 2D and 3D). 
	\item \textbf{adaptinit}: if \texttt{true}, the initial solution is  recomputed after the first mesh adaptation.
	\item \textbf{adaptmeshFF}: determines if mesh adaptation is used (\texttt{true}) or not (\texttt{false}).
	\item \textbf{useShift}: a Boolean indicating whether to use a shift when computing the BdG eigenvalues (see Eq. \eqref{eq-bdg-shift}).
	\item \textbf{Nadapt}: if mesh adaptation is used, the mesh is adaptated every \textbf{Nadapt} iterations during the continuation.
	\item \textbf{Nplot}: the wave function is plotted every \textbf{Nplot} iterations during the continuation.
	\item \textbf{Nsave}: the wave function is saved for Paraview or Tecplot every \textbf{Nsave} iterations during the continuation.
	\item \textbf{Nrst}: the wave function is saved for the BdG computation every \textbf{Nrst} iterations during the continuation.
	\item \textbf{tolerrF}: the tolerance value of $\epsilon_{\scriptscriptstyle F}$ in Eq. \eqref{eq-bdg-err1}.
	\item \textbf{tolNewton}: the tolerance value of $\epsilon_{\scriptscriptstyle q}$ in Eq. \eqref{eq-bdg-err1}.
	\item \textbf{shift}: the value of the shift $\sigma$ used when computing eigenvalues.
	\item \textbf{newtonMax}: the maximum number of Newton iterations.
\end{itemize}
\vspace{1cm}
\textbf{(2)} In the file \texttt{\$case.inc}, stored in the \texttt{INIT} directory, the parameters are:
\begin{itemize}
	\item General parameters for the case:\\
	$\bullet$ \textbf{dimension}: the dimension of the problem (1, 2 or 3).\\
	$\bullet$ \textbf{FEchoice}: the type of finite element used. Usually $P2$.\\
	$\bullet$ \textbf{nev}: the number of eigenvalues computed by the BdG code.
	\item Parameters used to restart a computation:\\
	$\bullet$ \textbf{restart}: a boolean indicating if the initial solution is a restart from a previous computation. If \texttt{true}, the solution and mesh stored in \textbf{fcaserestart} for the value of $\mu$ given by \textbf{murestart} will be used as  initial solution.\\
	$\bullet$ \textbf{murestart}: the initial value of $\mu$ in the case of a restart.\\
	$\bullet$ \textbf{fcaserestart}: the folder where the initial solution is stored in the case of a restart.
	\item Parameters of the continuation:\\
	$\bullet$ \textbf{kpol}, \textbf{lpol}, \textbf{mpol}: integers defining the initial state in the linear limit.\\
	$\bullet$ \textbf{startmu}: the initial value of $\mu$.\\
	$\bullet$ \textbf{endmu}: the final value of $\mu$.\\
	$\bullet$ \textbf{dmu}: the increment in $\mu$ during the continuation.\\
	$\bullet$ \textbf{facmu}: when using the linear limit, the initial value of $\mu$ is given by $\textbf{facmu}\cdot \mu_{\ket{klm}}$.
	\item Coefficients of the GP equation:\\
	$\bullet$ \textbf{beta}: the nonlinear coefficient (we set  $\beta=1$ in all test cases).\\
	$\bullet$ \textbf{ax}, \textbf{ay}, \textbf{az}: the frequencies of the trapping potential along the three axes.\\
	$\bullet$ \textbf{Ctrap}: a function defining the trapping potential.   
	\item Parameters for the mesh creation:\\
	$\bullet$ \textbf{Dx}: the distance between points on the mesh border.\\
	$\bullet$ \textbf{scaledom}: a coefficient used to control the size of the domain: the mesh radius is given by $\textbf{Rdom} = \textbf{scaledom}\ r_{\TF}$ where $r_\TF$ is the Thomas-Fermi radius.\\
	$\bullet$ \textbf{createMesh}: a macro creating the initial mesh \texttt{Th}.
	\item Parameters for the mesh adaptation:\\
	$\bullet$ \textbf{errU}: the interpolation error level.\\
	$\bullet$ \textbf{hmin}: the minimum length of a mesh element edge in the new mesh.\\
	$\bullet$ \textbf{hmax}: the maximum length of a mesh element edge in the new mesh.\\
	$\bullet$ \textbf{adaptratio}: the ratio for a prescribed smoothing of the metric. No smoothing is done if the value is less than $1.1$.
	\item Parameters for the initial solution:\\
	$\bullet$ \textbf{initname}: the name given to the initial solution.\\
	$\bullet$ \textbf{initcond}: a macro defining the initial solution for the \texttt{phi} variable.
	\item Definitions of the boundary conditions:\\
	$\bullet$ \textbf{BCGP}: the boundary conditions used in the GP code for Eqs. \eqref{eq-num-GP-stat-Newton} and \eqref{eq-num-GP2c-stat-Newton-1}-\eqref{eq-num-GP2c-stat-Newton-4}.\\
	$\bullet$ \textbf{BCBdG}: the boundary conditions used in the BdG code for Eqs. \eqref{eq-num-BdG-weak} and \eqref{eq-num-BdG2c-weak}.\\
	$\bullet$ \textbf{fcase}: the name given to the current computation.\\
	$\bullet$ \textbf{dircase}: the directory where the results are stored.
\end{itemize}
\vspace{1cm}
\textbf{(3)} In a two component case, some new parameters are defined in the \texttt{\$case.inc} file:
\begin{itemize}
	\item Parameters used to restart a computation:\\
	$\bullet$ \textbf{mu1restart}, \textbf{mu2restart}: initial values of $\mu_1$ and $\mu_2$ in the case of a restart.\\
	$\bullet$ \textbf{beta12restart}, \textbf{beta21restart} initial values of $\beta_{12}$ and $\beta_{21}$ in the case of a restart.
	\item Parameters of the continuation:\\
	$\bullet$ \textbf{startmu1}, \textbf{startmu2}: initial values of $\mu_1$ and $\mu_2$.\\
	$\bullet$ \textbf{endmu1}, \textbf{endmu2}: final values of $\mu_1$ and $\mu_2$.\\
	$\bullet$ \textbf{dmu1}, \textbf{dmu2}: increments of $\mu_1$ and $\mu_2$ during the continuation.\\
	$\bullet$ \textbf{startbeta12}, \textbf{startbeta21}: initial values of $\beta_{12}$ and $\beta_{21}$.\\
	$\bullet$ \textbf{endbeta12}, \textbf{endbeta21}: final values of $\beta_{12}$ and $\beta_{21}$.\\
	$\bullet$ \textbf{dbeta12}, \textbf{dbeta21}: increments of $\beta_{12}$ and $\beta_{21}$ during the continuation.
	\item Coefficients of the GP equation:\\
	$\bullet$ \textbf{beta11}, \textbf{beta12}: nonlinear coefficients $\beta_{11}$ and $\beta_{22}$.
	\item Parameters for the initial solution:\\
	$\bullet$ \textbf{initname1}: the name given to the initial solution for the first component.\\
	$\bullet$ \textbf{initname2}: the name given to the initial solution for the second component.\\
	$\bullet$ \textbf{initcond}: a macro defining the initial solution for \texttt{[phi1,phi2]} variables.\\
\end{itemize}

\subsection{Outputs}\label{sec-outputs-bdg}

When a computation starts, the \texttt{OUTPUT$\_$\$case} directory is created. It contains up to eight folders. The \texttt{RUNPARAM} directory contains a copy of the code and data files, allowing an easy identification of each case and preparing an eventual rerun of the same case. The other folders contains different output format files of the computed solution, to be visualised with Tecplot, Paraview or Gnuplot. The content of these subfolders depends on the case and on the computation parameters (differences in the two component code are given in parenthesis):
\begin{enumerate}
	\item The \texttt{Gnuplot} folder contains two files:\\
	$\bullet$ Informations about the stationary states are stored in the \texttt{GP\_results.dat} file. The columns are in order: the non-linear coefficient $\beta$ ($\beta_{12}$ and $\beta_{21}$), the imposed chemical potential $\mu$ ($\mu_1$ and $\mu_2$), the number of Newton iterations used for this value of $\mu$, the errors 
	$\epsilon_{\scriptscriptstyle F}$ and $\epsilon_{\scriptscriptstyle q}$
	\eqref{eq-bdg-err1}, the computed value of the chemical potential \eqref{eq-GP-mu} (computed values of $\mu_1$ and $\mu_2$), the number of atoms \eqref{eq-GP-N} (the number of atoms in the two components), the GP energy \eqref{eq-NRJ}, the mesh size, the number of degrees of freedom and the CPU time  to compute the stationary state.\\
	$\bullet$ BdG eigenvalues are stored in the \texttt{BdG\_results.dat} file. The columns are in order: the non-linear coefficient $\beta$ ($\beta_{12}$ and $\beta_{21}$), the imposed chemical potential $\mu$ ($\mu_1$ and $\mu_2$), the eigenvalue number between 0 and \textbf{nev}, the real and imaginary part of the eigenvalues, the Krein signature and its sign (the Krein signature and its sign for the two components), the residual \eqref{eq-bdg-residual} and the CPU time to compute the eigenvalues.
	\item The \texttt{Paraview} folder contains the wave functions stored as {\em .vtk} or {\em .vtu} and {\em .pvd} files:\\
	$\bullet$ \texttt{phi\_init.vtu} and \texttt{phi\_final.vtu} are the initial and final solutions.\\
	$\bullet$ \texttt{phi\_mu\_\$mu.vtu} contains the stationary wave function for a given value of $\mu$.\\
	$\bullet$ \texttt{phi\_mu1\_\$mu1\_mu2\_\$mu2.vtu} contains the stationary wave function for given values of $\mu_1$ and $\mu_2$ in the first continuation.\\
	$\bullet$ \texttt{phi\_beta12\_\$beta12\_beta21\_\$beta21.vtu} \enlargethispage{\baselineskip}contains the stationary wave function for given values of $\beta_{12}$ and $\beta_{21}$ in the second continuation.
	\item The \texttt{Paraview\_Eigenvectors} folder contains the eigenvectors stored as:\\
	$\bullet$ \texttt{eVec\_mu\_\$mu\_\$nev.vtu} in the one component code.\\
	$\bullet$ \texttt{eVec\_beta12\_\$beta12\_beta21\_\$beta21\_mu1\_\$mu1\_mu2\_\$nev.vtu} in the two component code.
	\item The \texttt{RST} folder contains the stationary states stored as \ff files. The names are:\\
	$\bullet$ \texttt{RST-\$mu.rst} or \texttt{RST-\$mu1-\$mu2-\$beta12-\$beta21.rst} for the data.\\
	$\bullet$ \texttt{RSTTh-\$mu} or \texttt{RSTTh-\$mu1-\$mu2-\$beta12-\$beta21} for the mesh files. The file extensions are {\em .mesh} (in 1D), {\em .msh} (in 2D) or {\em .meshb} (in 3D).
	\item The \texttt{Tecplot} folder contains the wave functions stored as {\em .dat} Tecplot files:\\
	$\bullet$ \texttt{phi\_init.dat} and \texttt{phi\_final.dat} are the initial and final solutions.\\
	$\bullet$ \texttt{phi\_mu\_\$mu.dat} contains the stationary wave function for a given value of $\mu$.\\
	$\bullet$ \texttt{phi\_mu1\_\$mu1\_mu2\_\$mu2.dat} contains the stationary wave function for given values of $\mu_1$ and $\mu_2$ in the first continuation.\\
	$\bullet$ \texttt{phi\_beta12\_\$beta12\_beta21\_\$beta21.dat} contains the stationary wave function for given values of $\beta_{12}$ and $\beta_{21}$ in the second continuation.
	\item The \texttt{Tecplot\_Eigenvectors} folder contains the eigenvectors stored in the Tecplot format:\\
	$\bullet$ \texttt{eVec\_mu\_\$mu\_\$nev.dat} in the one component code.\\
	$\bullet$ \texttt{eVec\_beta12\_\$beta12\_beta21\_\$beta21\_mu1\_\$mu1\_mu2\_\$nev.dat} in the two component code.
	\item The \texttt{Tecplot\_Eigenvalues} folder contains the eigenvalues stored in the Tecplot format. Filenames are \texttt{BdG\_results\_\$i.dat}. Each file contains the $i$-th smallest eigenvalue for each value of $\mu$ (or $\beta_{12}$ in the two-component code).
\end{enumerate}

\section{Summary and conclusions}\label{sec-conclusions}

The aim of the toolbox presented in this paper is the computation of stationary states and BdG modes of one- and two-component BECs in 1D and 2D. The use of mesh adaptation enables an efficient computation of stationary states by adapting the spatial discretization to the topology of the considered state. This makes possible the study of various 1D and 2D problems and even a simple 3D configuration without parallelization. 
The toolbox was created with \ff, a free and open-source finite element software for the study of partial differential equations. The method consists of two steps: (i) a Newton method, combined with a continuation on the chemical potential $\mu$ or the inter-component interaction parameters $\beta_{12}$ and $\beta_{21}$, is used to obtain the stationary state of the GP equation, (ii) the BdG modes are obtained by solving an eigenvalue problem with \texttt{ARPACK}. The numerical code was validated against test cases studied theoretically or numerically in the literature. All parameter files corresponding to these test cases are shared with the toolbox, allowing the user to obtain the results presented in this paper. These parameter files can be used as templates to initiate the study of other BEC states. We considered only one and two component systems with a cubic nonlinearity, but the toolbox could be easily modified to study other configurations such as quartic $\pm$ quadratic trapping potentials \citep{BEC-physV-2004-bretin}, dipolar interactions \citep{tang2022spectrally} or spinor condensates \citep{mithun2022existence}. Future extensions of the toolbox concern the implementation of domain decomposition methods and the use of  \texttt{PETSc} and \texttt{SLEPc}  libraries for this problem. The new parallel toolbox will make possible the computation (with a reasonable CPU time), of BdG modes for 3D BEC configurations without axial symmetry, such as BEC with $U$, $S$ vortices \citep{dan-2003-aft}, giant vortices \cite{dan-2005} or new computationally discovered  {\em exotic} states reported in very recent contributions \citep{panos-boulle-2022}.

 \section*{Acknowledgements}
 
 The authors acknowledge financial support from the French ANR grant ANR-18-CE46-0013 QUTE-HPC. Part of this work used computational resources provided by IDRIS (Institut du d{\'e}veloppement et des ressources en informatique scientifique) and CRIANN (Centre R{\'e}gional Informatique et d'Applications Num{\'e}riques de Normandie).
 The authors are grateful to P. K. Kevrekidis for stimulating discussions and warmly thank E. Charalampidis for his helpful input in validating numerical codes.





\end{document}